\documentclass{article}
\usepackage{iclr2027_conference,times}
\usepackage[T1]{fontenc}

\usepackage{hyperref}
\usepackage{url}

\usepackage{amsmath}
\usepackage{amssymb}
\usepackage{graphicx}
\usepackage{booktabs}
\usepackage{multirow}
\usepackage{array}
\usepackage{todonotes}
\usepackage{xcolor}
\usepackage{listings}
\usepackage{booktabs,multirow,graphicx,tabularx}
\usepackage{listings,xcolor,needspace,textcomp}
\usepackage{longtable}
\usepackage{booktabs,tabularx,array,fvextra}
\usepackage{booktabs}
\usepackage{multirow}
\newcolumntype{L}[1]{>{\raggedright\arraybackslash}p{#1}}

\iclrfinalcopy
\definecolor{Abdo}{RGB}{30, 100, 200}
\definecolor{zt}{RGB}{200, 30, 100}
\definecolor{Lars}{RGB}{200, 100, 30}

\newcommand{\meanstd}[2]{%
  \shortstack{#1\\[-1pt]{\scriptsize$\pm #2$}}%
}
\newcommand{\pairmeanstd}[4]{%
  \shortstack{#1\,/\,#3\\[-1pt]{\scriptsize$\pm #2\,/\,\pm #4$}}%
}

\title{Finding Emotions Where They Belong: Rethinking Audio Emotion Recognition through Masked Temporal Affective Grounding}
\author{
Abdelrahman Mohamed \\
Aalborg University \\
Pioneer Centre for AI
\And
Lars Kai Hansen \\
Technical University of Denmark \\
Pioneer Centre for AI
\And
Zheng-Hua Tan \\
Aalborg University \\
Pioneer Centre for AI
}

\begin{document}

\maketitle
\fancyhead{}
\renewcommand{\headrulewidth}{0pt}

\begin{abstract}
Audio emotion recognition (AER) typically assigns a single label to an entire recording, leaving the temporal scope of that label ambiguous when multiple speakers and affective events are present. We address this limitation by reformulating AER as a Temporal Affective Grounding (TAG) task that associates emotions with temporally bounded speech spans and vocal tone descriptions. To support this formulation, we curate temporally annotated versions of existing emotion recognition datasets and construct recordings containing two to four affective speech spans, including overlapping speech. Training in this longer format with a standard language-modeling objective can degrade both emotion recognition and temporal grounding performance, while tone descriptions can provide shortcuts for emotion prediction. To address these challenges, we introduce Masked Temporal Affective Grounding (M-TAG), a supervised training objective that combines full-sequence language modeling with emotion and timestamp cross-entropy losses under attention masking. The masking varies the context visible to emotion-label tokens to reduce reliance on shortcuts and improve generalization, while the timestamp loss incorporates a distance-aware weight to penalize larger temporal errors. We evaluate EMO-TAG, a model fine-tuned using our dataset and objective, on emotion recognition and affective temporal-grounding against three AER and audio-language baselines: Flamingo-Next, Audio-Reasoner, and AffectGPT. Our results show that existing models achieve limited affective temporal-grounding despite competitive emotion recognition performance. Across MELD and IEMOCAP, our model improves over our strongest baseline by an average of 10 percentage points in emotion accuracy on both the original and our refined test sets and by over 20 percentage points in Emotion-Duration F1 on our affective temporal-grounding benchmark. These improvements generalize to the MME-Emotion benchmark, where EMO-TAG achieves an average gain of five percentage points relative to the best baseline. Training and dataset-creation code and model checkpoints will be released upon publication.

\end{abstract}

\section{Introduction}
\label{sec:introduction}
Audio emotion recognition (AER) is commonly formulated as utterance-level
classification, where a model assigns a single emotion label to an entire
recording. Recent audio-language models (ALMs) extend this with
natural-language affective descriptions and rationales
\citep{zhao2025humanomni,lian2025affectgpt}, but the scope of the
predicted emotion still covers the whole sample. The generated
reasoning may describe broad acoustic or semantic cues without identifying
where the relevant cues occur, and the reasoning often does not align
with the predicted emotion \citep{rha2026emotion}. Therefore, a correct
predicted label does not necessarily indicate that the model has correctly
localized or interpreted the correct emotional cues in the input.
Additionally, the utterance-level assumption is poorly matched to the
content of commonly used emotion recognition benchmarks
\citep{busso2008iemocap,poria2019meld}. Inspection of these benchmarks
revealed that many recordings that are treated as a single sample contain
more than one speaker, overlapping speech, background laughter, noise,
or simply empty audio. Treating such recordings as ordinary single-label
examples leads to ambiguous evaluation and makes it unclear whether a
model recognizes affect from the intended speech, relies on unrelated
cues, or predicts the correct emotion despite insufficient audio evidence.

To address this limitation, we reformulate the AER problem as
\emph{Temporal Affective Grounding} (TAG).
Given an audio recording, a model generates a structured output
 of speech spans, where each span is a temporally bounded region of speech associated with a transcript, tone description, and emotion label. Here, temporal grounding refers to assigning an emotion label to a speech span. The boundaries indicate where the speech span of the emotion starts and ends, rather than the precise onset and offset of emotional
expression \citep{wang2023speech}.

To support this formulation, we curate a temporally annotated dataset
from popular public datasets
\citep{poria2019meld,busso2008iemocap,adigwe2018emotional}
and introduce a novel training objective. We filter and temporally annotate the datasets using voice activity detection (VAD) \citep{tan2020rvad} and then generate tone
descriptions of the filtered samples. We create refined single-label
training and testing datasets, multi-span training and testing datasets
containing two to four speech spans with known temporal and affective composition, and a challenging subset composed primarily of noisy samples,
such as those containing multiple speakers or overlapping speech, from
the original benchmarks.

Learning this structured task introduces a challenge beyond generating
the required format. Under a standard language-modeling objective,
emotion labels and timestamps account for only a small fraction of the
target sequence, so their contribution to the direction of the gradient
can be overshadowed by the more dominant transcription and tone tokens.
Furthermore, during teacher forcing \citep{williams1989learning}, the reference tone description
precedes the emotion label and can provide a shortcut for predicting it,
reducing the exposure to the audio. Timestamp prediction presents an
additional problem: categorical cross-entropy (CE) does not explicitly
distinguish a small temporal error from a distant prediction.

Therefore, we introduce \emph{Masked Temporal
Affective Grounding} (M-TAG), a supervised training objective that
combines full-sequence language modeling with masked emotion and
timestamp losses. The emotion loss provides additional
supervision on the first token of each emotion label, while the
timestamp loss weights the CE loss of timestamp tokens by the temporal distance
between the predicted and ground-truth time boundaries. When computing
these auxiliary losses, we selectively mask access to the preceding speech and tone tokens
to reduce the model's reliance on textual context. The language-modeling loss
continues to supervise the complete structured output using the full
causal context. Our training follows a curriculum that progresses from
single-span format learning to masked grounding and ultimately to multi-span
prediction.

We evaluate emotion recognition and affective temporal grounding against
audio-language and emotion-recognition baselines. On the constructed
benchmarks derived from MELD and IEMOCAP, our fine-tuned model, EMO-TAG, outperforms the strongest compared baseline on 
Emotion-Duration F1, a temporally-aware emotion recognition metric, by
23.12 and 35.17 percentage points, respectively. It further improves
emotion recognition on the original and refined single-span benchmarks, while also achieving the best emotion-recognition accuracy on
MME-Emotion benchmark \citep{zhang2026mme}.
In summary, our main contributions are as follows:
\begin{itemize}
\item We formulate AER as TAG, a structured prediction task that
    associates speech transcription, temporal boundaries, vocal tone,
    and emotion within individual speech spans.
\item We introduce M-TAG, a supervised objective combining language modeling with masked emotion and distance-weighted timestamp supervision for temporal affective grounding.
\item We construct temporally annotated single-span and multi-span
    data from existing AER datasets covering
    clean recordings, multiple affective events, and noisy subsets.
\end{itemize}

\begin{figure}[t]
\centering
\includegraphics[width=\linewidth]
{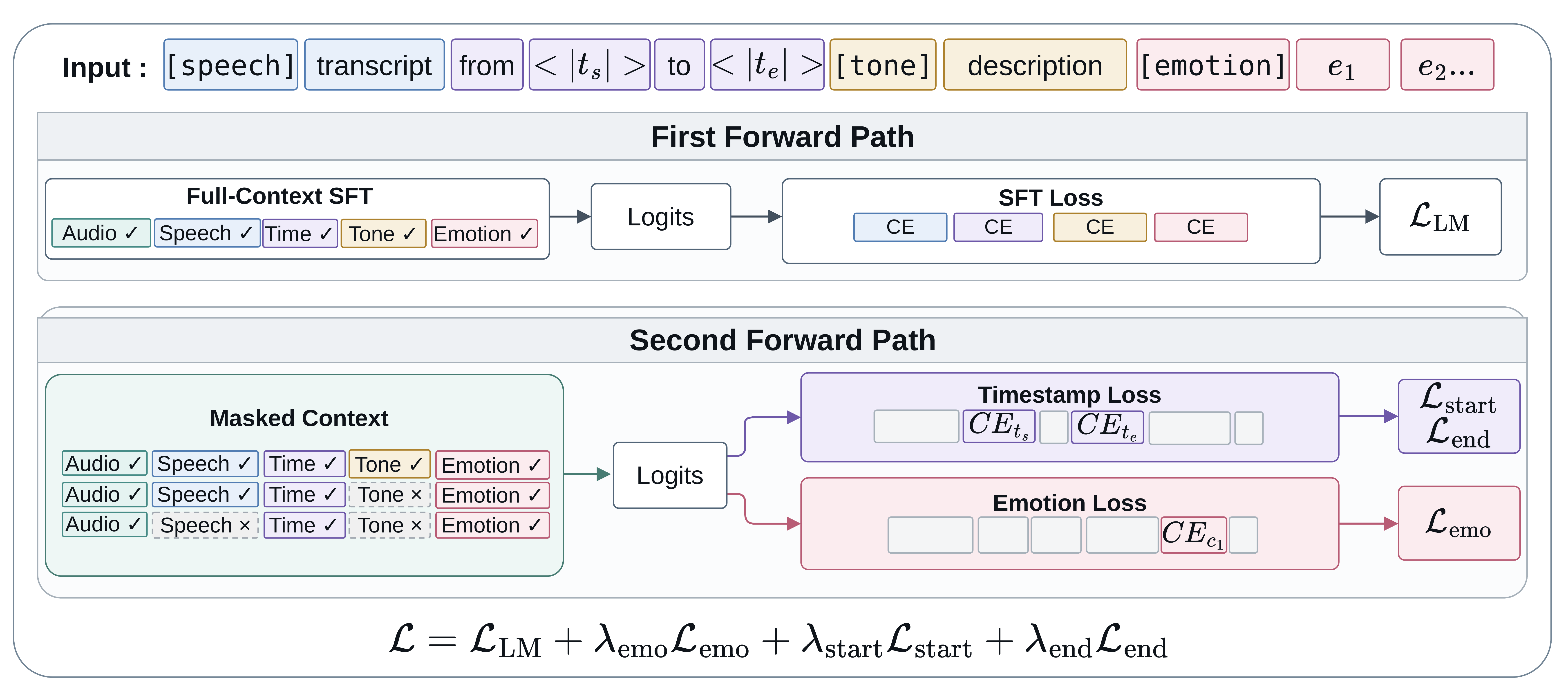}
\caption{Overview of our objective. The first forward path calculates the language modeling loss using the full sequence. The second forward pass calculate our auxiliary losses using varying context through attention masking. All logits are discarded in the second forward path except for the ones of timestamp tokens and the first token of the emotion label. }
\label{fig:method}
\end{figure}

\section{Related Work}
\noindent\textbf{Audio-language emotion recognition:}
General-purpose ALMs, such as Audio-Reasoner \citep{zhifei2025audio}, SALMONN \citep{tang2024salmonn} and Qwen2-Audio-Instruct \citep{chu2024qwen2}, support emotion recognition alongside speech transcription, audio captioning, and other audio-understanding tasks. More specialized models extend emotion recognition beyond categorical prediction. SECap \citep{xu2024secap} generates natural-language descriptions of emotions expressed in speech, while AffectGPT \citep{lian2025affectgpt} uses descriptive annotations to train multimodal models for both closed- and open-vocabulary emotion understanding. Other work integrates the acoustic cues as an input to the LLM. SpeechCueLLM \citep{wu2025beyond} converts utterance-level vocal characteristics into textual descriptions, which are passed with the input prompt, whereas VowelPrompt \citep{wang2026vowelprompt} focuses on the vowel letters and describes pitch and energy associated in the vowel segments, and use them to prompt the LLM. These methods aim to provide richer input descriptions and cues to the language model in order to improve the emotion recognition performance, while our model aims to train the LLM to be able to generate these additional cues.\\
\noindent\textbf{Emotion-aware training:}
Recent work considered extending token-level supervision to allow better emotion prediction and understanding.  AffectGPT-R1 \citep{lian2025affectgptr1} applies GRPO with emotion-wheel-based rewards \citep{plutchik1980general}, directly aligning training with open-vocabulary emotion evaluation. VowelPrompt \cite{wang2026vowelprompt} similarly combines SFT with RL to improve structured prosody-based reasoning, while Emotional Rationale Verifier \citep{rha2026emotion} further targets consistency between generated explanations and emotion predictions through an explanation-based reward. While these models use RL in order to improve performance, our approach focuses on leveraging the full capabilities of supervised learning with a carefully designed loss function. \\
\noindent\textbf{Temporal audio and emotion grounding:}
Audio Temporal grounding have been rapidly improving over the year. Audio Flamingo Next \citep{ghosh2026audioflamingonext} grounds intermediate reasoning steps in timestamps, while Timestamped Audio Captioning \citep{kumar2026tac} generates temporally localized descriptions of events in complex audio scenes. The closest work in emotion temporal grounding is Speech Emotion Diarization, introduced together with the Zaion Emotion Dataset (ZED) by \citet{wang2023speech}. ZED contains utterances with manually annotated precise emotion frames and aims to identify which emotion appears at each point in time. 
Our formulation differs in what the predicted boundaries represent. Creating a dataset large enough for ALLM fine-tuning with this level of annotation would require substantial human effort. We therefore adopt a simpler formulation that remains relevant to real-world settings. Rather than estimating the precise onset and offset of emotional expression within an utterance, our model aligns an emotion label with a temporally bounded speech span to which it represents.

\label{sec:related_work}

\section{Dataset Creation}
\label{sec:dataset_construction}
Our dataset build on existing public datasets, namely IEMOCAP \citep{busso2008iemocap}, MELD \citep{poria2019meld}, and Emov-DB \citep{adigwe2018emotional}. We filter these datasets from empty audio and ambiguous speech. We use a pipeline to check the audio, refines temporal boundaries, and reject inconsistent samples. Figure~\ref{fig:data_pipeline} in Appendix \ref{app:dataset_details} shows a complete overview of our pipeline.

\subsection{Audio Preprocessing}
\label{sec:benchmark_audit}
We use the voice activity detection (VAD) tool rVAD
\citep{tan2020rvad} to identify empty audio samples and segment
valid ones into speech and non-speech regions. We then address dataset-specific issues. The first major problem
was in MELD. As the dataset is derived from the sitcom  \emph{Friends}, many samples contain background laughter tracks. This can give models misleading cues
and affect the interpretation of these samples. We inspect the non-speech regions identified by VAD
and remove only those with high energy, aiming to remove laughter
tracks while preserving natural pauses and sighs. The second prominent issue was the presence of multiple speakers
and overlapping speech in the same recording in both IEMOCAP and MELD.
To address this, we transcribe each detected speech region with
WhisperX \citep{bain2023whisperx} and use forced alignment with the ground-truth transcript. We then check whether the words in each
segment are part of the original utterance. We consider a segment
unrelated if it contains no words matching the ground-truth
transcript. We then concatenate the remaining speech segments
and run WhisperX with forced alignment again to trim any remaining
irrelevant segments. If, after this final step, the transcript
contains two or more consecutive insertions
(extra words that are not part of the utterance) or
three or more replacements (different words that may have been
misrecognized by WhisperX), we reject the sample. Further details can be found in Appendix \ref{app:utterance_cleaning}

\subsection{Tone Annotation}
\label{sec:tone_annotation}
We use FlamingoNext \citep{ghosh2026audioflamingonext} to transcribe
each recording before describing the speaker's vocal tone. We retain
samples only when the generated transcription has at least $60\%$
text similarity to the reference transcript. Qwen3 extracts the speech
and tone descriptions from FlamingoNext's full output, converts
them into our target format, and checks consistency between the tone
and ground-truth emotion. Flagged samples are re-annotated by
FlamingoNext with the ground-truth emotion supplied in the prompt, samples
that remain inconsistent are discarded.\\
The initial tone descriptions vary in length and structure, which
degraded performance in our experiments. Inspired by
\citet{wu2025beyond,wang2026vowelprompt}, we refine and normalize
these descriptions using low-level acoustic features. We divide
each clip into five segments and extract features using openSMILE
\citep{eyben2010opensmile}. We also compute speaker-specific reference
statistics for features such as pitch, intensity, and speaking pace.
ChatGPT receives the initial tone description, segment-level features,
and speaker references to produce a concise, normalized tone
description. Tone examples are provided in Appendix \ref{app:tone_examples} and prompts are in Appendix \ref{app:tone_extraction_prompts}.

\subsection{Multi-Span Construction}
\label{sec:multispan_construction}

We construct two-, three-, and four-span examples by concatenating cleaned single-span recordings without replacement. For each dataset, the gap between consecutive spans is sampled uniformly between $0.5$ and $2.0$ seconds from silence regions extracted from recordings in the same dataset. This avoids inserting digital zeros and preserves dataset-specific background characteristics. We similarly construct overlapping samples by uniformly sampling an overlap between $0.5$ and $2.0$ seconds between segments. We select these segments from a pool of audio clips with a minimum duration of three seconds. More details can be found in Appendix \ref{app:ann}.

\section{Method}
\label{sec:method}

\subsection{Task Formulation}
\label{sec:task_formulation}
Given an audio recording $x$, TAG predicts
a sequence of speech spans $\mathcal{S}=(S_1,\ldots,S_K)$, where
$S_k=(\tau_k^{\mathrm{start}},\tau_k^{\mathrm{end}},u_k,c_k,z_k)$.
Here, $\tau_k^{\mathrm{start}}$ and $\tau_k^{\mathrm{end}}$ denote the
temporal boundaries of span $k$, $u_k$ is its transcription, $c_k$ is
a natural-language description of the vocal tone, and $z_k$ is the
emotion label. The boundaries identify the speech span to which the
emotion label applies. Our training samples contain
$K\in\{1,2,3,4\}$ spans. Each span is formatted as follows:
\begin{quote}
\small\ttfamily
<speakerX>: [speech] "transcript" from <|start|>s to\par <|end|>s, 
[tone] tone description, [emotion] emotion.
\end{quote}
Let $y=(y_1,\ldots,y_T)$ denote the complete serialized target sequence.
The model generates this sequence autoregressively as
$p_\theta(y\mid x)=\prod_{t=1}^{T}p_\theta(y_t\mid x,y_{<t})$.\\
Following \citet{kumar2026tac}, we represent temporal boundaries using
learnable timestamp tokens added to the model vocabulary. We quantize
timestamps with a step size of $\Delta=0.25$ seconds, covering recordings
up to $\tau_{\max}=30$ seconds. A continuous timestamp $\tau$ is mapped
to $Q(\tau)=\min\{\tau_{\max},
\Delta\,\operatorname{round}(\tau/\Delta)\}$.

\subsection{Masked Temporal Affective Grounding}
\label{sec:mtag_objective}

\paragraph{Full-context Supervision:}
We compute the language-modeling loss over the complete target sequence: $\mathcal{L}_{\mathrm{LM}} = -\frac{1}{T}\sum_{t=1}^{T}\log p_\theta(y_t\mid x,y_{<t})$.\\
As in vanilla language modeling, only target tokens contribute to the
loss, while prompt, audio, and padding tokens are ignored. 
This objective
provides the supervision for transcription, timestamps, tone descriptions, and
emotion labels in the target output format.

\paragraph{Masked Emotion Supervision.}
The auxiliary emotion loss places additional emphasis on the emotion
label of each span. There are two challenges when implementing this loss.
First, some emotion labels are represented by multiple tokens, so once
the first token is predicted, the subsequent tokens often have very small
CE losses. Averaging the CE loss over the complete label can dilute the
supervision signal for multi-token emotions compared
with single-token ones.
Since the first token of each emotion label in our vocabulary is unique,
we compute the auxiliary emotion loss using only the first token of
each emotion label.

The second challenge is that the tone description immediately precedes
the emotion label. This introduces a shortcut where the model can directly
predict the emotion label by focusing only on the tone description.
This issue becomes more prominent in our setting when we introduce
multiple spans. Inspecting attention weights during training revealed
that attention from emotion label positions to audio tokens weakens over
time, especially as the distance between these positions and the audio
tokens increases. To reduce this dependency, we vary the context available
when predicting the emotion label. With probability $p_{\mathrm{mask}}$,
we sample one of two strategies:
\emph{tone masking}, which blocks attention to tone tokens, or
\emph{tone-and-speech masking}, which blocks attention to both speech
and tone tokens. This encourages greater attention to more distant
tokens, including the audio tokens.
Let $q_k^{\mathrm{emo}}$ denote the position of the first emotion-label
token in span $k$, and let $M$ denote the sampled attention mask.
The per-span emotion loss is
$\ell_k^{\mathrm{emo}} = -\log p_\theta\!\left(y_{q_k^{\mathrm{emo}}}\mid x,y_{<q_k^{\mathrm{emo}}};M\right)$.

Context masking is applied in a second forward pass through the attention mask, similar to causal attention \citep{vaswani2017attention}. The selected tokens remain in
the input sequence, but attention from the emotion label to these
tokens is blocked. As shown in Figure \ref{fig:method}, the full-context forward
pass used for $\mathcal{L}_{\mathrm{LM}}$ is unaffected and provides the full-sequence supervision. This formulation does not only provide a
regularization effect but also allow datasets with only emotion
annotations to be included in training. This allows using emotion
recognition datasets without requiring additional speech or tone
annotations. We demonstrate this use case in
Appendix~\ref{app:additional_experiments}.

\paragraph{Masked Temporal Supervision:}
For our temporal loss, we supervise the start and end timestamp tokens
of each span. Timestamp prediction should depend only on the audio recording and previous timestamp tokens. However, masking the tone and speech in every training step would frequently cost an expensive extra forward pass. To avoid this, we use the same second forward pass and the attention mask $M$ of the emotion loss.
We additionally weight timestamp tokens CE relative to the
distance between the predicted and ground-truth boundaries. Let
$\hat{\tau}_k^b$ and $\tau_k^b$ denote the predicted and ground-truth
timestamps for boundary $b$ of span $k$. We define the absolute temporal
error $d_k^b$ and its distance-dependent weight as
\begin{equation}
d_k^b
=
\left|\hat{\tau}_k^b-\tau_k^b\right|,
\qquad
w_k^b
=
\operatorname{clip}\left(
\frac{d_k^b}{\delta},
c_{\min},
c_{\max}
\right).
\label{eq:timestamp_distance_weight}
\end{equation}
Here, $\delta$ is the distance pivot and $c_{\min}$ and $c_{\max}$ bound the coefficient. Let $q_k^b$ denote the target-token position of the boundary $b$.
The corresponding per-span temporal loss is
\begin{equation}
\ell_k^b
=
-w_k^b\log p_\theta\!\left(
y_{q_k^b}
\mid x,y_{<q_k^b};M
\right),
\qquad b\in\{\mathrm{start},\mathrm{end}\}.
\label{eq:masked_timestamp_loss}
\end{equation}

\paragraph{Full objective:}
For a sample containing $K$ spans, our auxiliary losses are aggregated as:
\begin{equation}
\mathcal{L}_{\mathrm{emo}}
=\sum_{k=1}^{K}\ell_k^{\mathrm{emo}},
\qquad
\mathcal{L}_{\mathrm{start}}
=\sum_{k=1}^{K}\ell_k^{\mathrm{start}},
\qquad
\mathcal{L}_{\mathrm{end}}
=\sum_{k=1}^{K}\ell_k^{\mathrm{end}}.
\label{eq:span_loss_aggregation}
\end{equation}
and the final training objective becomes:
\begin{equation}
\mathcal{L}
=
\mathcal{L}_{\mathrm{LM}}
+
\lambda_{\mathrm{emo}}\mathcal{L}_{\mathrm{emo}}
+
\lambda_{\mathrm{start}}\mathcal{L}_{\mathrm{start}}
+
\lambda_{\mathrm{end}}\mathcal{L}_{\mathrm{end}}.
\label{eq:mtag_loss}
\end{equation}
The $\lambda$ coefficients control the contributions of the full-sequence language
modeling loss, emotion recognition loss, and the two temporal boundaries loss.

\subsection{Training Curriculum}
\label{sec:training_curriculum}


\noindent\textbf{Stage I: Format learning.}
We first train on single-span samples using only
$\mathcal{L}_{\mathrm{LM}}$. This stage teaches the model to generate
our structured sequence of speech, timestamps, tone, and emotion.\\
\noindent\textbf{Stage II: Masked grounding.}
Once the model generation follows our format, we introduce the masked emotion and temporal losses as in Equation~\eqref{eq:mtag_loss} with $k=1$. This stage emphasizes emotion and
boundary accuracy while varying the context available to the auxiliary
predictions. We randomly add noise or silence in the range of $0.5$ to $1$ seconds to avoid the model simply predicting the start and the end of the audio clip.\\
\noindent\textbf{Stage III: Multi-span temporal learning.}
Finally, we introduce recordings containing two, three, and four speech
spans. The model learns to distinguish multiple affective events and associate
each predicted emotion with its corresponding temporal region. The
auxiliary losses are accumulated over the annotated spans as defined
in Equation~\eqref{eq:span_loss_aggregation}.

\section{Experiments}
\label{sec:experiments}
\noindent\textbf{Baselines:}
\label{sec:temporal_baseline}
We compare against FlamingoNext \citep{ghosh2026audioflamingonext}, Audio-Reasoner \citep{zhifei2025audio}, and AffectGPT \citep{lian2025affectgpt}. Both FlamingoNex and Audio-Reasoner can generate timestamped audio descriptions, while AffectGPT is specifically trained for emotion recognition. Since neither of the temporal models natively follows our structured output format, we retain each model's default prompting style and explicitly ask it to predict timestamps and an emotion from our predefined label set. We then pass the generated reasoning and the final response to Qwen3 \citep{yang2025qwen3}, which extracts the predicted emotion and temporal boundaries into our format. To avoid penalizing differences in generation style, as shown in Appendix \ref{app:prompt_evaluation_extractor}, the extraction prompt is deliberately permissive and extract any mention of the emotion and their timestamp if predicted.
Since AffectGPT is not trained to predict temporal spans, we evaluate it only on emotion recognition using the evaluation script and the final training checkpoint provided in its original codebase.\\
\noindent\textbf{Evaluation Datasets:}
\label{sec:evaluation_metrics}
We use 8 main emotions in our training and evaluation across all datasets, mainly neutral, happiness, sadness, anger, surprise, fear, disgust, and frustration. We use \emph{Raw} when we point to the original benchmark, \emph{Refined} when we point to our cleaned version of the benchmark, \emph{Noisy} when we point to the noisy samples of the original benchmark, and \emph{multi-span} when we point to our curated multiple spans dataset. More details on the datasets composition is in Appendix \ref{app:dataset_details}.
\begin{figure}[t]
\centering
\includegraphics[width=\linewidth]
{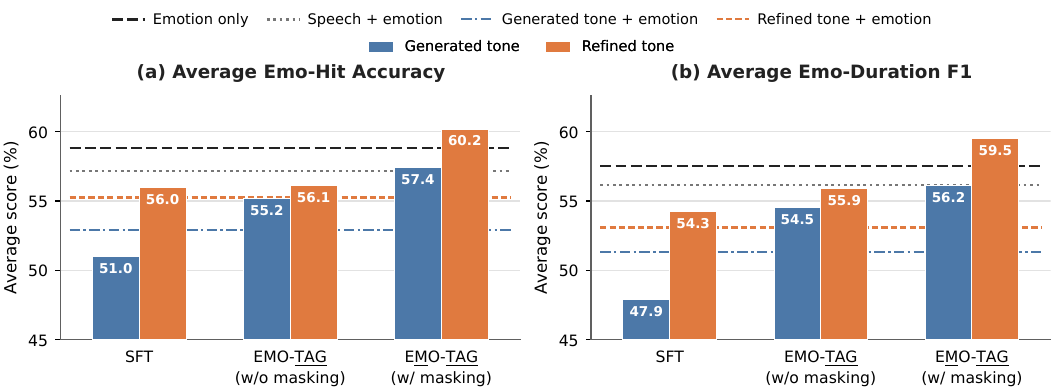}
\caption{Average performance across context settings. Bars compare generated tone (before low-level features) and refined tone (after low-level features) with full-context, while dashed lines show the partial-context SFT settings. All runs share the same random seed and hyperparameters.}
\label{fig:context_ablation}
\end{figure}

\subsection{Evaluation Metrics}
\label{sec:evaluation_metrics}
Our task requires evaluating \emph{what} emotion is
predicted and \emph{when} it is predicted. Therefore, we report three evaluation metrics: emotion segment accuracy, emotion hit accuracy, and Emotion-Duration F1. Emotion segment accuracy provides the conventional utterance-level
comparison, emotion hit accuracy checks if the model predicted the correct emotion in any predicted span, and Emotion-Duration F1 evaluates whether the correct
emotion is assigned to the correct temporal region.\\
\noindent\textbf{Emotion Segment Accuracy:}
Dedicatedly for single-span samples, we report standard emotion classification accuracy. Let $z_i$ denote the ground-truth emotion label of sample $i$, and let $\hat{z}_i$ denote the first valid emotion label generated by the model. Emotion segment accuracy is computed as: $\mathrm{Acc}_{\mathrm{emo}} = \frac{1}{N}\sum_{i=1}^{N}\mathbf{1}[\hat{z}_i=z_i]$, 
where $N$ is the total number of evaluated samples and
$\mathbf{1}[\cdot]$ equals one when the predicted emotion matches the ground-truth emotion
and zero otherwise.\\
\noindent\textbf{Emotion Hit Accuracy:}
For multi-span and single-span recordings containing overlapped or multiple speakers, a temporally-aware model would generate more than one emotion prediction. Evaluating only the first prediction can therefore penalize the model even when the ground-truth emotion is correctly identified in the appropriate temporal region.
We define an emotion hit for a ground-truth span when its emotion label appears among the predicted emotion labels associated with that span. Emotion hit accuracy is formulated as:
\begin{equation}
\mathrm{Hit}_{\mathrm{emo}}
=
\frac{1}{\sum_{i=1}^{N} K_i}
\sum_{i=1}^{N}
\sum_{k=1}^{K_i}
\mathbf{1}
\left[
z_{i,k}
\in
\widehat{\mathcal{Z}}_{i,k}
\right],
\label{eq:emotion_hit}
\end{equation}

where $K_i$ is the number of ground-truth spans in sample $i$, $z_{i,k}$ is the emotion of the $k$-th span, and $\widehat{\mathcal{Z}}_{i,k}$ denotes the set of predicted emotions associated with that span. For single-span samples, $K_i=1$, the metric checks whether the ground-truth emotion appears anywhere among the model's predicted emotion spans. For multi-span samples, the comparison is performed independently for each ground-truth span, preventing an emotion predicted for one span from being credited to another. \\
\noindent\textbf{Emotion-Duration F1:}
To evaluate temporal grounding, i.e., whether an emotion is assigned to the correct temporal speech region, we divide each recording into non-overlapping $0.25$-second frames, matching the resolution of our timestamp vocabulary. For each emotion class $c$, let $\mathcal{G}_c$ and $\mathcal{P}_c$ denote
the sets of ground-truth and predicted frames assigned to that emotion,
respectively. We compute duration precision and recall as

\begin{equation}
\begin{aligned}
P_{\mathrm{dur}}
&=\frac{\sum_c\lvert\mathcal{G}_c\cap\mathcal{P}_c\rvert}
{\sum_c\lvert\mathcal{P}_c\rvert},
&
R_{\mathrm{dur}}
&=\frac{\sum_c\lvert\mathcal{G}_c\cap\mathcal{P}_c\rvert}
{\sum_c\lvert\mathcal{G}_c\rvert}.
\end{aligned}
\label{eq:duration_precision}
\end{equation}

The intersection $\mathcal{G}_c \cap \mathcal{P}_c$ represents frames which both the predicted emotion label and its temporal location match the ground truth. Duration precision measures the proportion of predicted emotion frames that follows the ground-truth span, while duration recall measures the proportion of ground-truth emotion frames that are correctly recovered. Predictions with the correct temporal location but an incorrect emotion label do not contribute to the intersection and are therefore penalized. For example, if anger is annotated from $2.0$ to $4.0$ seconds but the model
predicts anger from $2.5$ to $3.5$ seconds, the overlapping interval is
correctly recovered, while the missed portions from $2.0$--$2.5$ and
$3.5$--$4.0$ seconds reduce recall. Predicting anger beyond the
annotated interval would reduce precision. The reported Emo-Duration F1 is calculated as the harmonic mean of
$P_{\mathrm{dur}}$ and $R_{\mathrm{dur}}$.








\subsection{Results}
\begin{table}[t]
\caption{Emotion recognition across original, refined, and noisy MELD
and IEMOCAP evaluation sets.  Scores are reported as Emotion Segment Accuracy / Emo-Hit Accuracy (\%). Results for EMO-TAG are averaged
over three runs, with sample standard deviations shown beneath the means.}
\label{tab:model_comparison_single}
\centering
\footnotesize
\setlength{\tabcolsep}{3.5pt}
\renewcommand{\arraystretch}{1.08}

\resizebox{\linewidth}{!}{%
\begin{tabular}{@{}lcccccc@{}}
\toprule
& \multicolumn{3}{c}{\textbf{IEMOCAP}}
& \multicolumn{3}{c}{\textbf{MELD}} \\
\cmidrule(lr){2-4}
\cmidrule(lr){5-7}

\textbf{Model}
& \textbf{Original}
& \textbf{Refined}
& \textbf{Noisy}
& \textbf{Original}
& \textbf{Refined}
& \textbf{Noisy} \\
\midrule

Audio-Reasoner
& 40.76 / 44.65
& 38.71 / 42.24
& 35.26 / 36.41
& 40.15 / 44.14
& 40.36 / 43.02
& 39.71 / {49.37} \\

AffectGPT
& 44.36 / 44.36
& 47.10 / 47.10
& 40.98 / 40.98
& 37.33 / 37.33
& 40.44 / 40.44
& 35.82 / 35.82 \\

FlamingoNext
& 37.58 / 49.22
& 36.49 / 43.55
& 39.37 / 47.45
& 44.52 / 53.45
& 48.91 / 54.91
& 33.82 / 45.21 \\

Qwen2-Audio-Instruct
& 43.52 / 43.52
& 41.94 / 41.94
& 43.66 / 43.66
& 38.88 / 39.08
& 36.68 / 36.68
& 30.15 / 30.15 \\

EMO-TAG (ours)
& \pairmeanstd{\textbf{65.62}}{1.53}{\textbf{66.39}}{1.52}
& \pairmeanstd{\textbf{65.32}}{0.99}{\textbf{65.78}}{1.06}
& \pairmeanstd{\textbf{63.51}}{1.70}{\textbf{63.67}}{1.56}
& \pairmeanstd{\textbf{54.65}}{0.37}{\textbf{55.73}}{0.33}
& \pairmeanstd{\textbf{57.91}}{0.98}{\textbf{58.12}}{0.91}
& \pairmeanstd{\textbf{44.12}}{2.16}{\textbf{51.47}}{1.08} \\

\bottomrule
\end{tabular}%
}
\end{table}
\noindent\textbf{Emotion Recognition:}
Table~\ref{tab:model_comparison_single} reports single-span emotion
recognition performance. On IEMOCAP, our model achieves the highest
emotion segment accuracy across the Original, Refined, and Noisy
sets, with mean accuracies of $65.62\%$, $65.32\%$, and $63.51\%$,
respectively. This exceeds AffectGPT by at least $18.22$ percentage
points, despite AffectGPT receiving both the utterance transcript
and audio as input. Our model also achieves the highest results on MELD Original and
Refined, with mean segment accuracies of $54.65\%$ and $57.91\%$, compared with $48.91\%$ and $44.52\%$ for the strongest baseline,
FlamingoNext. On MELD Noisy, it obtains an average segment accuracy
of $44.12\%$ and an Emo-Hit Accuracy of $51.47\%$, compared with
$39.71\%$ and $49.37\%$ for Audio-Reasoner. The gap between emotion
segment and emotion hit accuracy on the noisy subset of MELD
highlights the noisy nature of this subset. We further evaluate
emotion recognition on MME-Emotion \citep{zhang2026mme}.
Table~\ref{tab:model_comparison_multispan} compares our model
against the baseline results reported in this benchmark.
EMO-TAG achieves the highest mean emotion segment accuracy across all
three emotion recognition subsets, reaching $58.20\%$ on Noise,
$33.56\%$ on Wild, and $45.27\%$ on Lab. It outperforms
FlamingoNext, the strongest baseline across these subsets,
by $7.43$, $1.06$, and $7.64$ percentage points, respectively.
EMO-TAG also outperforms Audio-Reasoner on all three subsets,
despite Audio-Reasoner receiving both the audio and the
utterance transcript as input, whereas EMO-TAG uses only the audio.\\
\noindent\textbf{Emotion Grounding:}
Table~\ref{tab:model_comparison_multispan} evaluates whether models
can pair each ground-truth emotion with the appropriate temporal
speech span in recordings containing multiple affective events.
On MELD Multi-Span, EMO-TAG reaches an Emo-Duration F1 of $54.55\%$,
exceeding Audio-Reasoner, the strongest baseline at $31.43\%$, by
$23.12$ percentage points. The margin increases on IEMOCAP Multi-Span,
where EMO-TAG achieves $63.72\%$, compared with $28.55\%$ for
Audio-Reasoner, a gain of $35.17$ percentage points. EMO-TAG also obtains the highest Emo-Hit Accuracy on both datasets,
reaching $57.09\%$ on MELD and $61.22\%$ on IEMOCAP, compared with
the strongest baseline results of $46.03\%$ for Audio-Reasoner and
$42.53\%$ for FlamingoNext, respectively.\\
\noindent\textbf{Effect of Context:}
Figure~\ref{fig:context_ablation} shows the effect of varying the context used during SFT training and the impact of applying our masking objective.
Surprisingly, training with the emotion label alone is highly effective for emotion recognition and temporal grounding. Emotion-only training achieves Emo-Hit accuracy and Emo-Duration F1 of $58.8\%$ and $57.5\%$ averaged across MELD and IEMOCAP, respectively. Adding just speech leads to an average degradation in emotion recognition and temporal grounding of $1.7$ and $1.4$ percentage points. The original generated tone has a more pronounced negative effect, using it with the emotion label sharply degrades performance by $5.9$ and $6.2$ percentage points. However, refining the tone with the low-level features recovers $2.4$ and $1.8$ points of this degradation. Using SFT training with the full context leads to the worst performance when using the original generated tone. This performance significantly improves by $6.0$ and $6.4$ points when using the refined tone, achieving slightly higher performance than using refined tone alone. Introducing  our auxiliary loses notably improves the original tone SFT performance, with a modest improvement when switching to the refined tone. Nonetheless, the performance remains inferior to training with the emotion label only. With our full M-TAG objective, the performance gain is substantial across the board. Emo-TAG achieves an average improvement in Emo-Hit accuracy and Emo-Duration F1 over SFT training of $7.4$ and $8.3$ points with the original tone, and $4.2$ and $5.2$ points with the refined tone. Additionally, it achieves gains of $1.4$ and $2.0$ points over training with the emotion label only, making it the only setting that outperforms emotion-only training. Detailed ablation and qualitative results are provided in Appendix \ref{app:additional_ablations} and Appendix \ref{app:qualitative}.


\begin{table}[t]
\caption{Comparison on the MELD and IEMOCAP multi-span evaluation sets and the
MME-Emotion benchmarks. Avg. denotes the unweighted average across Noise, Wild, and Lab.
Results marked with $\ddagger$ are
taken from \citet{zhang2026mme}. EMO-TAG Results are averaged over three runs.}
\label{tab:model_comparison_multispan}
\centering
\footnotesize
\setlength{\tabcolsep}{2.5pt}
\renewcommand{\arraystretch}{1.08}

\begin{tabular*}{\linewidth}{@{\extracolsep{\fill}}lcccccccc@{}}
\toprule
& \multicolumn{4}{c}{\textbf{MME-Emotion}}
& \multicolumn{2}{c}{\textbf{MELD Multi-Span}}
& \multicolumn{2}{c}{\textbf{IEMOCAP Multi-Span}} \\
\cmidrule(lr){2-5}
\cmidrule(lr){6-7}
\cmidrule(lr){8-9}

\multirow{2}{*}{\textbf{Model}}
& \multicolumn{4}{c}{\textbf{Emo-Segment Accuracy} (\%)}
& \multirow{2}{*}{\shortstack{\textbf{Emo-Duration}\\\textbf{F1} (\%)}}
& \multirow{2}{*}{\shortstack{\textbf{Emo-Hit}\\\textbf{Acc.} (\%)}}
& \multirow{2}{*}{\shortstack{\textbf{Emo-Duration}\\\textbf{F1} (\%)}}
& \multirow{2}{*}{\shortstack{\textbf{Emo-Hit}\\\textbf{Acc.} (\%)}} \\

& \textbf{Noise}
& \textbf{Wild}
& \textbf{Lab}
& \textbf{Avg.}
& & & & \\
\midrule

Audio-Reasoner
& 47.60$^{\ddagger}$
& 28.80$^{\ddagger}$
& 32.20$^{\ddagger}$
& 36.20
& 31.43 & 46.03
& 28.55 & 34.68 \\

AffectGPT
& 16.80$^{\ddagger}$
& 8.30$^{\ddagger}$
& 18.00$^{\ddagger}$
& 14.37
& -- & --
& -- & -- \\

FlamingoNext
& 50.77 & 32.50 & 37.63
& 40.30
& 28.33 & 42.53
& 24.33 & 42.53 \\

Qwen2-Audio-Instruct
& 38.80$^{\ddagger}$
& 22.30$^{\ddagger}$
& 20.40$^{\ddagger}$
& 27.17
& 20.00 & 15.00
& 22.00 & 12.30 \\

EMO-TAG (ours)
& \meanstd{\textbf{58.20}}{1.95}
& \meanstd{\textbf{33.56}}{0.58}
& \meanstd{\textbf{45.27}}{2.69}
& \meanstd{\textbf{45.68}}{1.74}
& \meanstd{\textbf{54.55}}{1.38}
& \meanstd{\textbf{57.09}}{1.19}
& \meanstd{\textbf{63.72}}{2.42}
& \meanstd{\textbf{61.22}}{1.95} \\

\bottomrule
\end{tabular*}
\end{table}


\section{Discussion}
The multi-span results in Table \ref{tab:model_comparison_multispan} demonstrate the major gap with existing models between recognizing which emotions are present in a recording and assigning those emotions to the correct speech span. Existing audio-language models frequently recover a relevant emotion while achieving a substantially lower Emotion-Duration F1, indicating that these models did not attend to the correct speech span. EMO-TAG improves this grounding gap while achieving superior emotion-recognition performance. Our model consistently outperforms the compared models across all evaluated datasets, including the MME-Emotion subsets, despite using only audio, while some baselines additionally receive the utterance transcript. Furthermore, as shown in Table \ref{tab:dataset_statistics} in Appendix \ref{app:dataset_details}, we use only 12k unique training samples to achieve these results. This validates our TAG task formulation and the possibility of implementing it in a way that does not harm recognition performance. 

The context ablation in Figure \ref{fig:context_ablation} shows the counter-intuitive observation that providing more context does not necessarily produce a stronger model.
Emotion supervision alone achieved the best performance in our ablation, only second to the full objective, while the full-context SFT model gave the worst overall performance. The additional speech and tone description, although more informative to humans, seems to make learning this task harder. Additionally, as shown in Appendix \ref{app:tone_examples}, the initial tone descriptions provided more details about how each utterance was spoken, but varied considerably from one sample to another. Under the language-modeling objective, this inconsistency may have made the target descriptions more complicated to learn, contributing to lower performance. In contrast, the refined tone descriptions were more consistent and concrete, yielding much better performance. We therefore hypothesize a trade-off between the level of detail in the intermediate reasoning and the amount of available training data, more detailed reasoning may require more data to be learned reliably and may otherwise degrade performance.

One more takeaway from our experiments is the untapped potential of supervised learning. We argue that GRPO \citep{shao2024deepseekmath} should not be treated as the go-to approach when the goal is to improve a particular metric within a structured output. In our preliminary experiments, applying GRPO after SFT without our objective had modest performance improvements. Unlike hard reasoning tasks such as math or coding, our task requires concise generation, as there are not many ways to transcribe speech and describe a tone. RL finetuning encourages increasing the generation length \citep{yu2025dapo, liu2024understanding, mohamed2025devil}, and leads to unnecessary continuation in our case, resulting in less attention to the input audio. More details on our GRPO formulation and limitations are in Appendix \ref{app:grpo} and Appendix \ref{app:limitations}.

\section{Conclusion}
In this paper, we demonstrated the feasibility of formulating audio emotion recognition (AER) as a temporal affective grounding task. We achieved this by creating temporally annotated datasets containing a variety of span settings and introducing a masked training objective that emphasizes emotion recognition and temporal grounding performance without hindering the language-modeling objective. Our results showed the reliability of our  approach by improving both temporal grounding and emotion recognition against widely used baselines. Additionally, we identified and addressed limitations in the current AER training and evaluation datasets. In general, our findings show that explicitly modeling emotion recognition as a temporal task using our objective provides a more practical formulation without harming recognition performance.

\subsection*{Reproducibility statement}
To promote transparency and reproducibility, we plan to release our training and dataset-creation code, datasets (if copyright allows), model checkpoints, and Weights \& Biases (W\&B) artifacts containing full training details and evaluation results throughout the training process. By making these resources publicly available, we hope to encourage other researchers to do the same, helping make research more transparent and reproducible and allowing the field to move forward.
\subsection*{AI Use Statement}
AI tools, specifically ChatGPT, were used solely to assist with coding and manuscript refinement. For coding, ChatGPT was used to help implement functions according to clearly specified requirements and intended functionality. Generative AI was also used during data annotation, as described in Section \ref{sec:dataset_construction} and Appendix \ref{app:ann}. FlamingoNext generated the initial speech and tone descriptions, which were then extracted and structured with Qwen3. ChatGPT was used to refine the tone descriptions using the initial descriptions and acoustic features. Qwen3 was also used to convert baseline model outputs into a structured format for evaluation.
For manuscript preparation, ChatGPT was used to correct grammatical errors, improve wording, assist with table editing, and create early figure drafts. AI was not used to identify the research gap, formulate the proposed methodology, develop the underlying research ideas, interpret scientific contributions, or generate any novel aspects of the work. All research concepts, methodological decisions, and scientific contributions were developed by the authors.

% ...
%
% ...

\bibliography{iclr2027_conference}
\bibliographystyle{iclr2027_conference}
\newpage
%
%

\definecolor{appendixPromptBackground}{gray}{0.97}
\definecolor{appendixPromptBorder}{gray}{0.78}
\lstdefinestyle{appendixprompt}{
  basicstyle=\ttfamily\fontsize{8.5}{9.5}\selectfont,
  breaklines=true,
  breakatwhitespace=true,
  breakindent=1em,
  columns=fullflexible,
  keepspaces=true,
  showstringspaces=false,
  upquote=true,
  frame=single,
  framerule=0.3pt,
  framesep=5pt,
  xleftmargin=6pt,
  xrightmargin=6pt,
  backgroundcolor=\color{appendixPromptBackground},
  rulecolor=\color{appendixPromptBorder},
  aboveskip=4pt,
  belowskip=9pt,
  literate={’}{{\textquoteright}}1 {“}{{``}}1 {”}{{''}}1 {–}{{--}}1
}

\newcommand{\qualmodel}[1]{%
  \begingroup\setlength{\fboxsep}{4pt}%
  \colorbox{appendixPromptBackground}{%
    \parbox{\dimexpr\linewidth-8pt\relax}{\strut\textbf{#1}}}%
  \endgroup}

\clearpage
\appendix
\section*{Appendices}
\phantomsection
\label{app:contents}

\begingroup
\small
\setlength{\tabcolsep}{4pt}
\renewcommand{\arraystretch}{1.04}
\newcommand{\appmainentry}[2]{%
  \hyperref[#1]{\textbf{\ref*{#1}}} &
  \hyperref[#1]{\textbf{#2}} &
  \hyperref[#1]{\pageref*{#1}} \\}
\newcommand{\appsubentry}[2]{%
  \hspace*{0.7em}\hyperref[#1]{\ref*{#1}} &
  \hspace*{0.7em}\hyperref[#1]{#2} &
  \hyperref[#1]{\pageref*{#1}} \\}
\begin{tabularx}{\linewidth}{@{}lXr@{}}
\toprule
\textbf{Section} & \textbf{Contents} & \textbf{Page} \\
\midrule
\appmainentry{sec:implementation}{Implementation Details}
\appsubentry{app:implementation_model}{Model}
\appsubentry{app:utterance_cleaning}{Audio Preprocessing}
\appsubentry{app:ann}{Target Annotations}
\appsubentry{app:multi_span_design}{Multi-Span Design Choices}
\appmainentry{app:additional_ablations}{Training Ablation}
\appsubentry{app:context_and_masking}{Context Information and Masking}
\appsubentry{app:masking_probability}{Masking Probability}
\appsubentry{app:trainable_components}{Trainable Components}
\appmainentry{app:additional_experiments}{Additional Experiments}
\appsubentry{app:objective_data}{Emotion-Only Data}
\appsubentry{app:grpo}{GRPO Fine-tuning}
\appmainentry{app:dataset_details}{Datasets}
\appsubentry{app:quantitative_data}{Data Statistics}
\appsubentry{app:tone_examples}{Tone Descriptions}
\appmainentry{app:qualitative}{Qualitative Results}
\appsubentry{app:qualitative_000240}{Example 1: Emotion recognition and temporal localization}
\appsubentry{app:qualitative_000477}{Example 2: Emotion recognition with transcription differences}
\appsubentry{app:qualitative_000285}{Example 3: Recovering three emotion spans}
\appsubentry{app:qualitative_000026}{Example 4: Correct boundaries but incorrect emotions}
\appsubentry{app:qualitative_000658}{Example 5: Merging utterances with different emotions}
\appsubentry{app:qualitative_000089}{Example 6: Incomplete temporal coverage}
\appsubentry{app:iemocap_qualitative_000198}{Example 7: Emotion and temporal accuracy}
\appsubentry{app:iemocap_qualitative_000230}{Example 8: Short emotional utterances}
\appsubentry{app:iemocap_qualitative_000067}{Example 9: Three-emotion sequence}
\appsubentry{app:iemocap_qualitative_000380}{Example 10: Emotion confusion}
\appsubentry{app:iemocap_qualitative_000080}{Example 11: Merged utterances}
\appsubentry{app:iemocap_qualitative_000263}{Example 12: Premature onset}
\appmainentry{app:limitations}{Limitations}
\appmainentry{app:prompts}{Prompts}
\appsubentry{app:training_prompts}{Model Training and Context Ablations}
\appsubentry{app:evaluation_prompts}{Model Evaluation}
\appsubentry{app:tone_extraction_prompts}{Tone Extraction}
\appsubentry{app:tone_refinement_prompt}{Tone Refinement}
\bottomrule
\end{tabularx}
\endgroup
\clearpage

\section{Implementation Details}
\label{sec:implementation}
\subsection{Model}
\label{app:implementation_model}
We use Qwen2-Audio-7B-Instruct\footnote{\url{https://huggingface.co/Qwen/Qwen2-Audio-7B-Instruct}} as our backbone. We apply LoRA \citep{hu2021lora} to both the language-model and
audio-encoder components, while also fine-tuning the multimodal
projection layer. We use LoRA with rank $r=16$, scaling parameter
$\alpha=32$, and dropout $0.05$. The newly added timestamp-token
rows in the input embedding matrix and output head are also trainable. We use an effective batch size of $48$ and optimize
with AdamW with a learning rate of $10^{-5}$, using a cosine scheduler over $50$ epochs.

The masking is introduced after two epochs, while the multi-span training starts at epoch 13. The loss weights are set to $\lambda_{\mathrm{emo}}=1.5$,
$\lambda_{\mathrm{start}}=0.25$, and $\lambda_{\mathrm{end}}=0.25$.
The auxiliary loss weights are gradually increased to their respective
maximum values over the first three training epochs. We repeat this
warm-up over the first two epochs after introducing multi-span samples
to reduce training instability. For the auxiliary losses, we apply
context masking with probability $p_{\mathrm{mask}}=1.0$, using
tone-only masking in $60\%$ of cases and tone-and-speech masking in
the remaining $40\%$. For timestamp loss, we use $\delta$=1,  $c_{\min}$=0.25 and $c_{\max}$=3.0 We do not use early stopping and only evaluate the
final checkpoint. Sections~\ref{app:masking_probability}
and~\ref{app:trainable_components} discuss the masking probability
and LoRA configuration, respectively.

Training used eight AMD MI250X GPUs and took approximately
$46$ hours per run. The total compute used for training, ablations,
and unsuccessful experiments was approximately $30{,}000$ GPU-hours.


\subsection{Audio Preprocessing}
\label{app:utterance_cleaning}
We process audio as mono waveforms at 16\,kHz. For transcripts,
we lowercase the text, remove punctuation except apostrophes, and normalize
whitespace. Each candidate speech region is transcribed using WhisperX
with Whisper large-v3\footnote{\url{https://huggingface.co/openai/whisper-large-v3}}
and compared with the reference using token overlap.
For segments between selected speech regions, we compute RMS level,
peak amplitude, frame-level energy variation, and the proportion of voiced
frames. We retain intervals as natural pauses or sighs when their RMS and peak levels are
at most $-35$ and $-25$\,dBFS, respectively.
If the selected regions and intervening intervals can be retained
continuously, we crop a single interval. Otherwise, we concatenate the
retained regions and update their positions in the resulting audio clip.

To remove residual noise or extra speech at the boundaries, we use
WhisperX again for forced alignment. We crop from the start of the first aligned word to the end of the last aligned word, with small margins on both sides.
We compare transcriptions before and after trimming and accept the
refined crop only if it preserves the same transcription.

For MAFW and DEFW, in addition to removing empty samples, we remove non-English and unclear samples. Since there are no ground-truth annotations for these datasets, we compare transcriptions from FlamingoNext and Whisper, and if there is less than 60\% alignment, we discard the samples. 

\subsection{Target Annotations}
\label{app:ann}
\noindent\textbf{Emotion labels:} We standardize emotion labels across MELD, IEMOCAP, EMOVDB, MAFW, and DEFW using eight emotion categories: neutral, happiness, sadness, anger, surprise, fear, disgust, and frustration. Labels outside these categories are mapped to the closest emotion where appropriate; otherwise, the corresponding samples are removed. Specifically, we map both ``joy'' and ``excitement'' to ``happiness'' and remove samples labeled ``other,'' ``sleepy,'' ``worried,'' ``contempt,'' ``anxiety,'' ``helplessness,'' or ``disappointment.''


\noindent\textbf{Tone Extraction and Refinement}
\label{app:tone_creation}
The full FlamingoNext\footnote{\url{https://huggingface.co/nvidia/audio-flamingo-next-think-hf}} response is passed to Qwen3-14B\footnote{\url{https://huggingface.co/Qwen/Qwen3-14B}}, which returns a JSON
object with \texttt{speech} and \texttt{tone} fields. The extraction prompt in Section~\ref{app:tone_extraction_prompts}
asks Qwen3 to combine speech fragments and summarize
the main speaker's delivery. The reference emotion is included in the extraction input for consistency checking.

For tone refinement, each clip is divided into five segments and
processed with openSMILE. The segment-level features, speaker-specific
reference statistics, and initial tone description are supplied to ChatGPT
to produce the refined description. The refinement input includes speaker
references for characteristics such as pitch, intensity, and speaking pace.
Speaker-specific acoustic references help interpret features
relative to the speaker's usual delivery. For example, a high
absolute pitch does not necessarily indicate that the speaker has
raised their pitch. We provide these references during refinement
so that descriptions can account for individual vocal characteristics.
Segment-level measurements additionally provide evidence about
changes within an utterance that may be shadowed by a single
clip average. These inputs support more consistent descriptions of pitch, intensity, pace, and voice quality.

\subsection{Multi-Span Design Choices}
\label{app:multi_span_design}

\noindent\textbf{Sampling and silence intervals.}
Source utterances are sampled without replacement from our refined sample pool. Their order is randomized
without requiring an emotion change at every boundary. Consecutive
utterances with the same emotion remain separate annotated spans. Compositions that exceed the supported recording duration are resampled.
An interval classified as silence in our audio preprocessing is checked again
after sampling the actual waveform crop. We inspect its overall
energy, peaks, and the occurrence of active frames to reduce the
chance of introducing unannotated speech or background activity.
Silence is selected to match the sampled gap duration. When a
suitable single interval is unavailable, two intervals
with similar energy levels are joined to reach that duration.

\noindent\textbf{Speaker numbering and timestamp updates.}
Speaker tags are assigned within each constructed recording in
order of first appearance. The first distinct speaker identifier
receives \texttt{<speaker1>}, the next receives
\texttt{<speaker2>}, and so on. When a speaker identifier occurs
again, its original number is reused. For example, a sequence of speakers
$A,B,A$ is annotated as \texttt{<speaker1>},
\texttt{<speaker2>}, and \texttt{<speaker1>}. Numbering restarts
for each recording, so the same speaker may receive different
numbers in different examples.
We track the starting sample index of each utterance and calculate the temporal
offset $\delta_i$. A source span $[s_i,e_i]$ is then recorded as
$[s_i+\delta_i,e_i+\delta_i]$. Offsets also account for the inserted silence and overlap.

\clearpage
\section{Training Ablation}
\label{app:additional_ablations}
We examine how training context, loss components, masking probability, and trainable parameters affect performance.

\subsection{Context Information and Masking}
\label{app:context_and_masking}

Table~\ref{tab:context_loss_ablation} shows a per-dataset breakdown of the context ablation and our objective. 

\begin{table}[!htbp]
\caption{Effect of training context and loss components.
Single-span results are Emo-Hit Accuracy on the raw test sets.
Temporal metrics are Span F1 at a tolerance of 0.25\,s and mean
timestamp IoU (mIoU).}
\label{tab:context_loss_ablation}
\centering
\setlength{\tabcolsep}{3pt}
\renewcommand{\arraystretch}{1.12}
\vspace{4pt}

\resizebox{\linewidth}{!}{%
\begin{tabular}{@{}l*{10}{c}@{}}
\toprule
& \textbf{MELD}
& \textbf{IEMOCAP}
& \multicolumn{4}{c}{\textbf{MELD Multi-Span}}
& \multicolumn{4}{c}{\textbf{IEMOCAP Multi-Span}} \\
\cmidrule(lr){2-2}
\cmidrule(lr){3-3}
\cmidrule(lr){4-7}
\cmidrule(lr){8-11}

\textbf{Setting}
& \shortstack{\textbf{Emo-Hit}\\\textbf{Acc.}}
& \shortstack{\textbf{Emo-Hit}\\\textbf{Acc.}}
& \shortstack{\textbf{Emo-Duration}\\\textbf{F1}}
& \shortstack{\textbf{Emotion}\\\textbf{F1}}
& \shortstack{\textbf{Span F1}\\\textbf{@ 0.25\,s}}
& \textbf{mIoU}
& \shortstack{\textbf{Emo-Duration}\\\textbf{F1}}
& \shortstack{\textbf{Emotion}\\\textbf{F1}}
& \shortstack{\textbf{Span F1}\\\textbf{@ 0.25\,s}}
& \textbf{mIoU} \\
\midrule

\multicolumn{11}{@{}l}{\textit{Partial Context SFT}} \\
\addlinespace[1pt]

Emotion only
& 54.44 & \underline{66.36}
& 52.23 & 54.40 & -- & --
& 62.85 & 60.11 & -- & -- \\

Speech + Emotion
& 52.98 & 66.00
& 49.09 & 49.50 & -- & --
& \underline{63.25} & \underline{60.20} & -- & -- \\

Tone + Emotion
& 51.87 & 56.30
& 49.91 & 52.15 & -- & --
& 52.75 & 51.23 & -- & -- \\

Refined Tone + Emotion
& 53.60 & 60.12
& 49.98 & 53.44 & -- & --
& 56.21 & 53.86 & -- & -- \\
\midrule

\multicolumn{11}{@{}l}{\textit{Full context}} \\
\addlinespace[1pt]

SFT
& 53.91 & 61.09
& 50.78 & 53.58 & 94.49 & 95.11
& 57.76 & 55.39 & 74.65 & 86.71 \\

SFT+TAG
& 53.72 & 61.09
& 52.01 & 54.35 & 96.08 & 96.41
& 59.83 & 55.33 & 83.00 & 89.54 \\

SFT+M-TAG (EMO-TAG)
& \textbf{56.13} & \textbf{66.97}
& \textbf{56.60} & \textbf{58.79}
& \textbf{98.45} & \textbf{98.32}
& \textbf{65.65} & \textbf{62.26}
& \textbf{85.23} & \textbf{91.96} \\

\quad w/o emotion loss
& 54.02 & 61.39
& 51.47 & 54.90
& \underline{98.01} & \underline{97.68}
& 56.01 & 53.34
& \underline{83.69} & \underline{90.11} \\

\quad w/o temporal loss
& \underline{55.21} & 65.88
& \underline{52.83} & \underline{56.57}
& 95.38 & 95.89
& 61.64 & 59.83
& 73.08 & 87.22 \\

\bottomrule
\end{tabular}%
}
\end{table}

As was shown in the main paper, emotion-only training achieves surprisingly
strong performance, outperforming all other settings except our approach
and speech-plus-emotion training on IEMOCAP. The bottom part of the table
presents an ablation of our objective. Without context masking, adding
the auxiliary losses yields modest and inconsistent improvements over
SFT, suggesting that simply increasing the weight of emotion and
timestamp tokens is insufficient. Introducing context masking
considerably improves this performance.

As expected, removing the emotion loss mainly hurts emotion recognition,
while removing the timestamp loss reduces temporal accuracy. Since
Emotion-Duration F1 also depends on emotion accuracy, we additionally
report Span F1 @0.25 and mean intersection over union (mIoU) to assess
temporal performance. Span F1 @0.25 is the harmonic mean of span precision
and recall, with a prediction considered correct when both its start
and end times are within $0.25$ seconds of the reference boundaries.
mIoU measures the average intersection over union between predicted
and reference time spans.

\subsection{Masking Probability}
\label{app:masking_probability}

\begin{figure}[!htbp]
\centering
\includegraphics[width=\linewidth]{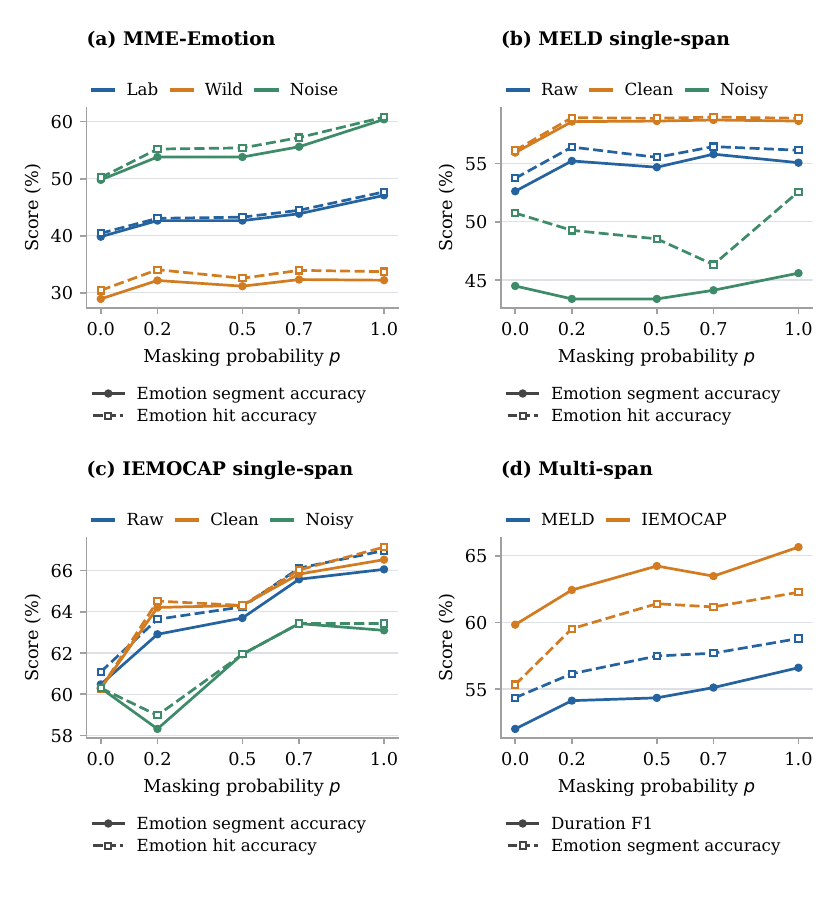}
\caption{Effect of context masking probability across emotion recognition
benchmarks. All runs use our full objective with encoder--decoder LoRA and a trainable
projection layer; $p=0$ disables context masking. Markers show the evaluated
probabilities $\{0,0.2,0.5,0.7,1\}$. All scores are percentages.}
\label{fig:masking_sweep}
\end{figure}

Figure~\ref{fig:masking_sweep} compares five context masking probabilities,
from $p=0$ (no masking) to $p=1$ (masking in every auxiliary-loss pass).

The effect of masking remains consistent across the evaluated probabilities.
We use $p=1$ for Emo-TAG in the main experiments.


\subsection{Trainable Components}
\label{app:trainable_components}

Table~\ref{tab:lora_projection_ablation} compares LoRA on the language model
alone with additional adaptation of the audio encoder and projection layer.

\begin{table}[!htbp]
\caption{Ablation of trainable components. Components are added cumulatively. All configurations use the full objective with
probability $p=1$. MME, MELD, and IEMOCAP report average emotion hit accuracy;
Multi-span reports average emotion-duration F1.}
\label{tab:lora_projection_ablation}
\centering
\footnotesize
\setlength{\tabcolsep}{4pt}
\renewcommand{\arraystretch}{1.08}

\begin{tabularx}{\linewidth}{@{}Xrrrr@{}}
\toprule
& \textbf{MELD} & \textbf{IEMOCAP} & \textbf{MME} & \textbf{Multi-span} \\
\textbf{Configuration} & Hit & Hit & Hit & Dur.\ F1 \\
\midrule
LLM only LoRA
& 54.03 & 63.40 & 42.35 & 54.11 \\
\quad + Audio-encoder LoRA
& 54.64 & 63.24 & 44.75 & 58.26 \\
\quad + Projection layer (Emo-TAG)
& \textbf{55.86} & \textbf{65.84}
& \textbf{47.40} & \textbf{61.13} \\
\bottomrule
\end{tabularx}
\end{table}

Adding audio-encoder LoRA improves multi-span and MME performance, with
little change on single-span IEMOCAP. Training the projection layer as well
gives the best results across all four settings.

\section{Additional Experiments}
\label{app:additional_experiments}

\subsection{Emotion-Only Data}
\label{app:objective_data}

Table~\ref{tab:objective_data_ablation} compares emotion-only and full
supervision when adding DEFW and MAFW training samples.
For emotion-only supervision, we use teacher forcing and apply the
masked affective loss only to the first token of each emotion label.
All other tokens in these samples contribute zero loss.

\begin{table}[!htbp]
\caption{Effect of training with additional data.
Single-span scores are mean Emo-Hit Accuracy over raw, clean, and noisy
test sets; multi-span scores are Emotion-Duration F1.
Multi-Span means using multi-span data in addition to single-span data.
Full means using the full training objective rather than only the
emotion loss for the additional data.}
\label{tab:objective_data_ablation}
\centering
\setlength{\tabcolsep}{3pt}
\renewcommand{\arraystretch}{1.0}
\vspace{4pt}

\begin{tabular*}{\linewidth}{@{\extracolsep{\fill}}lcccccc@{}}
\toprule
& \textbf{SFT}
& \textbf{Emo-TAG}
& \multicolumn{4}{c}{\textbf{Emo-TAG + DEFW + MAFW}} \\
\cmidrule(lr){4-7}
Supervision
& & & Emotion only & Full & Emotion only & Full \\
Training spans
& & & (Multi-Span) & (Multi-Span) & Single & Single \\
\midrule

\multicolumn{7}{@{}l}{\textit{MME-Emotion}} \\
\quad Lab
& 42.05 & \textbf{47.69} & 44.06 & 44.47 & \underline{45.27} & 42.05 \\
\quad Wild
& 31.83 & 33.72 & 35.33 & \textbf{37.00} & 35.44 & \underline{36.44} \\
\quad Noise
& 53.20 & 60.80 & \underline{61.20} & 59.40 & \textbf{61.60} & 56.60 \\

\midrule

\multicolumn{7}{@{}l}{\textit{Single-span}} \\
\quad MELD
& 51.91 & \textbf{55.86} & 54.56 & 53.61 & \underline{55.23} & 52.96 \\
\quad IEMOCAP
& 60.18 & \textbf{65.84} & 64.53 & \underline{64.97} & 64.65 & 64.40 \\
\quad DEFW
& 41.76 & 46.37 & 52.70 & \textbf{53.62} & 51.78 & \underline{53.36} \\
\quad MAFW
& 33.56 & 34.79 & \textbf{38.17} & 36.56 & 35.48 & \underline{38.10} \\

\addlinespace[5pt]
\multicolumn{7}{@{}l}{\textit{Multi-span}} \\
\quad MELD
& 50.78 & \textbf{56.60} & 53.82 & 52.66 & \underline{54.17} & 53.90 \\
\quad IEMOCAP
& 57.76 & \textbf{65.65} & \underline{63.86} & 63.58 & 63.57 & 62.68 \\
\bottomrule
\end{tabular*}
\end{table}

As shown in the table, the additional data from DEFW and MAFW improve
performance on their respective test sets, as well as on the Wild subset
of MME-Emotion and, with emotion-only supervision, its Noise subset.
However, adding these data slightly harms performance on IEMOCAP and
MELD, which may reflect differences between the nature of these datasets.

The important takeaway from this experiment is that using only emotion
supervision for these additional datasets maintains competitive performance
and sometimes even surpasses full supervision. This shows the flexibility
of our proposed objective.






\subsection{GRPO Fine-tuning}
\label{app:grpo}

We also ran a preliminary GRPO experiment, initializing the policy from
our SFT model and using a frozen copy as the reference policy. We continued
training the SFT LoRA adapter, keeping the same output format: speech,
timestamps, tone, and emotion.

For each generated response, we computed a weighted combination of format,
timestamp, emotion, speech, and tone rewards:
\begin{equation}
R =
0.10R_{\mathrm{fmt}}
+0.30R_{\mathrm{ts}}
+0.30R_{\mathrm{emo}}
+0.15R_{\mathrm{sp}}
+0.15R_{\mathrm{tone}}.
\label{eq:grpo_reward}
\end{equation}
The format reward measures compliance with the required output structure,
including the expected number of spans and the absence of additional text.
The timestamp reward decreases linearly with the sum of absolute start-
and end-time errors, reaching zero at a specified error tolerance.
Span scores are summed and divided by the larger of the predicted and
reference span counts, penalizing missing or additional spans.
The emotion reward measures the proportion of reference spans assigned
the correct emotion, comparing predictions and references in sequence order.
The speech reward gives full credit below a specified word error rate
threshold and decreases linearly to zero at a word error rate of one.
The tone reward uses the cosine similarity between the tone embeddings extracted with FlagEmbedding \footnote{https://huggingface.co/BAAI/bge-large-en-v1.5}

For each audio input, we sample three responses across eight
distributed processes and computed advantages using the pooled rewards with gradient accumulation of four.
We centered rewards by their group mean and normalized them by the group
standard deviation, with a lower bound on the denominator to avoid
amplifying small reward differences. The optimization
objective also included a KL penalty relative to the frozen SFT reference
policy. 

As shown in table \ref{tab:grpo_comparison}, GRPO yielded a modest improvement over the SFT model.
In early experiments, it also occasionally switched to Chinese during
generation.

\begin{table}[t]
\caption{Comparison of SFT, GRPO, and EMO-TAG.
Single-span scores are mean Emo-Hit Accuracy over the
Original, Refined, and Noisy sets; multi-span scores are
Emotion-Duration F1 on the Multiple sets.
MME scores are mean Emo-Segment Accuracy over Noise, Wild, and Lab.}
\label{tab:grpo_comparison}
\centering
\setlength{\tabcolsep}{5pt}
\renewcommand{\arraystretch}{1.08}
\begin{tabular}{@{}lccccc@{}}
\toprule
& \multicolumn{2}{c}{\textbf{IEMOCAP}}
& \multicolumn{2}{c}{\textbf{MELD}}
& \textbf{MME} \\
\cmidrule(lr){2-3}
\cmidrule(lr){4-5}
\textbf{Method}
& Single-span & Multi-span
& Single-span & Multi-span
& Average \\
\midrule
SFT
& 60.18 & 57.76 & 51.91 & 50.78 & 42.36 \\
SFT + GRPO
& 61.15 & 58.21 & 52.33 & 51.01 & 43.16 \\
EMO-TAG
& \textbf{65.84} & \textbf{65.65}
& \textbf{55.86} & \textbf{56.60}
& \textbf{47.40} \\
\bottomrule
\end{tabular}
\end{table}
\clearpage
\section{Datasets}
\label{app:dataset_details}
Our filtering pipeline is aggressive in order to remove as many low-quality samples as possible. This comes with the cost of losing many good samples, especially very short ones.
\subsection{Data Statistics}
\label{app:quantitative_data}
Table~\ref{tab:dataset_statistics} reports sample counts by span setting
and the number of samples removed from each dataset split.

\begin{table}[!htbp]
\caption{Dataset composition and filtering statistics. Top: composition of our
training and test datasets, where MELD represents a more natural setting and
EmoV-DB and IEMOCAP represent controlled and acted settings.
$^{*}$ Additional samples with overlapping speech spans.
Bottom: filtering statistics for the original datasets.}
\label{tab:dataset_statistics}
\centering
\footnotesize
\setlength{\tabcolsep}{4pt}
\renewcommand{\arraystretch}{1.08}

\textit{(a) Temporal dataset composition}
\vspace{3pt}

\begin{tabular*}{\linewidth}{@{\extracolsep{\fill}}lrrrrr@{}}
\toprule
\textbf{Spans} &
\textbf{MELD} &
\textbf{EmoV-DB} &
\textbf{IEMOCAP} &
\textbf{DFEW} &
\textbf{MAFW} \\
\midrule
1 & 6,818 & 2,565 & 3,484 & 1,911 & 1,311 \\
2 & 2,233 + 703$^{*}$ & 1,066 + 502$^{*}$ & 1,331 + 555$^{*}$ & 241 & 163 \\
3 & 1,606 + 520$^{*}$ & 822 + 411$^{*}$ & 967 + 411$^{*}$ & 188 & 132 \\
4 & 186 + 61$^{*}$ & 105 + 54$^{*}$ & 108 + 42$^{*}$ & 20 & 10 \\
\midrule
\textbf{Total} &
\textbf{10,843 + 1,284}$^{*}$ &
\textbf{4,558 + 967}$^{*}$ &
\textbf{5,890 + 1,008}$^{*}$ &
\textbf{2,360} &
\textbf{1,616} \\
\bottomrule
\end{tabular*}

\vspace{9pt}
\textit{(b) Filtering statistics}
\vspace{3pt}
\begin{tabular*}{\linewidth}{@{\extracolsep{\fill}}llrrr@{}}
\toprule
\textbf{Dataset} &
\textbf{Split} &
\textbf{Original} &
\textbf{Removed} &
\textbf{Retained (\%)} \\
\midrule
EmoV-DB
& Train & 5,256 & 2,691 & 48.80\% \\
\midrule
\multirow{2}{*}{MELD}
& Train & 9,989 & 3,171 & 68.26\% \\
& Test  & 2,610 & 541 & 79.27\% \\
\midrule
\multirow{2}{*}{IEMOCAP}
& Train & 5,810 & 2,326 & 59.97\% \\
& Test  & 1,623 & 631 & 61.12\% \\
\midrule
\multirow{2}{*}{DFEW}
& Train & 9,356 & 7,445 & 20.43\% \\
& Test  & 2,341 & 1,582 & 32.42\% \\
\midrule
\multirow{2}{*}{MAFW}
& Train & 7,333 & 6,022 & 17.88\% \\
& Test & 1,839 & 797 & 56.66\% \\
\bottomrule
\end{tabular*}
\end{table}

\begin{figure}
\centering
\includegraphics[width=0.95\linewidth]{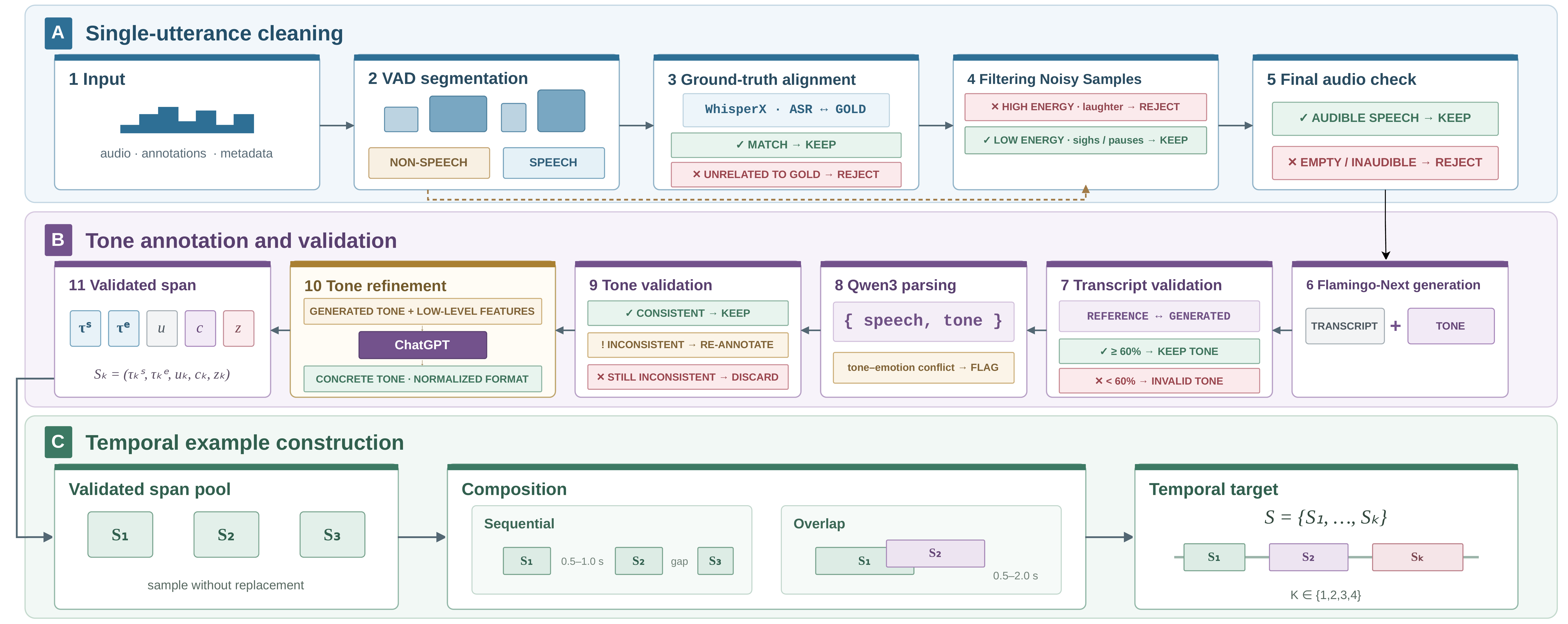}
\caption{Overview of the data-processing pipeline, from speech and tone
validation to the construction of cleaned single- and multi-span datasets.}
\label{fig:data_pipeline}
\end{figure}

Figure~\ref{fig:data_pipeline} summarizes the dataset cleaning and tone generation pipeline.

\subsection{Tone Descriptions}
\label{app:tone_examples}

Table~\ref{tab:tone_examples} compares five original tone descriptions with
their refined versions. The originals often describe emphasis on particular
words and changes in delivery. The refined versions use a more consistent
acoustic structure but sometimes lose these details.
\begingroup
\small
\setlength{\tabcolsep}{4pt}
\renewcommand{\arraystretch}{1.12}

\begin{longtable}{
    @{}p{0.13\linewidth}
    p{\dimexpr0.87\linewidth-2\tabcolsep\relax}@{}
}
\caption{Selected comparisons showing the complete \texttt{target\_text}
from the original and refined files.}
\label{tab:tone_examples}\\

\toprule
\textbf{Version} & \textbf{Complete target text} \\
\midrule
\endfirsthead

\multicolumn{2}{@{}l@{}}{%
    \tablename~\thetable\ (continued)} \\
\toprule
\textbf{Version} & \textbf{Complete target text} \\
\midrule
\endhead

\midrule
\multicolumn{2}{r@{}}{\textit{Continued on next page}} \\
\endfoot

\bottomrule
\endlastfoot

Sample &
\texttt{meld\_train\_dia613\_utt8} \\*

Original &
\texttt{[speech]} \texttt{<speaker1>}: "Yes!!\ \ Oh."
from \texttt{<|0.00|>s} to \texttt{<|1.23|>s},
\texttt{[tone]} Starts with a high pitch, fast, loud exclamation
of 'Yes!', then transitions to a soft, slow, low-pitched 'Oh.'
with subdued energy., \texttt{[emotion]} surprise. \\*

Refined &
\texttt{[speech]} \texttt{<speaker1>}: "Yes!!\ \ Oh."
from \texttt{<|0.00|>s} to \texttt{<|1.23|>s},
\texttt{[tone]} Deliberate, restrained delivery with high,
broadly varying pitch, strong, sustained loudness, slow,
continuous pacing, and unstable pitch control; pitch gradually
falls while loudness remains largely steady, \texttt{[emotion]} surprise. \\

\midrule

Sample &
\texttt{meld\_train\_dia1010\_utt9} \\*

Original &
\texttt{[speech]} \texttt{<speaker1>}: "No-o-o!\ \ No way!"
from \texttt{<|0.00|>s} to \texttt{<|3.39|>s},
\texttt{[tone]} Deep, drawn-out 'No' with strain and a sigh,
then rapid, loud 'No way' without pitch change, showing
escalating frustration and disbelief., \texttt{[emotion]} disgust. \\*

Refined &
\texttt{[speech]} \texttt{<speaker1>}: "No-o-o!\ \ No way!"
from \texttt{<|0.00|>s} to \texttt{<|3.39|>s},
\texttt{[tone]} Alert, startled delivery with high, narrowly
varying pitch, soft, even loudness, slow pacing with frequent
pauses, and breathy-leaning voicing; pitch gradually rises
while loudness remains largely steady,
\texttt{[emotion]} disgust. \\

\midrule

Sample &
\texttt{meld\_train\_dia791\_utt11} \\*

Original &
\texttt{[speech]} \texttt{<speaker1>}: "I’m not gonna be able
to do that for so long, and it’s so much fun!"
from \texttt{<|0.00|>s} to \texttt{<|4.25|>s},
\texttt{[tone]} Voice remains bright and forward, pitch rises
emphatically when discussing duration, peaks louder and faster
during 'so much fun!' with energetic intensity and no signs
of hesitation or distress.,
\texttt{[emotion]} joy. \\*

Refined &
\texttt{[speech]} \texttt{<speaker1>}: "I’m not gonna be able
to do that for so long, and it’s so much fun!"
from \texttt{<|0.00|>s} to \texttt{<|4.25|>s},
\texttt{[tone]} Animated, energetic delivery with mid-level,
broadly varying pitch, strong, strongly fluctuating loudness,
measured pacing with frequent pauses, and clear, stable
voicing; pitch gradually rises while loudness remains
largely steady, \texttt{[emotion]} joy. \\

\midrule

Sample &
\texttt{meld\_train\_dia116\_utt3} \\*

Original &
\texttt{[speech]} \texttt{<speaker1>}: "Wow, you ah, you really
like all this space stuff, huh?"
from \texttt{<|0.00|>s} to \texttt{<|4.41|>s},
\texttt{[tone]} Calm, conversational delivery with relaxed,
mildly curious mood. Starts with hesitation, steady mid-range
pitch, moderate pace, slight emphasis on 'really,' and rising
intonation on 'space stuff,' conveying casual interest and
gentle surprise., \texttt{[emotion]} surprise. \\*

Refined &
\texttt{[speech]} \texttt{<speaker1>}: "Wow, you ah, you really
like all this space stuff, huh?"
from \texttt{<|0.00|>s} to \texttt{<|4.41|>s},
\texttt{[tone]} Restrained, startled delivery with low, broadly
varying pitch, soft, even loudness, slow pacing with frequent
pauses, and firm, slightly pressed voicing; pitch and loudness
fluctuate unevenly across the utterance, \texttt{[emotion]} surprise. \\

\midrule

Sample &
\texttt{meld\_train\_dia139\_utt0} \\*

Original &
\texttt{[speech]} \texttt{<speaker1>}: "I really thought you
making a good point. I mean y'know, until you got cut off."
from \texttt{<|0.00|>s} to \texttt{<|3.79|>s},
\texttt{[tone]} Calm, neutral, and reflective with slight
emphasis on 'good point' and a marginal increase in volume
and speed when mentioning being cut off., \texttt{[emotion]} neutral. \\*

Refined &
\texttt{[speech]} \texttt{<speaker1>}: "I really thought you
making a good point. I mean y'know, until you got cut off."
from \texttt{<|0.00|>s} to \texttt{<|3.79|>s},
\texttt{[tone]} Calm, controlled delivery with low, moderately
varying pitch, soft, strongly fluctuating loudness, fast
pacing with frequent pauses, and clear, stable voicing;
pitch gradually falls while loudness remains largely steady, \texttt{[emotion]} neutral. \\

\end{longtable}
\endgroup
\clearpage

\section{Qualitative Results}
\label{app:qualitative}

\definecolor{qualCorrectBackground}{RGB}{220,242,221}
\definecolor{qualIncorrectBackground}{RGB}{255,222,222}
\definecolor{qualCorrectText}{RGB}{20,91,34}
\definecolor{qualIncorrectText}{RGB}{151,27,27}
\providecommand{\qcorrect}[1]{\begingroup\setlength{\fboxsep}{0.45pt}%
  \colorbox{qualCorrectBackground}{\textcolor{qualCorrectText}{\strut #1}}\endgroup}
\providecommand{\qincorrect}[1]{\begingroup\setlength{\fboxsep}{0.45pt}%
  \colorbox{qualIncorrectBackground}{\textcolor{qualIncorrectText}{\strut #1}}\endgroup}
\providecommand{\qualmodel}[1]{\begingroup\setlength{\fboxsep}{4pt}%
  \colorbox{black!4}{\parbox{\dimexpr\linewidth-8pt\relax}{\strut\textbf{#1}}}\endgroup}

\smallskip
\noindent\qcorrect{Green} marks exact reference matches; \qincorrect{red}
marks errors or extra predictions. Baseline models do not predict tone descriptions.
One important observation from the results is how multiple samples in the MELD benchmark have incorrect annotations. One such instance is \ref{app:qualitative_000026}. All models agree that the delivery is natural, which can also be verified by listening to the audio, but the annotation is surprise. Similarly, all models agree on the part missing from the ground-truth transcription in \ref{app:qualitative_000285}.
\subsection{Example 1: Emotion recognition and temporal localization}
\label{app:qualitative_000240}
\noindent\textbf{Success}\hfill{\footnotesize\texttt{MELD\_test\_concat\_000240\_k2\_1ea2e4937f}}\par
\smallskip
\noindent Emo-TAG identifies both emotions and their boundaries correctly. Audio-Reasoner also recognizes the emotions but misplaces the boundaries.

\begingroup
\small
\setlength{\tabcolsep}{5pt}
\renewcommand{\arraystretch}{1.12}
\setlength{\LTpre}{6pt}\setlength{\LTpost}{8pt}
\begin{longtable}{@{}>{\raggedright\arraybackslash}p{0.20\linewidth}>{\raggedright\arraybackslash}p{\dimexpr0.80\linewidth-2\tabcolsep\relax}@{}}
\noalign{\label{tab:qualitative_000240}}
\toprule
\textbf{Source} & \textbf{Emotion--span sequence} \\
\midrule
\endfirsthead
\multicolumn{2}{@{}l@{}}{\textit{Emotion--span summary (continued)}} \\
\toprule
\textbf{Source} & \textbf{Emotion--span sequence} \\
\midrule
\endhead
\midrule
\multicolumn{2}{r@{}}{\textit{Continued on next page}} \\
\endfoot
\bottomrule
\endlastfoot
\textbf{Reference} & \mbox{anger $[0.00, 2.00]$};\allowbreak\quad \mbox{sadness $[2.75, 4.25]$} \\ \addlinespace[2pt]
\textbf{Emo-TAG} & \mbox{\qcorrect{anger} $[\qcorrect{0.00}, \qcorrect{2.00}]$};\allowbreak\quad \mbox{\qcorrect{sadness} $[\qcorrect{2.75}, \qcorrect{4.25}]$} \\ \addlinespace[2pt]
\textbf{FlamingoNext} & \mbox{\qincorrect{frustration} $[\qincorrect{0.38}, \qincorrect{2.79}]$};\allowbreak\quad \mbox{\qcorrect{sadness} $[\qincorrect{5.13}, \qincorrect{6.00}]$} \\ \addlinespace[2pt]
\textbf{Audio-Reasoner} & \mbox{\qcorrect{anger} $[\qcorrect{0.00}, \qincorrect{3.47}]$};\allowbreak\quad \mbox{\qcorrect{sadness} $[\qincorrect{3.47}, \qincorrect{5.08}]$} \\ \addlinespace[2pt]
\end{longtable}
\endgroup

\begingroup
\small
\setlength{\tabcolsep}{6pt}
\renewcommand{\arraystretch}{1.12}
\setlength{\LTpre}{6pt}\setlength{\LTpost}{8pt}
\begin{longtable}{@{}>{\raggedright\arraybackslash}p{0.32\linewidth}>{\raggedright\arraybackslash}p{\dimexpr0.68\linewidth-2\tabcolsep\relax}@{}}

\noalign{\label{tab:qualitative_full_000240}}
\toprule
\textbf{Span / emotion / time} & \textbf{Speech and tone} \\
\midrule
\endfirsthead
\multicolumn{2}{@{}l@{}}{\textit{Example 1: annotations (continued)}} \\
\toprule
\textbf{Span / emotion / time} & \textbf{Speech and tone} \\
\midrule
\endhead
\midrule
\multicolumn{2}{r@{}}{\textit{Continued on next page}} \\
\endfoot
\bottomrule
\endlastfoot
\multicolumn{2}{@{}l@{}}{\qualmodel{Reference annotation}} \\*
{\footnotesize\textbf{Reference / 1}}\par{\footnotesize\texttt{<speaker1>}\quad\textbf{anger}}\par {\footnotesize\texttt{<|0.00|>s}--\texttt{<|2.00|>s}} & \textbf{Speech:} ``People have got to finish their stories!''\par\smallskip
\textbf{Tone:}  \\ \addlinespace[4pt]
{\footnotesize\textbf{Reference / 2}}\par{\footnotesize\texttt{<speaker2>}\quad\textbf{sadness}}\par {\footnotesize\texttt{<|2.75|>s}--\texttt{<|4.25|>s}} & \textbf{Speech:} ``Okay, I'm sorry.''\par\smallskip
\textbf{Tone:}  \\ \addlinespace[4pt]
\multicolumn{2}{@{}l@{}}{\qualmodel{Emo-TAG prediction}} \\*
{\footnotesize\textbf{Emo-TAG / 1}}\par{\footnotesize\texttt{<speaker1>}\quad\qcorrect{\textbf{anger}}}\par {\footnotesize\qcorrect{\texttt{<|0.00|>s}}--\qcorrect{\texttt{<|2.00|>s}}} & \textbf{Speech:} ``\qcorrect{People} \qcorrect{have} \qcorrect{got} \qcorrect{to} \qcorrect{finish} \qcorrect{their} \qcorrect{stories}!''\par\smallskip
\textbf{Tone:} High pitch, fast pace, and loud delivery with clear articulation, conveying urgency and frustration through heightened intensity \\ \addlinespace[4pt]
{\footnotesize\textbf{Emo-TAG / 2}}\par{\footnotesize\texttt{<speaker2>}\quad\qcorrect{\textbf{sadness}}}\par {\footnotesize\qcorrect{\texttt{<|2.75|>s}}--\qcorrect{\texttt{<|4.25|>s}}} & \textbf{Speech:} ``\qcorrect{Okay}, \qcorrect{I'm} \qcorrect{sorry}.''\par\smallskip
\textbf{Tone:} Very quickly, with a noticeably higher-than-normal pitch and a sense of mild strain or urgency rather than relaxed normal talking. She doesn't hesitate long between phrases, so the entire exchange feels rushed and somewhat pressured \\ \addlinespace[4pt]
\multicolumn{2}{@{}l@{}}{\qualmodel{FlamingoNext prediction}} \\*
{\footnotesize\textbf{FlamingoNext / 1}}\par{\footnotesize\texttt{<speaker1>}\quad\qincorrect{\textbf{frustration}}}\par {\footnotesize\qincorrect{\texttt{<|0.50|>s}}--\qincorrect{\texttt{<|2.75|>s}}} & \textbf{Speech:} ``\qcorrect{People} \qcorrect{have} \qcorrect{got} \qcorrect{to} \qcorrect{finish} \qcorrect{their} \qcorrect{stories}.''\par\smallskip
\textbf{Tone:}  \\ \addlinespace[4pt]
{\footnotesize\textbf{FlamingoNext / 2}}\par{\footnotesize\texttt{<speaker2>}\quad\qcorrect{\textbf{sadness}}}\par {\footnotesize\qincorrect{\texttt{<|5.25|>s}}--\qincorrect{\texttt{<|6.00|>s}}} & \textbf{Speech:} ``\qcorrect{Okay}, \qcorrect{I'm} \qcorrect{sorry}''\par\smallskip
\textbf{Tone:}  \\ \addlinespace[4pt]
\multicolumn{2}{@{}l@{}}{\qualmodel{Audio-Reasoner prediction}} \\*
{\footnotesize\textbf{Audio-Reasoner / 1}}\par{\footnotesize\texttt{<speaker1>}\quad\qcorrect{\textbf{anger}}}\par {\footnotesize\qcorrect{\texttt{<|0.00|>s}}--\qincorrect{\texttt{<|3.50|>s}}} & \textbf{Speech:} ``\qcorrect{People} \qcorrect{have} \qcorrect{got} \qcorrect{to} \qcorrect{finish} \qcorrect{their} \qcorrect{stories}!''\par\smallskip
\textbf{Tone:}  \\ \addlinespace[4pt]
{\footnotesize\textbf{Audio-Reasoner / 2}}\par{\footnotesize\texttt{<speaker2>}\quad\qcorrect{\textbf{sadness}}}\par {\footnotesize\qincorrect{\texttt{<|3.50|>s}}--\qincorrect{\texttt{<|5.00|>s}}} & \textbf{Speech:} ``\qcorrect{Okay}, \qcorrect{I'm} \qcorrect{sorry}.''\par\smallskip
\textbf{Tone:}  \\ \addlinespace[4pt]
\end{longtable}
\endgroup

\Needspace{12\baselineskip}
\subsection{Example 2: Emotion recognition with transcription differences}
\label{app:qualitative_000477}
\noindent\textbf{Success}\hfill{\footnotesize\texttt{MELD\_test\_concat\_000477\_k2\_948feb96e7}}\par
\smallskip
\noindent FlamingoNext transcribes ``How neat is that?'' and labels it happiness; the reference says ``how needy is that?'' and is labeled disgust. Emo-TAG gets both emotions and their boundaries right despite errors in its transcript.

\begingroup
\small
\setlength{\tabcolsep}{5pt}
\renewcommand{\arraystretch}{1.12}
\setlength{\LTpre}{6pt}\setlength{\LTpost}{8pt}
\begin{longtable}{@{}>{\raggedright\arraybackslash}p{0.20\linewidth}>{\raggedright\arraybackslash}p{\dimexpr0.80\linewidth-2\tabcolsep\relax}@{}}
\noalign{\label{tab:qualitative_000477}}
\toprule
\textbf{Source} & \textbf{Emotion--span sequence} \\
\midrule
\endfirsthead
\multicolumn{2}{@{}l@{}}{\textit{Emotion--span summary (continued)}} \\
\toprule
\textbf{Source} & \textbf{Emotion--span sequence} \\
\midrule
\endhead
\midrule
\multicolumn{2}{r@{}}{\textit{Continued on next page}} \\
\endfoot
\bottomrule
\endlastfoot
\textbf{Reference} & \mbox{disgust $[0.00, 1.25]$};\allowbreak\quad \mbox{neutral $[2.00, 3.50]$} \\ \addlinespace[2pt]
\textbf{Emo-TAG} & \mbox{\qcorrect{disgust} $[\qcorrect{0.00}, \qcorrect{1.25}]$};\allowbreak\quad \mbox{\qcorrect{neutral} $[\qcorrect{2.00}, \qcorrect{3.50}]$} \\ \addlinespace[2pt]
\textbf{FlamingoNext} & \mbox{\qincorrect{happiness} $[\qcorrect{0.00}, \qincorrect{1.00}]$};\allowbreak\quad \mbox{\qcorrect{neutral} $[\qincorrect{3.00}, \qincorrect{5.00}]$} \\ \addlinespace[2pt]
\textbf{Audio-Reasoner} & \mbox{\qincorrect{neutral} $[\qcorrect{0.00}, \qincorrect{2.08}]$};\allowbreak\quad \mbox{\qcorrect{neutral} $[\qincorrect{2.08}, \qincorrect{4.08}]$} \\ \addlinespace[2pt]
\end{longtable}
\endgroup

\begingroup
\small
\setlength{\tabcolsep}{6pt}
\renewcommand{\arraystretch}{1.12}
\setlength{\LTpre}{6pt}\setlength{\LTpost}{8pt}
\begin{longtable}{@{}>{\raggedright\arraybackslash}p{0.32\linewidth}>{\raggedright\arraybackslash}p{\dimexpr0.68\linewidth-2\tabcolsep\relax}@{}}

\noalign{\label{tab:qualitative_full_000477}}
\toprule
\textbf{Span / emotion / time} & \textbf{Speech and tone} \\
\midrule
\endfirsthead
\multicolumn{2}{@{}l@{}}{\textit{Example 2: annotations (continued)}} \\
\toprule
\textbf{Span / emotion / time} & \textbf{Speech and tone} \\
\midrule
\endhead
\midrule
\multicolumn{2}{r@{}}{\textit{Continued on next page}} \\
\endfoot
\bottomrule
\endlastfoot
\multicolumn{2}{@{}l@{}}{\qualmodel{Reference annotation}} \\*
{\footnotesize\textbf{Reference / 1}}\par{\footnotesize\texttt{<speaker1>}\quad\textbf{disgust}}\par {\footnotesize\texttt{<|0.00|>s}--\texttt{<|1.25|>s}} & \textbf{Speech:} ``I mean,... how needy is that?''\par\smallskip
\textbf{Tone:}  \\ \addlinespace[4pt]
{\footnotesize\textbf{Reference / 2}}\par{\footnotesize\texttt{<speaker2>}\quad\textbf{neutral}}\par {\footnotesize\texttt{<|2.00|>s}--\texttt{<|3.50|>s}} & \textbf{Speech:} ``Okay, well then bring her in.''\par\smallskip
\textbf{Tone:}  \\ \addlinespace[4pt]
\multicolumn{2}{@{}l@{}}{\qualmodel{Emo-TAG prediction}} \\*
{\footnotesize\textbf{Emo-TAG / 1}}\par{\footnotesize\texttt{<speaker1>}\quad\qcorrect{\textbf{disgust}}}\par {\footnotesize\qcorrect{\texttt{<|0.00|>s}}--\qcorrect{\texttt{<|1.25|>s}}} & \textbf{Speech:} ``\qincorrect{The} \qincorrect{boy}? \qcorrect{How} \qcorrect{needy} \qcorrect{is} \qcorrect{that}?''\par\smallskip
\textbf{Tone:} Noticeably higher-than-normal pitch, sounding startled and slightly incredulous instead of neutral or casual. Her speaking speed stays steady around mid-range, neither rushed nor drawn out, conveying mild surprise through inflection rather than dramatic signs \\ \addlinespace[4pt]
{\footnotesize\textbf{Emo-TAG / 2}}\par{\footnotesize\texttt{<speaker2>}\quad\qcorrect{\textbf{neutral}}}\par {\footnotesize\qcorrect{\texttt{<|2.00|>s}}--\qcorrect{\texttt{<|3.50|>s}}} & \textbf{Speech:} ``\qcorrect{Okay}, \qcorrect{well} \qcorrect{then} \qcorrect{bring} \qcorrect{her} \qincorrect{with} \qincorrect{you}.''\par\smallskip
\textbf{Tone:} Neutral, composed delivery with steady pace, even stress distribution, and no emotional inflection or vocal variation \\ \addlinespace[4pt]
\multicolumn{2}{@{}l@{}}{\qualmodel{FlamingoNext prediction}} \\*
{\footnotesize\textbf{FlamingoNext / 1}}\par{\footnotesize\texttt{<Speaker 1>}\quad\qincorrect{\textbf{happiness}}}\par {\footnotesize\qcorrect{\texttt{<|0.00|>s}}--\qincorrect{\texttt{<|1.00|>s}}} & \textbf{Speech:} ``\qcorrect{How} \qincorrect{neat} \qcorrect{is} \qcorrect{that}?''\par\smallskip
\textbf{Tone:}  \\ \addlinespace[4pt]
{\footnotesize\textbf{FlamingoNext / 2}}\par{\footnotesize\texttt{<Speaker 2>}\quad\qcorrect{\textbf{neutral}}}\par {\footnotesize\qincorrect{\texttt{<|3.00|>s}}--\qincorrect{\texttt{<|5.00|>s}}} & \textbf{Speech:} ``...\qincorrect{will} \qcorrect{then} \qcorrect{bring} \qcorrect{her}...''\par\smallskip
\textbf{Tone:}  \\ \addlinespace[4pt]
\multicolumn{2}{@{}l@{}}{\qualmodel{Audio-Reasoner prediction}} \\*
{\footnotesize\textbf{Audio-Reasoner / 1}}\par{\footnotesize\texttt{<speaker1>}\quad\qincorrect{\textbf{neutral}}}\par {\footnotesize\qcorrect{\texttt{<|0.00|>s}}--\qincorrect{\texttt{<|2.00|>s}}} & \textbf{Speech:} ``\qcorrect{How} \qcorrect{needy} \qcorrect{is} \qcorrect{that}?''\par\smallskip
\textbf{Tone:}  \\ \addlinespace[4pt]
{\footnotesize\textbf{Audio-Reasoner / 2}}\par{\footnotesize\texttt{<speaker2>}\quad\qcorrect{\textbf{neutral}}}\par {\footnotesize\qcorrect{\texttt{<|2.00|>s}}--\qincorrect{\texttt{<|4.00|>s}}} & \textbf{Speech:} ``\qcorrect{Okay}, \qcorrect{well} \qcorrect{then} \qcorrect{bring} \qcorrect{her}.''\par\smallskip
\textbf{Tone:}  \\ \addlinespace[4pt]
\end{longtable}
\endgroup

\Needspace{12\baselineskip}
\subsection{Example 3: Recovering three emotion spans}
\label{app:qualitative_000285}
\noindent\textbf{Success}\hfill{\footnotesize\texttt{MELD\_test\_concat\_000285\_k3\_6033afb40c}}\par
\smallskip
\noindent Emo-TAG recovers all three emotion spans but adds words to the final transcript. FlamingoNext splits the first utterance, while Audio-Reasoner misclassifies the first and last emotions.

\begingroup
\small
\setlength{\tabcolsep}{5pt}
\renewcommand{\arraystretch}{1.12}
\setlength{\LTpre}{6pt}\setlength{\LTpost}{8pt}
\begin{longtable}{@{}>{\raggedright\arraybackslash}p{0.20\linewidth}>{\raggedright\arraybackslash}p{\dimexpr0.80\linewidth-2\tabcolsep\relax}@{}}
\noalign{\label{tab:qualitative_000285}}
\toprule
\textbf{Source} & \textbf{Emotion--span sequence} \\
\midrule
\endfirsthead
\multicolumn{2}{@{}l@{}}{\textit{Emotion--span summary (continued)}} \\
\toprule
\textbf{Source} & \textbf{Emotion--span sequence} \\
\midrule
\endhead
\midrule
\multicolumn{2}{r@{}}{\textit{Continued on next page}} \\
\endfoot
\bottomrule
\endlastfoot
\textbf{Reference} & \mbox{disgust $[0.00, 3.25]$};\allowbreak\quad \mbox{neutral $[3.75, 7.75]$};\allowbreak\quad \mbox{neutral $[8.25, 10.50]$} \\ \addlinespace[2pt]
\textbf{Emo-TAG} & \mbox{\qcorrect{disgust} $[\qcorrect{0.00}, \qcorrect{3.25}]$};\allowbreak\quad \mbox{\qcorrect{neutral} $[\qcorrect{3.75}, \qcorrect{7.75}]$};\allowbreak\quad \mbox{\qcorrect{neutral} $[\qcorrect{8.25}, \qcorrect{10.50}]$} \\ \addlinespace[2pt]
\textbf{FlamingoNext} & \mbox{\qcorrect{disgust} $[\qincorrect{0.50}, \qincorrect{2.40}]$};\allowbreak\quad \mbox{\qincorrect{disgust} $[\qincorrect{3.60}, \qincorrect{7.80}]$};\allowbreak\quad \mbox{\qcorrect{neutral} $[\qincorrect{9.00}, \qincorrect{12.00}]$};\allowbreak\quad \mbox{\qincorrect{fear} $[\qincorrect{12.50}, \qincorrect{15.00}]$} \\ \addlinespace[2pt]
\textbf{Audio-Reasoner} & \mbox{\qincorrect{frustration} $[\qcorrect{0.00}, \qincorrect{2.98}]$};\allowbreak\quad \mbox{\qcorrect{neutral} $[\qincorrect{2.98}, \qincorrect{8.08}]$};\allowbreak\quad \mbox{\qincorrect{fear} $[\qincorrect{8.08}, \qincorrect{10.56}]$} \\ \addlinespace[2pt]
\end{longtable}
\endgroup

\begingroup
\small
\setlength{\tabcolsep}{6pt}
\renewcommand{\arraystretch}{1.12}
\setlength{\LTpre}{6pt}\setlength{\LTpost}{8pt}
\begin{longtable}{@{}>{\raggedright\arraybackslash}p{0.32\linewidth}>{\raggedright\arraybackslash}p{\dimexpr0.68\linewidth-2\tabcolsep\relax}@{}}

\noalign{\label{tab:qualitative_full_000285}}
\toprule
\textbf{Span / emotion / time} & \textbf{Speech and tone} \\
\midrule
\endfirsthead
\multicolumn{2}{@{}l@{}}{\textit{Example 3: annotations (continued)}} \\
\toprule
\textbf{Span / emotion / time} & \textbf{Speech and tone} \\
\midrule
\endhead
\midrule
\multicolumn{2}{r@{}}{\textit{Continued on next page}} \\
\endfoot
\bottomrule
\endlastfoot
\multicolumn{2}{@{}l@{}}{\qualmodel{Reference annotation}} \\*
{\footnotesize\textbf{Reference / 1}}\par{\footnotesize\texttt{<speaker1>}\quad\textbf{disgust}}\par {\footnotesize\texttt{<|0.00|>s}--\texttt{<|3.25|>s}} & \textbf{Speech:} ``No ... the leather sticks to my ass.''\par\smallskip
\textbf{Tone:}  \\ \addlinespace[4pt]
{\footnotesize\textbf{Reference / 2}}\par{\footnotesize\texttt{<speaker2>}\quad\textbf{neutral}}\par {\footnotesize\texttt{<|3.75|>s}--\texttt{<|7.75|>s}} & \textbf{Speech:} ``All-all right, well make sure you tell him that Joey Tribbiani stopped by to drop off all of these clothes.''\par\smallskip
\textbf{Tone:}  \\ \addlinespace[4pt]
{\footnotesize\textbf{Reference / 3}}\par{\footnotesize\texttt{<speaker3>}\quad\textbf{neutral}}\par {\footnotesize\texttt{<|8.25|>s}--\texttt{<|10.50|>s}} & \textbf{Speech:} ``I just have to clean that up.''\par\smallskip
\textbf{Tone:}  \\ \addlinespace[4pt]
\multicolumn{2}{@{}l@{}}{\qualmodel{Emo-TAG prediction}} \\*
{\footnotesize\textbf{Emo-TAG / 1}}\par{\footnotesize\texttt{<speaker1>}\quad\qcorrect{\textbf{disgust}}}\par {\footnotesize\qcorrect{\texttt{<|0.00|>s}}--\qcorrect{\texttt{<|3.25|>s}}} & \textbf{Speech:} ``\qcorrect{No}! \qcorrect{The} \qcorrect{leather} \qcorrect{sticks} \qcorrect{to} \qcorrect{my} \qcorrect{ass}!''\par\smallskip
\textbf{Tone:} Flat, resigned delivery with low pitch, moderate volume, slow pace, heavy drawl, and noticeable breathing, conveying irritation and weariness \\ \addlinespace[4pt]
{\footnotesize\textbf{Emo-TAG / 2}}\par{\footnotesize\texttt{<speaker2>}\quad\qcorrect{\textbf{neutral}}}\par {\footnotesize\qcorrect{\texttt{<|3.75|>s}}--\qcorrect{\texttt{<|7.75|>s}}} & \textbf{Speech:} ``\qcorrect{All} \qcorrect{right} \qcorrect{well}, \qcorrect{make} \qcorrect{sure} \qcorrect{you} \qcorrect{tell} \qcorrect{him} \qcorrect{that} \qcorrect{Joey} \qcorrect{Tribbiani} \qcorrect{stopped} \qcorrect{by} \qincorrect{and} \qincorrect{dropped} \qcorrect{off} \qcorrect{all} \qcorrect{these} \qcorrect{clothes}.''\par\smallskip
\textbf{Tone:} Steady, even manner with normal pitch and moderate volume, showing no emotional variation \\ \addlinespace[4pt]
{\footnotesize\textbf{Emo-TAG / 3}}\par{\footnotesize\texttt{<speaker3>}\quad\qcorrect{\textbf{neutral}}}\par {\footnotesize\qcorrect{\texttt{<|8.25|>s}}--\qcorrect{\texttt{<|10.50|>s}}} & \textbf{Speech:} ``\qincorrect{I} \qincorrect{don't} \qincorrect{mean} \qincorrect{to} \qincorrect{scare} \qincorrect{you}, \qcorrect{I} \qcorrect{just} \qcorrect{have} \qcorrect{to} \qcorrect{clean} \qcorrect{that} \qcorrect{up}.''\par\smallskip
\textbf{Tone:} Calm, steady, and controlled with moderate pace and normal volume, showing mild concern without panic or distress \\ \addlinespace[4pt]
\multicolumn{2}{@{}l@{}}{\qualmodel{FlamingoNext prediction}} \\*
{\footnotesize\textbf{FlamingoNext / 1}}\par{\footnotesize\texttt{<Speaker\_1>}\quad\qcorrect{\textbf{disgust}}}\par {\footnotesize\qincorrect{\texttt{<|0.50|>s}}--\qincorrect{\texttt{<|2.50|>s}}} & \textbf{Speech:} ``\qcorrect{No}.''\par\smallskip
\textbf{Tone:}  \\ \addlinespace[4pt]
{\footnotesize\textbf{FlamingoNext / 2}}\par{\footnotesize\texttt{<Speaker\_1>}\quad\qcorrect{\textbf{disgust}}}\par {\footnotesize\qincorrect{\texttt{<|3.50|>s}}--\qincorrect{\texttt{<|7.75|>s}}} & \textbf{Speech:} ``\qcorrect{The} \qcorrect{leather} \qcorrect{sticks} \qcorrect{to} \qcorrect{my} \qcorrect{ass}.''\par\smallskip
\textbf{Tone:}  \\ \addlinespace[4pt]
{\footnotesize\textbf{FlamingoNext / 3}}\par{\footnotesize\texttt{<Speaker\_1>}\quad\qcorrect{\textbf{neutral}}}\par {\footnotesize\qincorrect{\texttt{<|9.00|>s}}--\qincorrect{\texttt{<|12.00|>s}}} & \textbf{Speech:} ``\qcorrect{All} \qcorrect{right}, \qcorrect{well}, \qcorrect{make} \qcorrect{sure} \qcorrect{you} \qcorrect{tell} \qincorrect{them} \qcorrect{Joey} \qcorrect{Tribbiani} \qcorrect{stopped} \qcorrect{by} \qincorrect{and} \qincorrect{dropped} \qcorrect{off} \qcorrect{all} \qcorrect{these} \qcorrect{clothes}''\par\smallskip
\textbf{Tone:}  \\ \addlinespace[4pt]
{\footnotesize\textbf{FlamingoNext / 4}}\par{\footnotesize\texttt{<Speaker\_2>}\quad\qincorrect{\textbf{fear}}}\par {\footnotesize\qincorrect{\texttt{<|12.50|>s}}--\qincorrect{\texttt{<|15.00|>s}}} & \textbf{Speech:} ``\qincorrect{I} \qincorrect{didn't} \qincorrect{mean} \qincorrect{to} \qincorrect{scare} \qincorrect{ya}, \qcorrect{I} \qcorrect{just} \qcorrect{have} \qcorrect{to} \qcorrect{clean} \qcorrect{that} \qcorrect{up}.''\par\smallskip
\textbf{Tone:}  \\ \addlinespace[4pt]
\multicolumn{2}{@{}l@{}}{\qualmodel{Audio-Reasoner prediction}} \\*
{\footnotesize\textbf{Audio-Reasoner / 1}}\par{\footnotesize\texttt{<speaker1>}\quad\qincorrect{\textbf{frustration}}}\par {\footnotesize\qcorrect{\texttt{<|0.00|>s}}--\qincorrect{\texttt{<|3.00|>s}}} & \textbf{Speech:} ``\qcorrect{No}. \qcorrect{The} \qcorrect{leather} \qcorrect{sticks} \qcorrect{to} \qcorrect{my} \qcorrect{ass}.''\par\smallskip
\textbf{Tone:}  \\ \addlinespace[4pt]
{\footnotesize\textbf{Audio-Reasoner / 2}}\par{\footnotesize\texttt{<speaker2>}\quad\qcorrect{\textbf{neutral}}}\par {\footnotesize\qincorrect{\texttt{<|3.00|>s}}--\qincorrect{\texttt{<|8.00|>s}}} & \textbf{Speech:} ``\qincorrect{Alright}, \qcorrect{well}, \qcorrect{make} \qcorrect{sure} \qcorrect{you} \qcorrect{tell} \qcorrect{him} \qcorrect{that} \qcorrect{Joey} \qcorrect{Tribbiani} \qcorrect{stopped} \qcorrect{by} \qincorrect{and} \qincorrect{dropped} \qcorrect{off} \qcorrect{all} \qcorrect{these} \qcorrect{clothes}.''\par\smallskip
\textbf{Tone:}  \\ \addlinespace[4pt]
{\footnotesize\textbf{Audio-Reasoner / 3}}\par{\footnotesize\texttt{<speaker3>}\quad\qincorrect{\textbf{fear}}}\par {\footnotesize\qincorrect{\texttt{<|8.00|>s}}--\qcorrect{\texttt{<|10.50|>s}}} & \textbf{Speech:} ``\qincorrect{I} \qincorrect{didn't} \qincorrect{mean} \qincorrect{to} \qincorrect{scare} \qincorrect{you}. \qcorrect{I} \qcorrect{just} \qcorrect{have} \qcorrect{to} \qcorrect{clean} \qcorrect{that} \qcorrect{up}.''\par\smallskip
\textbf{Tone:}  \\ \addlinespace[4pt]
\end{longtable}
\endgroup

\Needspace{12\baselineskip}
\subsection{Example 4: Correct boundaries but incorrect emotions}
\label{app:qualitative_000026}
\noindent\textbf{Failure}\hfill{\footnotesize\texttt{MELD\_test\_concat\_000026\_k2\_f016dc1781}}\par
\smallskip
\noindent All three models label both utterances as neutral, missing surprise and sadness. Emo-TAG still places both boundaries correctly.

\begingroup
\small
\setlength{\tabcolsep}{5pt}
\renewcommand{\arraystretch}{1.12}
\setlength{\LTpre}{6pt}\setlength{\LTpost}{8pt}
\begin{longtable}{@{}>{\raggedright\arraybackslash}p{0.20\linewidth}>{\raggedright\arraybackslash}p{\dimexpr0.80\linewidth-2\tabcolsep\relax}@{}}
\noalign{\label{tab:qualitative_000026}}
\toprule
\textbf{Source} & \textbf{Emotion--span sequence} \\
\midrule
\endfirsthead
\multicolumn{2}{@{}l@{}}{\textit{Emotion--span summary (continued)}} \\
\toprule
\textbf{Source} & \textbf{Emotion--span sequence} \\
\midrule
\endhead
\midrule
\multicolumn{2}{r@{}}{\textit{Continued on next page}} \\
\endfoot
\bottomrule
\endlastfoot
\textbf{Reference} & \mbox{surprise $[0.00, 1.75]$};\allowbreak\quad \mbox{sadness $[2.75, 4.00]$} \\ \addlinespace[2pt]
\textbf{Emo-TAG} & \mbox{\qincorrect{neutral} $[\qcorrect{0.00}, \qcorrect{1.75}]$};\allowbreak\quad \mbox{\qincorrect{neutral} $[\qcorrect{2.75}, \qcorrect{4.00}]$} \\ \addlinespace[2pt]
\textbf{FlamingoNext} & \mbox{\qincorrect{neutral} $[\qincorrect{0.24}, \qincorrect{3.56}]$};\allowbreak\quad \mbox{\qincorrect{neutral} $[\qincorrect{4.78}, \qincorrect{5.94}]$};\allowbreak\quad \mbox{\qincorrect{neutral} $[\qincorrect{6.54}, \qincorrect{7.34}]$} \\ \addlinespace[2pt]
\textbf{Audio-Reasoner} & \mbox{\qincorrect{neutral} $[\qcorrect{0.00}, \qincorrect{2.98}]$};\allowbreak\quad \mbox{\qincorrect{neutral} $[\qincorrect{3.06}, \qincorrect{4.08}]$} \\ \addlinespace[2pt]
\end{longtable}
\endgroup

\begingroup
\small
\setlength{\tabcolsep}{6pt}
\renewcommand{\arraystretch}{1.12}
\setlength{\LTpre}{6pt}\setlength{\LTpost}{8pt}
\begin{longtable}{@{}>{\raggedright\arraybackslash}p{0.32\linewidth}>{\raggedright\arraybackslash}p{\dimexpr0.68\linewidth-2\tabcolsep\relax}@{}}

\noalign{\label{tab:qualitative_full_000026}}
\toprule
\textbf{Span / emotion / time} & \textbf{Speech and tone} \\
\midrule
\endfirsthead
\multicolumn{2}{@{}l@{}}{\textit{Example 4: annotations (continued)}} \\
\toprule
\textbf{Span / emotion / time} & \textbf{Speech and tone} \\
\midrule
\endhead
\midrule
\multicolumn{2}{r@{}}{\textit{Continued on next page}} \\
\endfoot
\bottomrule
\endlastfoot
\multicolumn{2}{@{}l@{}}{\qualmodel{Reference annotation}} \\*
{\footnotesize\textbf{Reference / 1}}\par{\footnotesize\texttt{<speaker1>}\quad\textbf{surprise}}\par {\footnotesize\texttt{<|0.00|>s}--\texttt{<|1.75|>s}} & \textbf{Speech:} ``Ooh, do I sense a little bit of resentment?''\par\smallskip
\textbf{Tone:}  \\ \addlinespace[4pt]
{\footnotesize\textbf{Reference / 2}}\par{\footnotesize\texttt{<speaker2>}\quad\textbf{sadness}}\par {\footnotesize\texttt{<|2.75|>s}--\texttt{<|4.00|>s}} & \textbf{Speech:} ``No. Sorry.''\par\smallskip
\textbf{Tone:}  \\ \addlinespace[4pt]
\multicolumn{2}{@{}l@{}}{\qualmodel{Emo-TAG prediction}} \\*
{\footnotesize\textbf{Emo-TAG / 1}}\par{\footnotesize\texttt{<speaker1>}\quad\qincorrect{\textbf{neutral}}}\par {\footnotesize\qcorrect{\texttt{<|0.00|>s}}--\qcorrect{\texttt{<|1.75|>s}}} & \textbf{Speech:} ``\qcorrect{Do} \qcorrect{I} \qcorrect{sense} \qcorrect{a} \qcorrect{little} \qcorrect{bit} \qcorrect{of} \qcorrect{resentment}?''\par\smallskip
\textbf{Tone:} Calm, conversational delivery with subtle rising intonation and deliberate pace, suggesting mild curiosity and\textbackslash{}u5C0F\textbackslash{}u5FC3\textbackslash{}u601D \\ \addlinespace[4pt]
{\footnotesize\textbf{Emo-TAG / 2}}\par{\footnotesize\texttt{<speaker2>}\quad\qincorrect{\textbf{neutral}}}\par {\footnotesize\qcorrect{\texttt{<|2.75|>s}}--\qcorrect{\texttt{<|4.00|>s}}} & \textbf{Speech:} ``\qcorrect{No}. \qcorrect{Sorry}.''\par\smallskip
\textbf{Tone:} calm, controlled, soft-to-moderate volume, steady cadence, no emotional inflection \\ \addlinespace[4pt]
\multicolumn{2}{@{}l@{}}{\qualmodel{FlamingoNext prediction}} \\*
{\footnotesize\textbf{FlamingoNext / 1}}\par{\footnotesize\texttt{<Speaker\_0>}\quad\qincorrect{\textbf{neutral}}}\par {\footnotesize\qincorrect{\texttt{<|0.25|>s}}--\qincorrect{\texttt{<|3.50|>s}}} & \textbf{Speech:} ``\qcorrect{Do} \qcorrect{I} \qcorrect{sense} \qcorrect{a} \qcorrect{little} \qcorrect{bit} \qcorrect{of} \qcorrect{resentment}?''\par\smallskip
\textbf{Tone:}  \\ \addlinespace[4pt]
{\footnotesize\textbf{FlamingoNext / 2}}\par{\footnotesize\texttt{<Speaker\_1>}\quad\qincorrect{\textbf{neutral}}}\par {\footnotesize\qincorrect{\texttt{<|4.75|>s}}--\qincorrect{\texttt{<|6.00|>s}}} & \textbf{Speech:} ``\qcorrect{No}.''\par\smallskip
\textbf{Tone:}  \\ \addlinespace[4pt]
{\footnotesize\textbf{FlamingoNext / 3}}\par{\footnotesize\texttt{<Speaker\_1>}\quad\qincorrect{\textbf{neutral}}}\par {\footnotesize\qincorrect{\texttt{<|6.50|>s}}--\qincorrect{\texttt{<|7.25|>s}}} & \textbf{Speech:} ``\qcorrect{Sorry}''\par\smallskip
\textbf{Tone:}  \\ \addlinespace[4pt]
\multicolumn{2}{@{}l@{}}{\qualmodel{Audio-Reasoner prediction}} \\*
{\footnotesize\textbf{Audio-Reasoner / 1}}\par{\footnotesize\texttt{<speaker1>}\quad\qincorrect{\textbf{neutral}}}\par {\footnotesize\qcorrect{\texttt{<|0.00|>s}}--\qincorrect{\texttt{<|3.00|>s}}} & \textbf{Speech:} ``\qcorrect{Do} \qcorrect{I} \qcorrect{sense} \qcorrect{a} \qcorrect{little} \qcorrect{bit} \qcorrect{of} \qcorrect{resentment}?''\par\smallskip
\textbf{Tone:}  \\ \addlinespace[4pt]
{\footnotesize\textbf{Audio-Reasoner / 2}}\par{\footnotesize\texttt{<speaker2>}\quad\qincorrect{\textbf{neutral}}}\par {\footnotesize\qincorrect{\texttt{<|3.00|>s}}--\qcorrect{\texttt{<|4.00|>s}}} & \textbf{Speech:} ``\qcorrect{No}, \qcorrect{sorry}.''\par\smallskip
\textbf{Tone:}  \\ \addlinespace[4pt]
\end{longtable}
\endgroup

\Needspace{12\baselineskip}
\subsection{Example 5: Merging utterances with different emotions}
\label{app:qualitative_000658}
\noindent\textbf{Failure}\hfill{\footnotesize\texttt{MELD\_test\_concat\_000658\_k3\_65d90b2c2e}}\par
\smallskip
\noindent Emo-TAG merges the last two utterances into one surprise span, missing the change to neutral and including the gap between them. The baselines produce duplicate or extra spans, and all three models misclassify the first emotion.

\begingroup
\small
\setlength{\tabcolsep}{5pt}
\renewcommand{\arraystretch}{1.12}
\setlength{\LTpre}{6pt}\setlength{\LTpost}{8pt}
\begin{longtable}{@{}>{\raggedright\arraybackslash}p{0.20\linewidth}>{\raggedright\arraybackslash}p{\dimexpr0.80\linewidth-2\tabcolsep\relax}@{}}
\noalign{\label{tab:qualitative_000658}}
\toprule
\textbf{Source} & \textbf{Emotion--span sequence} \\
\midrule
\endfirsthead
\multicolumn{2}{@{}l@{}}{\textit{Emotion--span summary (continued)}} \\
\toprule
\textbf{Source} & \textbf{Emotion--span sequence} \\
\midrule
\endhead
\midrule
\multicolumn{2}{r@{}}{\textit{Continued on next page}} \\
\endfoot
\bottomrule
\endlastfoot
\textbf{Reference} & \mbox{happiness $[0.00, 4.25]$};\allowbreak\quad \mbox{surprise $[5.25, 6.00]$};\allowbreak\quad \mbox{neutral $[6.75, 7.00]$} \\ \addlinespace[2pt]
\textbf{Emo-TAG} & \mbox{\qincorrect{neutral} $[\qcorrect{0.00}, \qcorrect{4.25}]$};\allowbreak\quad \mbox{\qcorrect{surprise} $[\qincorrect{5.00}, \qincorrect{7.00}]$} \\ \addlinespace[2pt]
\textbf{FlamingoNext} & \mbox{\qincorrect{neutral} $[\qcorrect{0.00}, \qincorrect{5.00}]$};\allowbreak\quad \mbox{\qcorrect{surprise} $[\qincorrect{5.00}, \qincorrect{5.00}]$};\allowbreak\quad \mbox{\qcorrect{neutral} $[\qincorrect{6.00}, \qcorrect{7.00}]$};\allowbreak\quad \mbox{\qincorrect{neutral} $[\qincorrect{6.00}, \qincorrect{7.00}]$} \\ \addlinespace[2pt]
\textbf{Audio-Reasoner} & \mbox{\qincorrect{neutral} $[\qcorrect{0.00}, \qincorrect{5.47}]$};\allowbreak\quad \mbox{\qcorrect{surprise} $[\qincorrect{5.47}, \qincorrect{6.08}]$};\allowbreak\quad \mbox{\qcorrect{neutral} $[\qincorrect{6.08}, \qincorrect{6.78}]$};\allowbreak\quad \mbox{\qincorrect{surprise} $[\qincorrect{6.80}, \qincorrect{7.08}]$};\allowbreak\quad \mbox{\qincorrect{neutral} $[\qincorrect{7.08}, \qincorrect{7.78}]$};\allowbreak\quad \mbox{\qincorrect{surprise} $[\qincorrect{7.80}, \qincorrect{8.08}]$};\allowbreak\quad \mbox{\qincorrect{neutral} $[\qincorrect{8.08}, \qincorrect{8.78}]$};\allowbreak\quad \mbox{\qincorrect{surprise} $[\qincorrect{8.80}, \qincorrect{9.08}]$} \\ \addlinespace[2pt]
\end{longtable}
\endgroup

\begingroup
\small
\setlength{\tabcolsep}{6pt}
\renewcommand{\arraystretch}{1.12}
\setlength{\LTpre}{6pt}\setlength{\LTpost}{8pt}
\begin{longtable}{@{}>{\raggedright\arraybackslash}p{0.32\linewidth}>{\raggedright\arraybackslash}p{\dimexpr0.68\linewidth-2\tabcolsep\relax}@{}}

\noalign{\label{tab:qualitative_full_000658}}
\toprule
\textbf{Span / emotion / time} & \textbf{Speech and tone} \\
\midrule
\endfirsthead
\multicolumn{2}{@{}l@{}}{\textit{Example 5: annotations (continued)}} \\
\toprule
\textbf{Span / emotion / time} & \textbf{Speech and tone} \\
\midrule
\endhead
\midrule
\multicolumn{2}{r@{}}{\textit{Continued on next page}} \\
\endfoot
\bottomrule
\endlastfoot
\multicolumn{2}{@{}l@{}}{\qualmodel{Reference annotation}} \\*
{\footnotesize\textbf{Reference / 1}}\par{\footnotesize\texttt{<speaker1>}\quad\textbf{happiness}}\par {\footnotesize\texttt{<|0.00|>s}--\texttt{<|4.25|>s}} & \textbf{Speech:} ``Unless, we give her all gifts she can use after she's done being pregnant.''\par\smallskip
\textbf{Tone:}  \\ \addlinespace[4pt]
{\footnotesize\textbf{Reference / 2}}\par{\footnotesize\texttt{<speaker2>}\quad\textbf{surprise}}\par {\footnotesize\texttt{<|5.25|>s}--\texttt{<|6.00|>s}} & \textbf{Speech:} ``Oh my god.''\par\smallskip
\textbf{Tone:}  \\ \addlinespace[4pt]
{\footnotesize\textbf{Reference / 3}}\par{\footnotesize\texttt{<speaker3>}\quad\textbf{neutral}}\par {\footnotesize\texttt{<|6.75|>s}--\texttt{<|7.00|>s}} & \textbf{Speech:} ``You think?''\par\smallskip
\textbf{Tone:}  \\ \addlinespace[4pt]
\multicolumn{2}{@{}l@{}}{\qualmodel{Emo-TAG prediction}} \\*
{\footnotesize\textbf{Emo-TAG / 1}}\par{\footnotesize\texttt{<speaker1>}\quad\qincorrect{\textbf{neutral}}}\par {\footnotesize\qcorrect{\texttt{<|0.00|>s}}--\qcorrect{\texttt{<|4.25|>s}}} & \textbf{Speech:} ``\qcorrect{Unless} \qcorrect{we} \qcorrect{give} \qcorrect{her} \qcorrect{all} \qcorrect{gifts} \qcorrect{she} \qcorrect{can} \qcorrect{use} \qcorrect{after} \qcorrect{she's} \qcorrect{done} \qcorrect{being} \qcorrect{pregnant}.''\par\smallskip
\textbf{Tone:} Steady, even delivery with consistent pitch and moderate volume, showing no emotional inflection or variation in pace \\ \addlinespace[4pt]
{\footnotesize\textbf{Emo-TAG / 2}}\par{\footnotesize\texttt{<speaker2>}\quad\qcorrect{\textbf{surprise}}}\par {\footnotesize\qincorrect{\texttt{<|5.00|>s}}--\qincorrect{\texttt{<|7.00|>s}}} & \textbf{Speech:} ``\qcorrect{Oh} \qcorrect{my} \qcorrect{God}! \qincorrect{What'd} \qcorrect{you} \qcorrect{think}?''\par\smallskip
\textbf{Tone:} Startled exclamation followed by urgent, inquiry-filled questioning with rising intonation, conveying shock and urgency \\ \addlinespace[4pt]
\multicolumn{2}{@{}l@{}}{\qualmodel{FlamingoNext prediction}} \\*
{\footnotesize\textbf{FlamingoNext / 1}}\par{\footnotesize\texttt{<Speaker\_1>}\quad\qincorrect{\textbf{neutral}}}\par {\footnotesize\qcorrect{\texttt{<|0.00|>s}}--\qincorrect{\texttt{<|5.00|>s}}} & \textbf{Speech:} ``\qcorrect{Unless} \qcorrect{we} \qcorrect{give} \qcorrect{her} \qcorrect{all} \qcorrect{gifts} \qincorrect{that} \qcorrect{she} \qcorrect{can} \qcorrect{use} \qcorrect{after} \qcorrect{she's} \qcorrect{done} \qcorrect{being} \qcorrect{pregnant}.''\par\smallskip
\textbf{Tone:}  \\ \addlinespace[4pt]
{\footnotesize\textbf{FlamingoNext / 2}}\par{\footnotesize\texttt{<Speaker\_2>}\quad\qcorrect{\textbf{surprise}}}\par {\footnotesize\qincorrect{\texttt{<|5.00|>s}}--\qincorrect{\texttt{<|5.00|>s}}} & \textbf{Speech:} ``\qcorrect{Oh} \qcorrect{my} \qcorrect{god}.''\par\smallskip
\textbf{Tone:}  \\ \addlinespace[4pt]
{\footnotesize\textbf{FlamingoNext / 3}}\par{\footnotesize\texttt{<Speaker\_1>}\quad\qcorrect{\textbf{neutral}}}\par {\footnotesize\qincorrect{\texttt{<|6.00|>s}}--\qcorrect{\texttt{<|7.00|>s}}} & \textbf{Speech:} ``\qincorrect{Do} \qcorrect{you} \qcorrect{think}?''\par\smallskip
\textbf{Tone:}  \\ \addlinespace[4pt]
{\footnotesize\qincorrect{\textbf{FlamingoNext / 4}}}\par{\footnotesize\texttt{<Speaker\_2>}\quad\qincorrect{\textbf{neutral}}}\par {\footnotesize\qincorrect{\texttt{<|6.00|>s}}--\qincorrect{\texttt{<|7.00|>s}}} & \textbf{Speech:} ``\qincorrect{Do} \qincorrect{you} \qincorrect{think}?''\par\smallskip
\textbf{Tone:}  \\ \addlinespace[4pt]
\multicolumn{2}{@{}l@{}}{\qualmodel{Audio-Reasoner prediction}} \\*
{\footnotesize\textbf{Audio-Reasoner / 1}}\par{\footnotesize\texttt{<speaker1>}\quad\qincorrect{\textbf{neutral}}}\par {\footnotesize\qcorrect{\texttt{<|0.00|>s}}--\qincorrect{\texttt{<|5.50|>s}}} & \textbf{Speech:} ``\qcorrect{Unless} \qcorrect{we} \qcorrect{give} \qcorrect{her} \qcorrect{all} \qcorrect{gifts} \qincorrect{that} \qcorrect{she} \qcorrect{can} \qcorrect{use} \qcorrect{after} \qcorrect{she's} \qcorrect{done} \qcorrect{being} \qcorrect{pregnant}.''\par\smallskip
\textbf{Tone:}  \\ \addlinespace[4pt]
{\footnotesize\textbf{Audio-Reasoner / 2}}\par{\footnotesize\texttt{<speaker2>}\quad\qcorrect{\textbf{surprise}}}\par {\footnotesize\qincorrect{\texttt{<|5.50|>s}}--\qcorrect{\texttt{<|6.00|>s}}} & \textbf{Speech:} ``\qcorrect{Oh} \qcorrect{my} \qcorrect{God}!''\par\smallskip
\textbf{Tone:}  \\ \addlinespace[4pt]
{\footnotesize\textbf{Audio-Reasoner / 3}}\par{\footnotesize\texttt{<speaker3>}\quad\qcorrect{\textbf{neutral}}}\par {\footnotesize\qincorrect{\texttt{<|6.00|>s}}--\qincorrect{\texttt{<|6.75|>s}}} & \textbf{Speech:} ``\qincorrect{Do} \qcorrect{you} \qcorrect{think}?''\par\smallskip
\textbf{Tone:}  \\ \addlinespace[4pt]
{\footnotesize\qincorrect{\textbf{Audio-Reasoner / 4}}}\par{\footnotesize\texttt{<speaker4>}\quad\qincorrect{\textbf{surprise}}}\par {\footnotesize\qincorrect{\texttt{<|6.75|>s}}--\qincorrect{\texttt{<|7.00|>s}}} & \textbf{Speech:} ``\qincorrect{Oh} \qincorrect{my} \qincorrect{God}!''\par\smallskip
\textbf{Tone:}  \\ \addlinespace[4pt]
{\footnotesize\qincorrect{\textbf{Audio-Reasoner / 5}}}\par{\footnotesize\texttt{<speaker5>}\quad\qincorrect{\textbf{neutral}}}\par {\footnotesize\qincorrect{\texttt{<|7.00|>s}}--\qincorrect{\texttt{<|7.75|>s}}} & \textbf{Speech:} ``\qincorrect{Do} \qincorrect{you} \qincorrect{think}?''\par\smallskip
\textbf{Tone:}  \\ \addlinespace[4pt]
{\footnotesize\qincorrect{\textbf{Audio-Reasoner / 6}}}\par{\footnotesize\texttt{<speaker6>}\quad\qincorrect{\textbf{surprise}}}\par {\footnotesize\qincorrect{\texttt{<|7.75|>s}}--\qincorrect{\texttt{<|8.00|>s}}} & \textbf{Speech:} ``\qincorrect{Oh} \qincorrect{my} \qincorrect{God}!''\par\smallskip
\textbf{Tone:}  \\ \addlinespace[4pt]
{\footnotesize\qincorrect{\textbf{Audio-Reasoner / 7}}}\par{\footnotesize\texttt{<speaker7>}\quad\qincorrect{\textbf{neutral}}}\par {\footnotesize\qincorrect{\texttt{<|8.00|>s}}--\qincorrect{\texttt{<|8.75|>s}}} & \textbf{Speech:} ``\qincorrect{Do} \qincorrect{you} \qincorrect{think}?''\par\smallskip
\textbf{Tone:}  \\ \addlinespace[4pt]
{\footnotesize\qincorrect{\textbf{Audio-Reasoner / 8}}}\par{\footnotesize\texttt{<speaker8>}\quad\qincorrect{\textbf{surprise}}}\par {\footnotesize\qincorrect{\texttt{<|8.75|>s}}--\qincorrect{\texttt{<|9.00|>s}}} & \textbf{Speech:} ``\qincorrect{Oh} \qincorrect{my} \qincorrect{God}!''\par\smallskip
\textbf{Tone:}  \\ \addlinespace[4pt]
{\footnotesize\qincorrect{\textbf{Audio-Reasoner / 9}}}\par{\footnotesize\texttt{<speaker9>}\quad\qincorrect{\textbf{unknown}}}\par \qincorrect{\textit{Time not supplied}} & \textbf{Speech:} ``\qincorrect{Do} \qincorrect{you} \qincorrect{think}?''\par\smallskip
\textbf{Tone:}  \\ \addlinespace[4pt]
\end{longtable}
\endgroup

\Needspace{12\baselineskip}
\subsection{Example 6: Incomplete temporal coverage}
\label{app:qualitative_000089}
\noindent\textbf{Failure}\hfill{\footnotesize\texttt{MELD\_test\_concat\_000089\_k3\_045055ccdc}}\par
\smallskip
\noindent Emo-TAG gets the emotion sequence right but ends the second span 2.25 seconds early. Both baselines miss the initial anger and add spans beyond the final reference utterance.

\begingroup
\small
\setlength{\tabcolsep}{5pt}
\renewcommand{\arraystretch}{1.12}
\setlength{\LTpre}{6pt}\setlength{\LTpost}{8pt}
\begin{longtable}{@{}>{\raggedright\arraybackslash}p{0.20\linewidth}>{\raggedright\arraybackslash}p{\dimexpr0.80\linewidth-2\tabcolsep\relax}@{}}
\noalign{\label{tab:qualitative_000089}}
\toprule
\textbf{Source} & \textbf{Emotion--span sequence} \\
\midrule
\endfirsthead
\multicolumn{2}{@{}l@{}}{\textit{Emotion--span summary (continued)}} \\
\toprule
\textbf{Source} & \textbf{Emotion--span sequence} \\
\midrule
\endhead
\midrule
\multicolumn{2}{r@{}}{\textit{Continued on next page}} \\
\endfoot
\bottomrule
\endlastfoot
\textbf{Reference} & \mbox{anger $[0.00, 4.75]$};\allowbreak\quad \mbox{neutral $[5.50, 10.25]$};\allowbreak\quad \mbox{neutral $[10.75, 11.25]$} \\ \addlinespace[2pt]
\textbf{Emo-TAG} & \mbox{\qcorrect{anger} $[\qcorrect{0.00}, \qcorrect{4.75}]$};\allowbreak\quad \mbox{\qcorrect{neutral} $[\qcorrect{5.50}, \qincorrect{8.00}]$};\allowbreak\quad \mbox{\qcorrect{neutral} $[\qcorrect{10.75}, \qcorrect{11.25}]$} \\ \addlinespace[2pt]
\textbf{FlamingoNext} & \mbox{\qincorrect{neutral} $[\qcorrect{0.00}, \qincorrect{2.00}]$};\allowbreak\quad \mbox{\qincorrect{neutral} $[\qincorrect{3.00}, \qincorrect{4.00}]$};\allowbreak\quad \mbox{\qincorrect{neutral} $[\qincorrect{6.00}, \qincorrect{7.00}]$};\allowbreak\quad \mbox{\qincorrect{surprise} $[\qincorrect{8.00}, \qincorrect{19.00}]$} \\ \addlinespace[2pt]
\textbf{Audio-Reasoner} & \mbox{\qincorrect{neutral} $[\qcorrect{0.00}, \qincorrect{2.00}]$};\allowbreak\quad \mbox{\qincorrect{neutral} $[\qincorrect{2.00}, \qincorrect{8.00}]$};\allowbreak\quad \mbox{\qcorrect{neutral} $[\qincorrect{8.00}, \qincorrect{10.00}]$};\allowbreak\quad \mbox{\qcorrect{neutral} $[\qincorrect{10.00}, \qincorrect{14.00}]$};\allowbreak\quad \mbox{\qcorrect{neutral} $[\qincorrect{14.00}, \qincorrect{18.00}]$};\allowbreak\quad \mbox{\qcorrect{neutral} $[\qincorrect{18.00}, \qincorrect{20.00}]$};\allowbreak\quad \mbox{\qincorrect{neutral} $[\qincorrect{20.00}, \qincorrect{21.00}]$};\allowbreak\quad \mbox{\qincorrect{neutral} $[\qincorrect{21.00}, \qincorrect{22.00}]$};\allowbreak\quad \mbox{\qincorrect{neutral} $[\qincorrect{22.00}, \qincorrect{23.00}]$} \\ \addlinespace[2pt]
\end{longtable}
\endgroup

\begingroup
\small
\setlength{\tabcolsep}{6pt}
\renewcommand{\arraystretch}{1.12}
\setlength{\LTpre}{6pt}\setlength{\LTpost}{8pt}
\begin{longtable}{@{}>{\raggedright\arraybackslash}p{0.32\linewidth}>{\raggedright\arraybackslash}p{\dimexpr0.68\linewidth-2\tabcolsep\relax}@{}}
\noalign{\label{tab:qualitative_full_000089}}
\toprule
\textbf{Span / emotion / time} & \textbf{Speech and tone} \\
\midrule
\endfirsthead
\multicolumn{2}{@{}l@{}}{\textit{Example 6: annotations (continued)}} \\
\toprule
\textbf{Span / emotion / time} & \textbf{Speech and tone} \\
\midrule
\endhead
\midrule
\multicolumn{2}{r@{}}{\textit{Continued on next page}} \\
\endfoot
\bottomrule
\endlastfoot
\multicolumn{2}{@{}l@{}}{\qualmodel{Reference annotation}} \\*
{\footnotesize\textbf{Reference / 1}}\par{\footnotesize\texttt{<speaker1>}\quad\textbf{anger}}\par {\footnotesize\texttt{<|0.00|>s}--\texttt{<|4.75|>s}} & \textbf{Speech:} ``Oh yeah? If I can't, I'll knock five bucks off the price off the unit.''\par\smallskip
\textbf{Tone:}  \\ \addlinespace[4pt]
{\footnotesize\textbf{Reference / 2}}\par{\footnotesize\texttt{<speaker2>}\quad\textbf{neutral}}\par {\footnotesize\texttt{<|5.50|>s}--\texttt{<|10.25|>s}} & \textbf{Speech:} ``Ohh, well. Y'know we were always together, in fact the had a nickname for the three of us.''\par\smallskip
\textbf{Tone:}  \\ \addlinespace[4pt]
{\footnotesize\textbf{Reference / 3}}\par{\footnotesize\texttt{<speaker3>}\quad\textbf{neutral}}\par {\footnotesize\texttt{<|10.75|>s}--\texttt{<|11.25|>s}} & \textbf{Speech:} ``Diverse.''\par\smallskip
\textbf{Tone:}  \\ \addlinespace[4pt]
\multicolumn{2}{@{}l@{}}{\qualmodel{Emo-TAG prediction}} \\*
{\footnotesize\textbf{Emo-TAG / 1}}\par{\footnotesize\texttt{<speaker1>}\quad\qcorrect{\textbf{anger}}}\par {\footnotesize\qcorrect{\texttt{<|0.00|>s}}--\qcorrect{\texttt{<|4.75|>s}}} & \textbf{Speech:} ``\qcorrect{Oh} \qcorrect{yeah}? \qcorrect{If} \qcorrect{I} \qcorrect{can't}, \qcorrect{I'll} \qcorrect{knock} \qcorrect{five} \qcorrect{bucks} \qcorrect{off} \qcorrect{the} \qcorrect{price} \qincorrect{of} \qcorrect{the} \qcorrect{unit}.''\par\smallskip
\textbf{Tone:} Starts with high pitch, fast-paced excitement, shifts to calm explanation, then ends with low-pitched, slow, grand statement \\ \addlinespace[4pt]
{\footnotesize\textbf{Emo-TAG / 2}}\par{\footnotesize\texttt{<speaker2>}\quad\qcorrect{\textbf{neutral}}}\par {\footnotesize\qcorrect{\texttt{<|5.50|>s}}--\qincorrect{\texttt{<|8.00|>s}}} & \textbf{Speech:} ``\qincorrect{Oh} \qcorrect{well}, \qcorrect{we} \qcorrect{were} \qcorrect{always} \qcorrect{together}. \qincorrect{You} \qincorrect{know}, \qcorrect{the} \qcorrect{three} \qcorrect{of} \qcorrect{us}.''\par\smallskip
\textbf{Tone:} Calm, steady delivery with steady pitch and volume, slight pauses between phrases, reflective and sincere affect \\ \addlinespace[4pt]
{\footnotesize\textbf{Emo-TAG / 3}}\par{\footnotesize\texttt{<speaker3>}\quad\qcorrect{\textbf{neutral}}}\par {\footnotesize\qcorrect{\texttt{<|10.75|>s}}--\qcorrect{\texttt{<|11.25|>s}}} & \textbf{Speech:} ``\qincorrect{Diversity}.''\par\smallskip
\textbf{Tone:} Neutral, steady delivery with moderate pace, normal volume, and clear articulation; no emotional inflection or vocal strain detected \\ \addlinespace[4pt]
\multicolumn{2}{@{}l@{}}{\qualmodel{FlamingoNext prediction}} \\*
{\footnotesize\textbf{FlamingoNext / 1}}\par{\footnotesize\texttt{<speaker1>}\quad\qincorrect{\textbf{neutral}}}\par {\footnotesize\qcorrect{\texttt{<|0.00|>s}}--\qincorrect{\texttt{<|2.00|>s}}} & \textbf{Speech:} ``\qcorrect{Oh} \qcorrect{yeah}?''\par\smallskip
\textbf{Tone:}  \\ \addlinespace[4pt]
{\footnotesize\textbf{FlamingoNext / 2}}\par{\footnotesize\texttt{<speaker1>}\quad\qincorrect{\textbf{neutral}}}\par {\footnotesize\qincorrect{\texttt{<|3.00|>s}}--\qincorrect{\texttt{<|4.00|>s}}} & \textbf{Speech:} ``\qcorrect{If} \qcorrect{I} \qcorrect{can't}.''\par\smallskip
\textbf{Tone:}  \\ \addlinespace[4pt]
{\footnotesize\textbf{FlamingoNext / 3}}\par{\footnotesize\texttt{<speaker1>}\quad\qincorrect{\textbf{neutral}}}\par {\footnotesize\qincorrect{\texttt{<|6.00|>s}}--\qincorrect{\texttt{<|7.00|>s}}} & \textbf{Speech:} ``\qcorrect{I'll} \qcorrect{knock} \qcorrect{five} \qcorrect{bucks} \qcorrect{off} \qcorrect{the} \qcorrect{price} \qincorrect{of} \qcorrect{the} \qcorrect{unit}.''\par\smallskip
\textbf{Tone:}  \\ \addlinespace[4pt]
{\footnotesize\textbf{FlamingoNext / 4}}\par{\footnotesize\texttt{<speaker1>}\quad\qincorrect{\textbf{surprise}}}\par {\footnotesize\qincorrect{\texttt{<|8.00|>s}}--\qincorrect{\texttt{<|19.00|>s}}} & \textbf{Speech:} ``\qcorrect{We} \qcorrect{were} \qcorrect{always} \qcorrect{together}... \qcorrect{In} \qcorrect{fact}, \qcorrect{the} \qincorrect{other} \qincorrect{kids} \qcorrect{had} \qcorrect{a} \qcorrect{nickname} \qcorrect{for} \qcorrect{the} \qcorrect{three} \qcorrect{of} \qcorrect{us}... \qincorrect{Divorce}.''\par\smallskip
\textbf{Tone:}  \\ \addlinespace[4pt]
\multicolumn{2}{@{}l@{}}{\qualmodel{Audio-Reasoner prediction}} \\*
{\footnotesize\textbf{Audio-Reasoner / 1}}\par{\footnotesize\texttt{<speaker1>}\quad\qincorrect{\textbf{neutral}}}\par {\footnotesize\qcorrect{\texttt{<|0.00|>s}}--\qincorrect{\texttt{<|2.00|>s}}} & \textbf{Speech:} ``\qcorrect{Oh} \qcorrect{yeah}.''\par\smallskip
\textbf{Tone:}  \\ \addlinespace[4pt]
{\footnotesize\textbf{Audio-Reasoner / 2}}\par{\footnotesize\texttt{<speaker2>}\quad\qincorrect{\textbf{neutral}}}\par {\footnotesize\qincorrect{\texttt{<|2.00|>s}}--\qincorrect{\texttt{<|8.00|>s}}} & \textbf{Speech:} ``\qcorrect{If} \qcorrect{I} \qcorrect{can't}, \qcorrect{I'll} \qcorrect{knock} \qcorrect{five} \qcorrect{bucks} \qcorrect{off} \qcorrect{the} \qcorrect{price} \qincorrect{of} \qcorrect{the} \qcorrect{unit}.''\par\smallskip
\textbf{Tone:}  \\ \addlinespace[4pt]
{\footnotesize\textbf{Audio-Reasoner / 3}}\par{\footnotesize\texttt{<speaker3>}\quad\qcorrect{\textbf{neutral}}}\par {\footnotesize\qincorrect{\texttt{<|8.00|>s}}--\qincorrect{\texttt{<|10.00|>s}}} & \textbf{Speech:} ``\qincorrect{Oh} \qcorrect{well}.''\par\smallskip
\textbf{Tone:}  \\ \addlinespace[4pt]
{\footnotesize\textbf{Audio-Reasoner / 4}}\par{\footnotesize\texttt{<speaker4>}\quad\qcorrect{\textbf{neutral}}}\par {\footnotesize\qincorrect{\texttt{<|10.00|>s}}--\qincorrect{\texttt{<|14.00|>s}}} & \textbf{Speech:} ``\qincorrect{No}, \qcorrect{we} \qcorrect{were} \qcorrect{always} \qcorrect{together}.''\par\smallskip
\textbf{Tone:}  \\ \addlinespace[4pt]
{\footnotesize\textbf{Audio-Reasoner / 5}}\par{\footnotesize\texttt{<speaker5>}\quad\qcorrect{\textbf{neutral}}}\par {\footnotesize\qincorrect{\texttt{<|14.00|>s}}--\qincorrect{\texttt{<|18.00|>s}}} & \textbf{Speech:} ``\qcorrect{In} \qcorrect{fact}, \qcorrect{the} \qincorrect{other} \qincorrect{kids} \qcorrect{had} \qcorrect{a} \qcorrect{nickname} \qcorrect{for} \qcorrect{the} \qcorrect{three} \qcorrect{of} \qcorrect{us}.''\par\smallskip
\textbf{Tone:}  \\ \addlinespace[4pt]
{\footnotesize\textbf{Audio-Reasoner / 6}}\par{\footnotesize\texttt{<speaker6>}\quad\qcorrect{\textbf{neutral}}}\par {\footnotesize\qincorrect{\texttt{<|18.00|>s}}--\qincorrect{\texttt{<|20.00|>s}}} & \textbf{Speech:} ``\qincorrect{Divided}.''\par\smallskip
\textbf{Tone:}  \\ \addlinespace[4pt]
{\footnotesize\qincorrect{\textbf{Audio-Reasoner / 7}}}\par{\footnotesize\texttt{<speaker7>}\quad\qincorrect{\textbf{neutral}}}\par {\footnotesize\qincorrect{\texttt{<|20.00|>s}}--\qincorrect{\texttt{<|21.00|>s}}} & \textbf{Speech:} ``\qincorrect{Oh}.''\par\smallskip
\textbf{Tone:}  \\ \addlinespace[4pt]
{\footnotesize\qincorrect{\textbf{Audio-Reasoner / 8}}}\par{\footnotesize\texttt{<speaker8>}\quad\qincorrect{\textbf{neutral}}}\par {\footnotesize\qincorrect{\texttt{<|21.00|>s}}--\qincorrect{\texttt{<|22.00|>s}}} & \textbf{Speech:} ``\qincorrect{Yeah}.''\par\smallskip
\textbf{Tone:}  \\ \addlinespace[4pt]
{\footnotesize\qincorrect{\textbf{Audio-Reasoner / 9}}}\par{\footnotesize\texttt{<speaker9>}\quad\qincorrect{\textbf{neutral}}}\par {\footnotesize\qincorrect{\texttt{<|22.00|>s}}--\qincorrect{\texttt{<|23.00|>s}}} & \textbf{Speech:} ``\qincorrect{Oh}.''\par\smallskip
\textbf{Tone:}  \\ \addlinespace[4pt]
\end{longtable}
\endgroup

\Needspace{12\baselineskip}
\subsection{Example 7: Emotion and temporal accuracy}
\label{app:iemocap_qualitative_000198}
\noindent\textbf{Success}\hfill{\footnotesize\texttt{IEMOCAP\_train\_concat\_000198\_k2\_674c73dbf8}}\par
\smallskip\noindent Emo-TAG matches both emotions and their boundaries. Audio-Reasoner shifts the boundary between them, while FlamingoNext splits the first utterance.

\begingroup
\small
\setlength{\tabcolsep}{6pt}
\renewcommand{\arraystretch}{1.12}
\setlength{\LTpre}{6pt}\setlength{\LTpost}{8pt}
\begin{longtable}{@{}>{\raggedright\arraybackslash}p{0.2\linewidth}>{\raggedright\arraybackslash}p{\dimexpr0.8\linewidth-2\tabcolsep\relax}@{}}
\noalign{\label{tab:iemocap_qualitative_000198}}
\toprule
\textbf{Source} & \textbf{Emotion--span sequence} \\
\midrule
\endfirsthead
\multicolumn{2}{@{}l@{}}{\textit{Emotion--span summary (continued)}} \\
\toprule
\textbf{Source} & \textbf{Emotion--span sequence} \\
\midrule
\endhead
\midrule
\multicolumn{2}{r@{}}{\textit{Continued on next page}} \\
\endfoot
\bottomrule
\endlastfoot
\textbf{Reference} & \mbox{anger $[0.00, 4.25]$};\allowbreak\quad \mbox{neutral $[5.00, 6.75]$} \\ \addlinespace[2pt]
\textbf{Emo-TAG} & \mbox{\qcorrect{anger} $[\qcorrect{0.00}, \qcorrect{4.25}]$};\allowbreak\quad \mbox{\qcorrect{neutral} $[\qcorrect{5.00}, \qcorrect{6.75}]$} \\ \addlinespace[2pt]
\textbf{FlamingoNext} & \mbox{\qcorrect{anger} $[\qincorrect{1.00}, \qincorrect{5.00}]$};\allowbreak\quad \mbox{\qcorrect{anger} $[\qincorrect{6.00}, \qincorrect{9.00}]$};\allowbreak\quad \mbox{\qcorrect{neutral} $[\qincorrect{9.00}, \qincorrect{11.00}]$} \\ \addlinespace[2pt]
\textbf{Audio-Reasoner} & \mbox{\qcorrect{anger} $[\qcorrect{0.00}, \qincorrect{4.75}]$};\allowbreak\quad \mbox{\qcorrect{neutral} $[\qincorrect{4.75}, \qcorrect{6.75}]$} \\ \addlinespace[2pt]
\end{longtable}
\endgroup

\begingroup
\small
\setlength{\tabcolsep}{6pt}
\renewcommand{\arraystretch}{1.12}
\setlength{\LTpre}{6pt}\setlength{\LTpost}{8pt}
\begin{longtable}{@{}>{\raggedright\arraybackslash}p{0.32\linewidth}>{\raggedright\arraybackslash}p{\dimexpr0.68\linewidth-2\tabcolsep\relax}@{}}
\noalign{\label{tab:iemocap_qualitative_full_000198}}
\toprule
\textbf{Span / emotion / time} & \textbf{Speech and tone} \\
\midrule
\endfirsthead
\multicolumn{2}{@{}l@{}}{\textit{Example 7: annotations (continued)}} \\
\toprule
\textbf{Span / emotion / time} & \textbf{Speech and tone} \\
\midrule
\endhead
\midrule
\multicolumn{2}{r@{}}{\textit{Continued on next page}} \\
\endfoot
\bottomrule
\endlastfoot
\multicolumn{2}{@{}l@{}}{\qualmodel{Reference}} \\*
{\footnotesize\textbf{Reference / 1}}\par{\footnotesize\texttt{<speaker1>}\quad \textbf{anger}}\par{\footnotesize\texttt{<|0.00|>s}--\texttt{<|4.25|>s}} & \textbf{Speech:} ``I have been a good son for too long. A good sucker. I'm through with it.''\par\smallskip
\textbf{Tone:}  \\ \addlinespace[4pt]
{\footnotesize\textbf{Reference / 2}}\par{\footnotesize\texttt{<speaker2>}\quad \textbf{neutral}}\par{\footnotesize\texttt{<|5.00|>s}--\texttt{<|6.75|>s}} & \textbf{Speech:} ``Um- How many people do you want to live with?''\par\smallskip
\textbf{Tone:}  \\ \addlinespace[4pt]
\multicolumn{2}{@{}l@{}}{\qualmodel{Emo-TAG}} \\*
{\footnotesize\textbf{Emo-TAG / 1}}\par{\footnotesize\texttt{<speaker1>}\quad \qcorrect{\textbf{anger}}}\par{\footnotesize\qcorrect{\texttt{<|0.00|>s}}--\qcorrect{\texttt{<|4.25|>s}}} & \textbf{Speech:} ``\qcorrect{I} \qcorrect{have} \qcorrect{been} \qcorrect{a} \qcorrect{good} \qcorrect{son} \qcorrect{for} \qcorrect{too} \qcorrect{long}. \qcorrect{A} \qcorrect{good} \qcorrect{sucker}. \qcorrect{I'm} \qcorrect{through} \qcorrect{with} \qcorrect{it}.''\par\smallskip
\textbf{Tone:} Low pitch, fast pace, and high-volume delivery with clear signs of anger, conveying defiant frustration and decisiveness \\ \addlinespace[4pt]
{\footnotesize\textbf{Emo-TAG / 2}}\par{\footnotesize\texttt{<speaker2>}\quad \qcorrect{\textbf{neutral}}}\par{\footnotesize\qcorrect{\texttt{<|5.00|>s}}--\qcorrect{\texttt{<|6.75|>s}}} & \textbf{Speech:} ``\qcorrect{How} \qcorrect{many} \qcorrect{people} \qcorrect{do} \qcorrect{you} \qcorrect{want} \qcorrect{to} \qcorrect{live} \qcorrect{with}?''\par\smallskip
\textbf{Tone:} Calm but tensely, with slight hesitation on 'with,' rising intonation, and restrained urgency suggesting mild impatience beneath neutral phrasing \\ \addlinespace[4pt]
\multicolumn{2}{@{}l@{}}{\qualmodel{FlamingoNext}} \\*
{\footnotesize\textbf{FlamingoNext / 1}}\par{\footnotesize\texttt{<Speaker\_0>}\quad \qcorrect{\textbf{anger}}}\par{\footnotesize\qincorrect{\texttt{<|1.00|>s}}--\qincorrect{\texttt{<|5.00|>s}}} & \textbf{Speech:} ``\qcorrect{I} \qcorrect{have} \qcorrect{been} \qcorrect{a} \qcorrect{good} \qcorrect{son} \qcorrect{for} \qcorrect{too} \qcorrect{long}''\par\smallskip
\textbf{Tone:}  \\ \addlinespace[4pt]
{\footnotesize\textbf{FlamingoNext / 2}}\par{\footnotesize\texttt{<Speaker\_0>}\quad \qcorrect{\textbf{anger}}}\par{\footnotesize\qincorrect{\texttt{<|6.00|>s}}--\qincorrect{\texttt{<|9.00|>s}}} & \textbf{Speech:} ``\qcorrect{a} \qcorrect{good} \qcorrect{sucker}!''\par\smallskip
\textbf{Tone:}  \\ \addlinespace[4pt]
{\footnotesize\textbf{FlamingoNext / 3}}\par{\footnotesize\texttt{<Speaker\_1>}\quad \qcorrect{\textbf{neutral}}}\par{\footnotesize\qincorrect{\texttt{<|9.00|>s}}--\qincorrect{\texttt{<|11.00|>s}}} & \textbf{Speech:} ``\qcorrect{How} \qcorrect{many} \qcorrect{people} \qcorrect{do} \qcorrect{you} \qcorrect{want} \qcorrect{to} \qcorrect{live} \qcorrect{with}?''\par\smallskip
\textbf{Tone:}  \\ \addlinespace[4pt]
\multicolumn{2}{@{}l@{}}{\qualmodel{Audio-Reasoner}} \\*
{\footnotesize\textbf{Audio-Reasoner / 1}}\par{\footnotesize\texttt{<speaker1>}\quad \qcorrect{\textbf{anger}}}\par{\footnotesize\qcorrect{\texttt{<|0.00|>s}}--\qincorrect{\texttt{<|4.75|>s}}} & \textbf{Speech:} ``\qincorrect{I've} \qcorrect{been} \qcorrect{a} \qcorrect{good} \qcorrect{son} \qcorrect{for} \qcorrect{too} \qcorrect{long}. \qcorrect{A} \qcorrect{good} \qcorrect{sucker}. \qcorrect{I'm} \qcorrect{through} \qcorrect{with} \qcorrect{it}.''\par\smallskip
\textbf{Tone:}  \\ \addlinespace[4pt]
{\footnotesize\textbf{Audio-Reasoner / 2}}\par{\footnotesize\texttt{<speaker2>}\quad \qcorrect{\textbf{neutral}}}\par{\footnotesize\qincorrect{\texttt{<|4.75|>s}}--\qcorrect{\texttt{<|6.75|>s}}} & \textbf{Speech:} ``\qcorrect{How} \qcorrect{many} \qcorrect{people} \qcorrect{do} \qcorrect{you} \qcorrect{want} \qcorrect{to} \qcorrect{live} \qcorrect{with}?''\par\smallskip
\textbf{Tone:}  \\ \addlinespace[4pt]
\end{longtable}
\endgroup

\Needspace{10\baselineskip}
\Needspace{12\baselineskip}
\subsection{Example 8: Short emotional utterances}
\label{app:iemocap_qualitative_000230}
\noindent\textbf{Success}\hfill{\footnotesize\texttt{IEMOCAP\_train\_concat\_000230\_k2\_9a9af06851}}\par
\smallskip\noindent Emo-TAG correctly identifies frustration followed by sadness and matches both spans. The baselines predict sadness followed by neutral.

\begingroup
\small
\setlength{\tabcolsep}{6pt}
\renewcommand{\arraystretch}{1.12}
\setlength{\LTpre}{6pt}\setlength{\LTpost}{8pt}
\begin{longtable}{@{}>{\raggedright\arraybackslash}p{0.2\linewidth}>{\raggedright\arraybackslash}p{\dimexpr0.8\linewidth-2\tabcolsep\relax}@{}}
\noalign{\label{tab:iemocap_qualitative_000230}}
\toprule
\textbf{Source} & \textbf{Emotion--span sequence} \\
\midrule
\endfirsthead
\multicolumn{2}{@{}l@{}}{\textit{Emotion--span summary (continued)}} \\
\toprule
\textbf{Source} & \textbf{Emotion--span sequence} \\
\midrule
\endhead
\midrule
\multicolumn{2}{r@{}}{\textit{Continued on next page}} \\
\endfoot
\bottomrule
\endlastfoot
\textbf{Reference} & \mbox{frustration $[0.00, 1.75]$};\allowbreak\quad \mbox{sadness $[2.00, 2.75]$} \\ \addlinespace[2pt]
\textbf{Emo-TAG} & \mbox{\qcorrect{frustration} $[\qcorrect{0.00}, \qcorrect{1.75}]$};\allowbreak\quad \mbox{\qcorrect{sadness} $[\qcorrect{2.00}, \qcorrect{2.75}]$} \\ \addlinespace[2pt]
\textbf{FlamingoNext} & \mbox{\qincorrect{sadness} $[\qcorrect{0.00}, \qincorrect{1.00}]$};\allowbreak\quad \mbox{\qincorrect{neutral} $[\qcorrect{2.00}, \qincorrect{3.00}]$} \\ \addlinespace[2pt]
\textbf{Audio-Reasoner} & \mbox{\qincorrect{sadness} $[\qcorrect{0.00}, \qincorrect{2.00}]$};\allowbreak\quad \mbox{\qincorrect{neutral} $[\qcorrect{2.00}, \qincorrect{3.00}]$} \\ \addlinespace[2pt]
\end{longtable}
\endgroup

\begingroup
\small
\setlength{\tabcolsep}{6pt}
\renewcommand{\arraystretch}{1.12}
\setlength{\LTpre}{6pt}\setlength{\LTpost}{8pt}
\begin{longtable}{@{}>{\raggedright\arraybackslash}p{0.32\linewidth}>{\raggedright\arraybackslash}p{\dimexpr0.68\linewidth-2\tabcolsep\relax}@{}}
\noalign{\label{tab:iemocap_qualitative_full_000230}}
\toprule
\textbf{Span / emotion / time} & \textbf{Speech and tone} \\
\midrule
\endfirsthead
\multicolumn{2}{@{}l@{}}{\textit{Example 8: annotations (continued)}} \\
\toprule
\textbf{Span / emotion / time} & \textbf{Speech and tone} \\
\midrule
\endhead
\midrule
\multicolumn{2}{r@{}}{\textit{Continued on next page}} \\
\endfoot
\bottomrule
\endlastfoot
\multicolumn{2}{@{}l@{}}{\qualmodel{Reference}} \\*
{\footnotesize\textbf{Reference / 1}}\par{\footnotesize\texttt{<speaker1>}\quad \textbf{frustration}}\par{\footnotesize\texttt{<|0.00|>s}--\texttt{<|1.75|>s}} & \textbf{Speech:} ``This is humiliating.''\par\smallskip
\textbf{Tone:}  \\ \addlinespace[4pt]
{\footnotesize\textbf{Reference / 2}}\par{\footnotesize\texttt{<speaker1>}\quad \textbf{sadness}}\par{\footnotesize\texttt{<|2.00|>s}--\texttt{<|2.75|>s}} & \textbf{Speech:} ``I didn't say that.''\par\smallskip
\textbf{Tone:}  \\ \addlinespace[4pt]
\multicolumn{2}{@{}l@{}}{\qualmodel{Emo-TAG}} \\*
{\footnotesize\textbf{Emo-TAG / 1}}\par{\footnotesize\texttt{<speaker1>}\quad \qcorrect{\textbf{frustration}}}\par{\footnotesize\qcorrect{\texttt{<|0.00|>s}}--\qcorrect{\texttt{<|1.75|>s}}} & \textbf{Speech:} ``\qcorrect{This} \qcorrect{is} \qcorrect{humiliating}.''\par\smallskip
\textbf{Tone:} Voice remains high pitch, loud, and fast pace with tense urgency, reflecting clear frustration and irritation \\ \addlinespace[4pt]
{\footnotesize\textbf{Emo-TAG / 2}}\par{\footnotesize\texttt{<speaker2>}\quad \qcorrect{\textbf{sadness}}}\par{\footnotesize\qcorrect{\texttt{<|2.00|>s}}--\qcorrect{\texttt{<|2.75|>s}}} & \textbf{Speech:} ``\qcorrect{I} \qcorrect{didn't} \qcorrect{say} \qcorrect{that}.''\par\smallskip
\textbf{Tone:} Low-pitched voice, speaks softly, moves slowly with hesitant rhythm, conveying subdued introspection and quiet sorrow through measured pace and restrained delivery \\ \addlinespace[4pt]
\multicolumn{2}{@{}l@{}}{\qualmodel{FlamingoNext}} \\*
{\footnotesize\textbf{FlamingoNext / 1}}\par{\footnotesize\texttt{<speaker1>}\quad \qincorrect{\textbf{sadness}}}\par{\footnotesize\qcorrect{\texttt{<|0.00|>s}}--\qincorrect{\texttt{<|1.00|>s}}} & \textbf{Speech:} ``\qcorrect{This} \qcorrect{is} \qcorrect{humiliating}.''\par\smallskip
\textbf{Tone:}  \\ \addlinespace[4pt]
{\footnotesize\textbf{FlamingoNext / 2}}\par{\footnotesize\texttt{<speaker2>}\quad \qincorrect{\textbf{neutral}}}\par{\footnotesize\qcorrect{\texttt{<|2.00|>s}}--\qincorrect{\texttt{<|3.00|>s}}} & \textbf{Speech:} ``\qcorrect{I} \qcorrect{didn't} \qcorrect{say} \qcorrect{that}.''\par\smallskip
\textbf{Tone:}  \\ \addlinespace[4pt]
\multicolumn{2}{@{}l@{}}{\qualmodel{Audio-Reasoner}} \\*
{\footnotesize\textbf{Audio-Reasoner / 1}}\par{\footnotesize\texttt{<speaker1>}\quad \qincorrect{\textbf{sadness}}}\par{\footnotesize\qcorrect{\texttt{<|0.00|>s}}--\qincorrect{\texttt{<|2.00|>s}}} & \textbf{Speech:} ``\qcorrect{This} \qcorrect{is} \qcorrect{humiliating}.''\par\smallskip
\textbf{Tone:}  \\ \addlinespace[4pt]
{\footnotesize\textbf{Audio-Reasoner / 2}}\par{\footnotesize\texttt{<speaker2>}\quad \qincorrect{\textbf{neutral}}}\par{\footnotesize\qcorrect{\texttt{<|2.00|>s}}--\qincorrect{\texttt{<|3.00|>s}}} & \textbf{Speech:} ``\qcorrect{I} \qcorrect{didn't} \qcorrect{say} \qcorrect{that}.''\par\smallskip
\textbf{Tone:}  \\ \addlinespace[4pt]
\end{longtable}
\endgroup

\Needspace{10\baselineskip}
\Needspace{12\baselineskip}
\subsection{Example 9: Three-emotion sequence}
\label{app:iemocap_qualitative_000067}
\noindent\textbf{Success}\hfill{\footnotesize\texttt{IEMOCAP\_train\_concat\_000067\_k3\_93e3a64564}}\par
\smallskip\noindent Emo-TAG follows the sequence from frustration to happiness to sadness and matches all three spans, despite transcription errors.

\begingroup
\small
\setlength{\tabcolsep}{6pt}
\renewcommand{\arraystretch}{1.12}
\setlength{\LTpre}{6pt}\setlength{\LTpost}{8pt}
\begin{longtable}{@{}>{\raggedright\arraybackslash}p{0.2\linewidth}>{\raggedright\arraybackslash}p{\dimexpr0.8\linewidth-2\tabcolsep\relax}@{}}
\noalign{\label{tab:iemocap_qualitative_000067}}
\toprule
\textbf{Source} & \textbf{Emotion--span sequence} \\
\midrule
\endfirsthead
\multicolumn{2}{@{}l@{}}{\textit{Emotion--span summary (continued)}} \\
\toprule
\textbf{Source} & \textbf{Emotion--span sequence} \\
\midrule
\endhead
\midrule
\multicolumn{2}{r@{}}{\textit{Continued on next page}} \\
\endfoot
\bottomrule
\endlastfoot
\textbf{Reference} & \mbox{frustration $[0.00, 3.50]$};\allowbreak\quad \mbox{happiness $[4.25, 7.50]$};\allowbreak\quad \mbox{sadness $[8.50, 10.75]$} \\ \addlinespace[2pt]
\textbf{Emo-TAG} & \mbox{\qcorrect{frustration} $[\qcorrect{0.00}, \qcorrect{3.50}]$};\allowbreak\quad \mbox{\qcorrect{happiness} $[\qcorrect{4.25}, \qcorrect{7.50}]$};\allowbreak\quad \mbox{\qcorrect{sadness} $[\qcorrect{8.50}, \qcorrect{10.75}]$} \\ \addlinespace[2pt]
\textbf{FlamingoNext} & \mbox{\qincorrect{neutral} $[\qcorrect{0.00}, \qincorrect{3.00}]$};\allowbreak\quad \mbox{\qincorrect{neutral} $[\qincorrect{1.00}, \qincorrect{2.00}]$};\allowbreak\quad \mbox{\qincorrect{surprise} $[\qincorrect{4.00}, \qincorrect{7.00}]$};\allowbreak\quad \mbox{\qincorrect{disgust} $[\qincorrect{8.00}, \qincorrect{9.00}]$};\allowbreak\quad \mbox{\qincorrect{sadness} $[\qincorrect{9.00}, \qincorrect{10.00}]$} \\ \addlinespace[2pt]
\textbf{Audio-Reasoner} & \mbox{\qcorrect{frustration} $[\qcorrect{0.00}, \qincorrect{4.00}]$};\allowbreak\quad \mbox{\qincorrect{surprise} $[\qincorrect{4.00}, \qincorrect{7.00}]$};\allowbreak\quad \mbox{\qincorrect{neutral} $[\qincorrect{7.25}, \qincorrect{9.00}]$} \\ \addlinespace[2pt]
\end{longtable}
\endgroup

\begingroup
\small
\setlength{\tabcolsep}{6pt}
\renewcommand{\arraystretch}{1.12}
\setlength{\LTpre}{6pt}\setlength{\LTpost}{8pt}
\begin{longtable}{@{}>{\raggedright\arraybackslash}p{0.32\linewidth}>{\raggedright\arraybackslash}p{\dimexpr0.68\linewidth-2\tabcolsep\relax}@{}}
\noalign{\label{tab:iemocap_qualitative_full_000067}}
\toprule
\textbf{Span / emotion / time} & \textbf{Speech and tone} \\
\midrule
\endfirsthead
\multicolumn{2}{@{}l@{}}{\textit{Example 9: annotations (continued)}} \\
\toprule
\textbf{Span / emotion / time} & \textbf{Speech and tone} \\
\midrule
\endhead
\midrule
\multicolumn{2}{r@{}}{\textit{Continued on next page}} \\
\endfoot
\bottomrule
\endlastfoot
\multicolumn{2}{@{}l@{}}{\qualmodel{Reference}} \\*
{\footnotesize\textbf{Reference / 1}}\par{\footnotesize\texttt{<speaker1>}\quad \textbf{frustration}}\par{\footnotesize\texttt{<|0.00|>s}--\texttt{<|3.50|>s}} & \textbf{Speech:} ``I'm growing quite tired and bored of this conversation.''\par\smallskip
\textbf{Tone:}  \\ \addlinespace[4pt]
{\footnotesize\textbf{Reference / 2}}\par{\footnotesize\texttt{<speaker2>}\quad \textbf{happiness}}\par{\footnotesize\texttt{<|4.25|>s}--\texttt{<|7.50|>s}} & \textbf{Speech:} ``oh, God I kissed Annie. I kissed Annie. I can't How long have I waited to do that?''\par\smallskip
\textbf{Tone:}  \\ \addlinespace[4pt]
{\footnotesize\textbf{Reference / 3}}\par{\footnotesize\texttt{<speaker2>}\quad \textbf{sadness}}\par{\footnotesize\texttt{<|8.50|>s}--\texttt{<|10.75|>s}} & \textbf{Speech:} ``And I guess that included you.''\par\smallskip
\textbf{Tone:}  \\ \addlinespace[4pt]
\multicolumn{2}{@{}l@{}}{\qualmodel{Emo-TAG}} \\*
{\footnotesize\textbf{Emo-TAG / 1}}\par{\footnotesize\texttt{<speaker1>}\quad \qcorrect{\textbf{frustration}}}\par{\footnotesize\qcorrect{\texttt{<|0.00|>s}}--\qcorrect{\texttt{<|3.50|>s}}} & \textbf{Speech:} ``\qincorrect{I} \qincorrect{am} \qcorrect{growing} \qcorrect{quite} \qcorrect{tired} \qcorrect{and} \qcorrect{bored} \qcorrect{of} \qcorrect{this} \qcorrect{conversation}.''\par\smallskip
\textbf{Tone:} Mildly frustrated and weary, with a steady, conversational pace and slight strain, indicating irritation rather than intense anger or sadness \\ \addlinespace[4pt]
{\footnotesize\textbf{Emo-TAG / 2}}\par{\footnotesize\texttt{<speaker2>}\quad \qcorrect{\textbf{happiness}}}\par{\footnotesize\qcorrect{\texttt{<|4.25|>s}}--\qcorrect{\texttt{<|7.50|>s}}} & \textbf{Speech:} ``\qcorrect{God} \qincorrect{you're} \qincorrect{standing}. \qcorrect{How} \qcorrect{long} \qcorrect{have} \qcorrect{I} \qcorrect{waited} \qcorrect{to} \qcorrect{do} \qcorrect{that}?''\par\smallskip
\textbf{Tone:} High pitch, fast pace, energetic, with quick articulation and rising intonation, showing excitement and urgency \\ \addlinespace[4pt]
{\footnotesize\textbf{Emo-TAG / 3}}\par{\footnotesize\texttt{<speaker2>}\quad \qcorrect{\textbf{sadness}}}\par{\footnotesize\qcorrect{\texttt{<|8.50|>s}}--\qcorrect{\texttt{<|10.75|>s}}} & \textbf{Speech:} ``\qincorrect{Can} \qcorrect{I} \qcorrect{guess} \qcorrect{that} \qcorrect{included} \qcorrect{you}?''\par\smallskip
\textbf{Tone:} Soft, hesitant start with high pitch, escalating to faster pace and louder volume, showing nervous tension and anxiety \\ \addlinespace[4pt]
\multicolumn{2}{@{}l@{}}{\qualmodel{FlamingoNext}} \\*
{\footnotesize\textbf{FlamingoNext / 1}}\par{\footnotesize\texttt{<speaker1>}\quad \qincorrect{\textbf{neutral}}}\par{\footnotesize\qcorrect{\texttt{<|0.00|>s}}--\qincorrect{\texttt{<|3.00|>s}}} & \textbf{Speech:} ``\qincorrect{she} \qincorrect{is} \qcorrect{growing} \qcorrect{quite} \qcorrect{tired} \qcorrect{and} \qcorrect{bored} \qcorrect{of} \qcorrect{this} \qcorrect{conversation}''\par\smallskip
\textbf{Tone:}  \\ \addlinespace[4pt]
{\footnotesize\qincorrect{\textbf{FlamingoNext / 2}}}\par{\footnotesize\texttt{<speaker2>}\quad \qincorrect{\textbf{neutral}}}\par{\footnotesize\qincorrect{\texttt{<|1.00|>s}}--\qincorrect{\texttt{<|2.00|>s}}} & \textbf{Speech:} ``''\par\smallskip
\textbf{Tone:}  \\ \addlinespace[4pt]
{\footnotesize\textbf{FlamingoNext / 3}}\par{\footnotesize\texttt{<speaker3>}\quad \qincorrect{\textbf{surprise}}}\par{\footnotesize\qincorrect{\texttt{<|4.00|>s}}--\qincorrect{\texttt{<|7.00|>s}}} & \textbf{Speech:} ``''\par\smallskip
\textbf{Tone:}  \\ \addlinespace[4pt]
{\footnotesize\textbf{FlamingoNext / 4}}\par{\footnotesize\texttt{<speaker4>}\quad \qincorrect{\textbf{disgust}}}\par{\footnotesize\qincorrect{\texttt{<|8.00|>s}}--\qincorrect{\texttt{<|9.00|>s}}} & \textbf{Speech:} ``''\par\smallskip
\textbf{Tone:}  \\ \addlinespace[4pt]
{\footnotesize\qincorrect{\textbf{FlamingoNext / 5}}}\par{\footnotesize\texttt{<speaker5>}\quad \qincorrect{\textbf{sadness}}}\par{\footnotesize\qincorrect{\texttt{<|9.00|>s}}--\qincorrect{\texttt{<|10.00|>s}}} & \textbf{Speech:} ``''\par\smallskip
\textbf{Tone:}  \\ \addlinespace[4pt]
\multicolumn{2}{@{}l@{}}{\qualmodel{Audio-Reasoner}} \\*
{\footnotesize\textbf{Audio-Reasoner / 1}}\par{\footnotesize\texttt{<speaker1>}\quad \qcorrect{\textbf{frustration}}}\par{\footnotesize\qcorrect{\texttt{<|0.00|>s}}--\qincorrect{\texttt{<|4.00|>s}}} & \textbf{Speech:} ``\qincorrect{I} \qincorrect{am} \qcorrect{growing} \qcorrect{quite} \qcorrect{tired} \qcorrect{and} \qcorrect{bored} \qcorrect{of} \qcorrect{this} \qcorrect{conversation}.''\par\smallskip
\textbf{Tone:}  \\ \addlinespace[4pt]
{\footnotesize\textbf{Audio-Reasoner / 2}}\par{\footnotesize\texttt{<speaker2>}\quad \qincorrect{\textbf{surprise}}}\par{\footnotesize\qincorrect{\texttt{<|4.00|>s}}--\qincorrect{\texttt{<|7.00|>s}}} & \textbf{Speech:} ``\qcorrect{God}, \qincorrect{you're} \qincorrect{standing}. \qcorrect{How} \qcorrect{long} \qcorrect{have} \qcorrect{I} \qcorrect{waited} \qcorrect{to} \qcorrect{do} \qcorrect{that}?''\par\smallskip
\textbf{Tone:}  \\ \addlinespace[4pt]
{\footnotesize\textbf{Audio-Reasoner / 3}}\par{\footnotesize\texttt{<speaker3>}\quad \qincorrect{\textbf{neutral}}}\par{\footnotesize\qincorrect{\texttt{<|7.25|>s}}--\qincorrect{\texttt{<|9.00|>s}}} & \textbf{Speech:} ``\qcorrect{And} \qcorrect{I} \qcorrect{guess} \qincorrect{it} \qcorrect{included} \qcorrect{you}.''\par\smallskip
\textbf{Tone:}  \\ \addlinespace[4pt]
\end{longtable}
\endgroup

\Needspace{10\baselineskip}
\Needspace{12\baselineskip}
\subsection{Example 10: Emotion confusion}
\label{app:iemocap_qualitative_000380}
\noindent\textbf{Failure}\hfill{\footnotesize\texttt{IEMOCAP\_train\_concat\_000380\_k2\_9373a1a283}}\par
\smallskip\noindent Emo-TAG mistakes the neutral utterance for frustration despite matching both spans. FlamingoNext identifies both emotions but misplaces their boundaries.

\begingroup
\small
\setlength{\tabcolsep}{6pt}
\renewcommand{\arraystretch}{1.12}
\setlength{\LTpre}{6pt}\setlength{\LTpost}{8pt}
\begin{longtable}{@{}>{\raggedright\arraybackslash}p{0.2\linewidth}>{\raggedright\arraybackslash}p{\dimexpr0.8\linewidth-2\tabcolsep\relax}@{}}
\noalign{\label{tab:iemocap_qualitative_000380}}
\toprule
\textbf{Source} & \textbf{Emotion--span sequence} \\
\midrule
\endfirsthead
\multicolumn{2}{@{}l@{}}{\textit{Emotion--span summary (continued)}} \\
\toprule
\textbf{Source} & \textbf{Emotion--span sequence} \\
\midrule
\endhead
\midrule
\multicolumn{2}{r@{}}{\textit{Continued on next page}} \\
\endfoot
\bottomrule
\endlastfoot
\textbf{Reference} & \mbox{neutral $[0.00, 2.25]$};\allowbreak\quad \mbox{sadness $[3.25, 4.25]$} \\ \addlinespace[2pt]
\textbf{Emo-TAG} & \mbox{\qincorrect{frustration} $[\qcorrect{0.00}, \qcorrect{2.25}]$};\allowbreak\quad \mbox{\qcorrect{sadness} $[\qcorrect{3.25}, \qcorrect{4.25}]$} \\ \addlinespace[2pt]
\textbf{FlamingoNext} & \mbox{\qcorrect{neutral} $[\qcorrect{0.00}, \qincorrect{3.00}]$};\allowbreak\quad \mbox{\qcorrect{sadness} $[\qincorrect{4.00}, \qincorrect{5.00}]$} \\ \addlinespace[2pt]
\textbf{Audio-Reasoner} & \mbox{\qcorrect{neutral} $[\qcorrect{0.00}, \qincorrect{4.00}]$};\allowbreak\quad \mbox{\qincorrect{neutral} $[\qincorrect{4.00}, \qincorrect{6.00}]$};\allowbreak\quad \mbox{\qincorrect{neutral} $[\qincorrect{6.00}, \qincorrect{7.00}]$} \\ \addlinespace[2pt]
\end{longtable}
\endgroup

\begingroup
\small
\setlength{\tabcolsep}{6pt}
\renewcommand{\arraystretch}{1.12}
\setlength{\LTpre}{6pt}\setlength{\LTpost}{8pt}
\begin{longtable}{@{}>{\raggedright\arraybackslash}p{0.32\linewidth}>{\raggedright\arraybackslash}p{\dimexpr0.68\linewidth-2\tabcolsep\relax}@{}}
\noalign{\label{tab:iemocap_qualitative_full_000380}}
\toprule
\textbf{Span / emotion / time} & \textbf{Speech and tone} \\
\midrule
\endfirsthead
\multicolumn{2}{@{}l@{}}{\textit{Example 10: annotations (continued)}} \\
\toprule
\textbf{Span / emotion / time} & \textbf{Speech and tone} \\
\midrule
\endhead
\midrule
\multicolumn{2}{r@{}}{\textit{Continued on next page}} \\
\endfoot
\bottomrule
\endlastfoot
\multicolumn{2}{@{}l@{}}{\qualmodel{Reference}} \\*
{\footnotesize\textbf{Reference / 1}}\par{\footnotesize\texttt{<speaker1>}\quad \textbf{neutral}}\par{\footnotesize\texttt{<|0.00|>s}--\texttt{<|2.25|>s}} & \textbf{Speech:} ``Are you taking the late fee off of my account then?''\par\smallskip
\textbf{Tone:}  \\ \addlinespace[4pt]
{\footnotesize\textbf{Reference / 2}}\par{\footnotesize\texttt{<speaker2>}\quad \textbf{sadness}}\par{\footnotesize\texttt{<|3.25|>s}--\texttt{<|4.25|>s}} & \textbf{Speech:} ``He cried hard?''\par\smallskip
\textbf{Tone:}  \\ \addlinespace[4pt]
\multicolumn{2}{@{}l@{}}{\qualmodel{Emo-TAG}} \\*
{\footnotesize\textbf{Emo-TAG / 1}}\par{\footnotesize\texttt{<speaker1>}\quad \qincorrect{\textbf{frustration}}}\par{\footnotesize\qcorrect{\texttt{<|0.00|>s}}--\qcorrect{\texttt{<|2.25|>s}}} & \textbf{Speech:} ``\qcorrect{Are} \qcorrect{you} \qcorrect{taking} \qcorrect{the} \qcorrect{late} \qcorrect{fee} \qcorrect{off} \qcorrect{of} \qcorrect{my} \qcorrect{account} \qcorrect{then}?''\par\smallskip
\textbf{Tone:} Slightly raised pitch, quickened tempo, and subtle tension suggest mild frustration, with controlled delivery maintaining polite composure \\ \addlinespace[4pt]
{\footnotesize\textbf{Emo-TAG / 2}}\par{\footnotesize\texttt{<speaker2>}\quad \qcorrect{\textbf{sadness}}}\par{\footnotesize\qcorrect{\texttt{<|3.25|>s}}--\qcorrect{\texttt{<|4.25|>s}}} & \textbf{Speech:} ``\qcorrect{He} \qcorrect{cried} \qcorrect{hard}.''\par\smallskip
\textbf{Tone:} Low pitch, soft-spoken, slow pace with steady volume, conveying quiet sorrow through subdued vocal delivery \\ \addlinespace[4pt]
\multicolumn{2}{@{}l@{}}{\qualmodel{FlamingoNext}} \\*
{\footnotesize\textbf{FlamingoNext / 1}}\par{\footnotesize\texttt{<speaker1>}\quad \qcorrect{\textbf{neutral}}}\par{\footnotesize\qcorrect{\texttt{<|0.00|>s}}--\qincorrect{\texttt{<|3.00|>s}}} & \textbf{Speech:} ``\qcorrect{Are} \qcorrect{you} \qcorrect{taking} \qcorrect{the} \qcorrect{late} \qcorrect{fee} \qcorrect{off} \qcorrect{my} \qcorrect{account} \qcorrect{then}?''\par\smallskip
\textbf{Tone:}  \\ \addlinespace[4pt]
{\footnotesize\textbf{FlamingoNext / 2}}\par{\footnotesize\texttt{<speaker2>}\quad \qcorrect{\textbf{sadness}}}\par{\footnotesize\qincorrect{\texttt{<|4.00|>s}}--\qincorrect{\texttt{<|5.00|>s}}} & \textbf{Speech:} ``\qcorrect{He} \qcorrect{cried} \qcorrect{hard}.''\par\smallskip
\textbf{Tone:}  \\ \addlinespace[4pt]
\multicolumn{2}{@{}l@{}}{\qualmodel{Audio-Reasoner}} \\*
{\footnotesize\textbf{Audio-Reasoner / 1}}\par{\footnotesize\texttt{<speaker1>}\quad \qcorrect{\textbf{neutral}}}\par{\footnotesize\qcorrect{\texttt{<|0.00|>s}}--\qincorrect{\texttt{<|4.00|>s}}} & \textbf{Speech:} ``\qcorrect{Are} \qcorrect{you} \qcorrect{taking} \qcorrect{the} \qcorrect{late} \qcorrect{fee} \qcorrect{off} \qcorrect{of} \qcorrect{my} \qcorrect{account} \qcorrect{then}?''\par\smallskip
\textbf{Tone:}  \\ \addlinespace[4pt]
{\footnotesize\textbf{Audio-Reasoner / 2}}\par{\footnotesize\texttt{<speaker2>}\quad \qincorrect{\textbf{neutral}}}\par{\footnotesize\qincorrect{\texttt{<|4.00|>s}}--\qincorrect{\texttt{<|6.00|>s}}} & \textbf{Speech:} ``\qcorrect{He} \qcorrect{cried} \qcorrect{hard}.''\par\smallskip
\textbf{Tone:}  \\ \addlinespace[4pt]
{\footnotesize\qincorrect{\textbf{Audio-Reasoner / 3}}}\par{\footnotesize\texttt{<speaker3>}\quad \qincorrect{\textbf{neutral}}}\par{\footnotesize\qincorrect{\texttt{<|6.00|>s}}--\qincorrect{\texttt{<|7.00|>s}}} & \textbf{Speech:} ``\qincorrect{He} \qincorrect{cried} \qincorrect{hard}.''\par\smallskip
\textbf{Tone:}  \\ \addlinespace[4pt]
\end{longtable}
\endgroup

\Needspace{10\baselineskip}
\Needspace{12\baselineskip}
\subsection{Example 11: Merged utterances}
\label{app:iemocap_qualitative_000080}
\noindent\textbf{Failure}\hfill{\footnotesize\texttt{IEMOCAP\_train\_concat\_000080\_k3\_73b9cca132}}\par
\smallskip\noindent Emo-TAG combines the first two utterances into one frustration span, missing the change from neutral to frustration.

\begingroup
\small
\setlength{\tabcolsep}{6pt}
\renewcommand{\arraystretch}{1.12}
\setlength{\LTpre}{6pt}\setlength{\LTpost}{8pt}
\begin{longtable}{@{}>{\raggedright\arraybackslash}p{0.2\linewidth}>{\raggedright\arraybackslash}p{\dimexpr0.8\linewidth-2\tabcolsep\relax}@{}}
\noalign{\label{tab:iemocap_qualitative_000080}}
\toprule
\textbf{Source} & \textbf{Emotion--span sequence} \\
\midrule
\endfirsthead
\multicolumn{2}{@{}l@{}}{\textit{Emotion--span summary (continued)}} \\
\toprule
\textbf{Source} & \textbf{Emotion--span sequence} \\
\midrule
\endhead
\midrule
\multicolumn{2}{r@{}}{\textit{Continued on next page}} \\
\endfoot
\bottomrule
\endlastfoot
\textbf{Reference} & \mbox{neutral $[0.00, 1.25]$};\allowbreak\quad \mbox{frustration $[2.00, 3.25]$};\allowbreak\quad \mbox{frustration $[3.75, 5.75]$} \\ \addlinespace[2pt]
\textbf{Emo-TAG} & \mbox{\qincorrect{frustration} $[\qcorrect{0.00}, \qincorrect{3.25}]$};\allowbreak\quad \mbox{\qcorrect{frustration} $[\qcorrect{3.75}, \qcorrect{5.75}]$} \\ \addlinespace[2pt]
\textbf{FlamingoNext} & \mbox{\qcorrect{neutral} $[\qcorrect{0.00}, \qincorrect{1.00}]$};\allowbreak\quad \mbox{\qcorrect{neutral} $[\qincorrect{2.00}, \qincorrect{3.00}]$};\allowbreak\quad \mbox{\qincorrect{neutral} $[\qincorrect{4.00}, \qincorrect{5.00}]$};\allowbreak\quad \mbox{\qincorrect{anger} \qincorrect{(untimed)}};\allowbreak\quad \mbox{\qincorrect{neutral} \qincorrect{(untimed)}} \\ \addlinespace[2pt]
\textbf{Audio-Reasoner} & \mbox{\qincorrect{frustration} $[\qcorrect{0.00}, \qincorrect{3.00}]$};\allowbreak\quad \mbox{\qincorrect{anger} $[\qincorrect{3.00}, \qincorrect{3.50}]$};\allowbreak\quad \mbox{\qincorrect{neutral} $[\qincorrect{3.50}, \qincorrect{4.00}]$};\allowbreak\quad \mbox{\qincorrect{neutral} $[\qincorrect{4.00}, \qincorrect{4.00}]$};\allowbreak\quad \mbox{\qincorrect{neutral} $[\qincorrect{4.00}, \qincorrect{4.00}]$};\allowbreak\quad \mbox{\qincorrect{neutral} $[\qincorrect{4.00}, \qincorrect{4.25}]$};\allowbreak\quad \mbox{\qincorrect{neutral} $[\qincorrect{4.25}, \qincorrect{4.25}]$};\allowbreak\quad \mbox{\qincorrect{neutral} $[\qincorrect{4.25}, \qincorrect{4.25}]$};\allowbreak\quad \mbox{\qincorrect{unknown} $[\qincorrect{4.25}, \qincorrect{4.25}]$} \\ \addlinespace[2pt]
\end{longtable}
\endgroup

\begingroup
\small
\setlength{\tabcolsep}{6pt}
\renewcommand{\arraystretch}{1.12}
\setlength{\LTpre}{6pt}\setlength{\LTpost}{8pt}
\begin{longtable}{@{}>{\raggedright\arraybackslash}p{0.32\linewidth}>{\raggedright\arraybackslash}p{\dimexpr0.68\linewidth-2\tabcolsep\relax}@{}}
\noalign{\label{tab:iemocap_qualitative_full_000080}}
\toprule
\textbf{Span / emotion / time} & \textbf{Speech and tone} \\
\midrule
\endfirsthead
\multicolumn{2}{@{}l@{}}{\textit{Example 11: annotations (continued)}} \\
\toprule
\textbf{Span / emotion / time} & \textbf{Speech and tone} \\
\midrule
\endhead
\midrule
\multicolumn{2}{r@{}}{\textit{Continued on next page}} \\
\endfoot
\bottomrule
\endlastfoot
\multicolumn{2}{@{}l@{}}{\qualmodel{Reference}} \\*
{\footnotesize\textbf{Reference / 1}}\par{\footnotesize\texttt{<speaker1>}\quad \textbf{neutral}}\par{\footnotesize\texttt{<|0.00|>s}--\texttt{<|1.25|>s}} & \textbf{Speech:} ``oh, She's back at work. yeah.''\par\smallskip
\textbf{Tone:}  \\ \addlinespace[4pt]
{\footnotesize\textbf{Reference / 2}}\par{\footnotesize\texttt{<speaker1>}\quad \textbf{frustration}}\par{\footnotesize\texttt{<|2.00|>s}--\texttt{<|3.25|>s}} & \textbf{Speech:} ``And you couldn't find it?''\par\smallskip
\textbf{Tone:}  \\ \addlinespace[4pt]
{\footnotesize\textbf{Reference / 3}}\par{\footnotesize\texttt{<speaker2>}\quad \textbf{frustration}}\par{\footnotesize\texttt{<|3.75|>s}--\texttt{<|5.75|>s}} & \textbf{Speech:} ``It's not fair.''\par\smallskip
\textbf{Tone:}  \\ \addlinespace[4pt]
\multicolumn{2}{@{}l@{}}{\qualmodel{Emo-TAG}} \\*
{\footnotesize\textbf{Emo-TAG / 1}}\par{\footnotesize\texttt{<speaker1>}\quad \qincorrect{\textbf{frustration}}}\par{\footnotesize\qcorrect{\texttt{<|0.00|>s}}--\qincorrect{\texttt{<|3.25|>s}}} & \textbf{Speech:} ``\qcorrect{Oh}, \qcorrect{she's} \qcorrect{back} \qcorrect{at} \qcorrect{work}. \qcorrect{And} \qcorrect{you} \qcorrect{couldn't} \qcorrect{find} \qcorrect{it}.''\par\smallskip
\textbf{Tone:} A subtle frustration tinged with resignation, conveyed through low pitch, slow delivery and moderate volume \\ \addlinespace[4pt]
{\footnotesize\textbf{Emo-TAG / 2}}\par{\footnotesize\texttt{<speaker2>}\quad \qcorrect{\textbf{frustration}}}\par{\footnotesize\qcorrect{\texttt{<|3.75|>s}}--\qcorrect{\texttt{<|5.75|>s}}} & \textbf{Speech:} ``\qcorrect{It's} \qcorrect{not} \qcorrect{fair}.''\par\smallskip
\textbf{Tone:} High-pitched voice with strong emphasis on 'not,' fast pace, slight rising intonation, and steady volume, conveying clear frustration and exasperation \\ \addlinespace[4pt]
\multicolumn{2}{@{}l@{}}{\qualmodel{FlamingoNext}} \\*
{\footnotesize\textbf{FlamingoNext / 1}}\par{\footnotesize\texttt{<speaker1>}\quad \qcorrect{\textbf{neutral}}}\par{\footnotesize\qcorrect{\texttt{<|0.00|>s}}--\qincorrect{\texttt{<|1.00|>s}}} & \textbf{Speech:} ``\qcorrect{Oh}''\par\smallskip
\textbf{Tone:}  \\ \addlinespace[4pt]
{\footnotesize\textbf{FlamingoNext / 2}}\par{\footnotesize\texttt{<speaker1>}\quad \qcorrect{\textbf{neutral}}}\par{\footnotesize\qincorrect{\texttt{<|2.00|>s}}--\qincorrect{\texttt{<|3.00|>s}}} & \textbf{Speech:} ``\qcorrect{she's} \qcorrect{back} \qcorrect{at} \qcorrect{work}.''\par\smallskip
\textbf{Tone:}  \\ \addlinespace[4pt]
{\footnotesize\textbf{FlamingoNext / 3}}\par{\footnotesize\texttt{<speaker1>}\quad \qincorrect{\textbf{neutral}}}\par{\footnotesize\qincorrect{\texttt{<|4.00|>s}}--\qincorrect{\texttt{<|5.00|>s}}} & \textbf{Speech:} ``\qcorrect{And} \qcorrect{you} \qcorrect{couldn't} \qcorrect{find} \qincorrect{her}?''\par\smallskip
\textbf{Tone:}  \\ \addlinespace[4pt]
{\footnotesize\textbf{FlamingoNext / 4}}\par{\footnotesize\texttt{<speaker2>}\quad \qincorrect{\textbf{anger}}}\par{\footnotesize\qincorrect{\textit{Time not supplied}}} & \textbf{Speech:} ``\qcorrect{It's} \qcorrect{not} \qcorrect{fair}!''\par\smallskip
\textbf{Tone:}  \\ \addlinespace[4pt]
{\footnotesize\qincorrect{\textbf{FlamingoNext / 5}}}\par{\footnotesize\texttt{<speaker1>}\quad \qincorrect{\textbf{neutral}}}\par{\footnotesize\qincorrect{\textit{Time not supplied}}} & \textbf{Speech:} ``\qincorrect{Well}...''\par\smallskip
\textbf{Tone:}  \\ \addlinespace[4pt]
\multicolumn{2}{@{}l@{}}{\qualmodel{Audio-Reasoner}} \\*
{\footnotesize\textbf{Audio-Reasoner / 1}}\par{\footnotesize\texttt{<speaker1>}\quad \qincorrect{\textbf{frustration}}}\par{\footnotesize\qcorrect{\texttt{<|0.00|>s}}--\qincorrect{\texttt{<|3.00|>s}}} & \textbf{Speech:} ``\qcorrect{Oh}, \qcorrect{she's} \qcorrect{back} \qcorrect{at} \qcorrect{work}. \qcorrect{And} \qcorrect{you} \qcorrect{couldn't} \qcorrect{find} \qcorrect{it}.''\par\smallskip
\textbf{Tone:}  \\ \addlinespace[4pt]
{\footnotesize\textbf{Audio-Reasoner / 2}}\par{\footnotesize\texttt{<speaker2>}\quad \qincorrect{\textbf{anger}}}\par{\footnotesize\qincorrect{\texttt{<|3.00|>s}}--\qincorrect{\texttt{<|3.50|>s}}} & \textbf{Speech:} ``\qcorrect{It's} \qcorrect{not} \qcorrect{fair}.''\par\smallskip
\textbf{Tone:}  \\ \addlinespace[4pt]
{\footnotesize\qincorrect{\textbf{Audio-Reasoner / 3}}}\par{\footnotesize\texttt{<speaker3>}\quad \qincorrect{\textbf{neutral}}}\par{\footnotesize\qincorrect{\texttt{<|3.50|>s}}--\qincorrect{\texttt{<|4.00|>s}}} & \textbf{Speech:} ``\qincorrect{Well}, \qincorrect{you} \qincorrect{have} \qincorrect{to} \qincorrect{be}.''\par\smallskip
\textbf{Tone:}  \\ \addlinespace[4pt]
{\footnotesize\qincorrect{\textbf{Audio-Reasoner / 4}}}\par{\footnotesize\texttt{<speaker4>}\quad \qincorrect{\textbf{neutral}}}\par{\footnotesize\qincorrect{\texttt{<|4.00|>s}}--\qincorrect{\texttt{<|4.00|>s}}} & \textbf{Speech:} ``\qincorrect{Clippers}.''\par\smallskip
\textbf{Tone:}  \\ \addlinespace[4pt]
{\footnotesize\qincorrect{\textbf{Audio-Reasoner / 5}}}\par{\footnotesize\texttt{<speaker5>}\quad \qincorrect{\textbf{neutral}}}\par{\footnotesize\qincorrect{\texttt{<|4.00|>s}}--\qincorrect{\texttt{<|4.00|>s}}} & \textbf{Speech:} ``\qincorrect{Clippers}.''\par\smallskip
\textbf{Tone:}  \\ \addlinespace[4pt]
{\footnotesize\qincorrect{\textbf{Audio-Reasoner / 6}}}\par{\footnotesize\texttt{<speaker6>}\quad \qincorrect{\textbf{neutral}}}\par{\footnotesize\qincorrect{\texttt{<|4.00|>s}}--\qincorrect{\texttt{<|4.25|>s}}} & \textbf{Speech:} ``\qincorrect{Clippers}.''\par\smallskip
\textbf{Tone:}  \\ \addlinespace[4pt]
{\footnotesize\qincorrect{\textbf{Audio-Reasoner / 7}}}\par{\footnotesize\texttt{<speaker7>}\quad \qincorrect{\textbf{neutral}}}\par{\footnotesize\qincorrect{\texttt{<|4.25|>s}}--\qincorrect{\texttt{<|4.25|>s}}} & \textbf{Speech:} ``\qincorrect{Clippers}.''\par\smallskip
\textbf{Tone:}  \\ \addlinespace[4pt]
{\footnotesize\qincorrect{\textbf{Audio-Reasoner / 8}}}\par{\footnotesize\texttt{<speaker8>}\quad \qincorrect{\textbf{neutral}}}\par{\footnotesize\qincorrect{\texttt{<|4.25|>s}}--\qincorrect{\texttt{<|4.25|>s}}} & \textbf{Speech:} ``\qincorrect{Clippers}.''\par\smallskip
\textbf{Tone:}  \\ \addlinespace[4pt]
{\footnotesize\qincorrect{\textbf{Audio-Reasoner / 9}}}\par{\footnotesize\texttt{<speaker9>}\quad \qincorrect{\textbf{unknown}}}\par{\footnotesize\qincorrect{\texttt{<|4.25|>s}}--\qincorrect{\texttt{<|4.25|>s}}} & \textbf{Speech:} ``\qincorrect{Clippers}.''\par\smallskip
\textbf{Tone:}  \\ \addlinespace[4pt]
\end{longtable}
\endgroup

\Needspace{10\baselineskip}
\Needspace{12\baselineskip}
\subsection{Example 12: Premature onset}
\label{app:iemocap_qualitative_000263}
\noindent\textbf{Failure}\hfill{\footnotesize\texttt{IEMOCAP\_train\_concat\_000263\_k3\_2a1b8f8fcc}}\par
\smallskip\noindent Emo-TAG starts the happiness span 1.75 seconds early, creating a 1.25-second overlap with the preceding neutral span.

\begingroup
\small
\setlength{\tabcolsep}{6pt}
\renewcommand{\arraystretch}{1.12}
\setlength{\LTpre}{6pt}\setlength{\LTpost}{8pt}
\begin{longtable}{@{}>{\raggedright\arraybackslash}p{0.2\linewidth}>{\raggedright\arraybackslash}p{\dimexpr0.8\linewidth-2\tabcolsep\relax}@{}}
\noalign{\label{tab:iemocap_qualitative_000263}}
\toprule
\textbf{Source} & \textbf{Emotion--span sequence} \\
\midrule
\endfirsthead
\multicolumn{2}{@{}l@{}}{\textit{Emotion--span summary (continued)}} \\
\toprule
\textbf{Source} & \textbf{Emotion--span sequence} \\
\midrule
\endhead
\midrule
\multicolumn{2}{r@{}}{\textit{Continued on next page}} \\
\endfoot
\bottomrule
\endlastfoot
\textbf{Reference} & \mbox{neutral $[0.00, 4.50]$};\allowbreak\quad \mbox{happiness $[5.00, 6.50]$};\allowbreak\quad \mbox{neutral $[7.25, 8.75]$} \\ \addlinespace[2pt]
\textbf{Emo-TAG} & \mbox{\qcorrect{neutral} $[\qcorrect{0.00}, \qcorrect{4.50}]$};\allowbreak\quad \mbox{\qcorrect{happiness} $[\qincorrect{3.25}, \qcorrect{6.50}]$};\allowbreak\quad \mbox{\qcorrect{neutral} $[\qcorrect{7.25}, \qcorrect{8.75}]$} \\ \addlinespace[2pt]
\textbf{FlamingoNext} & \mbox{\qincorrect{surprise} $[\qincorrect{0.50}, \qincorrect{3.75}]$};\allowbreak\quad \mbox{\qcorrect{neutral} $[\qincorrect{3.00}, \qincorrect{7.00}]$};\allowbreak\quad \mbox{\qincorrect{neutral} $[\qincorrect{7.00}, \qincorrect{7.25}]$};\allowbreak\quad \mbox{\qincorrect{happiness} $[\qincorrect{8.00}, \qincorrect{9.50}]$} \\ \addlinespace[2pt]
\textbf{Audio-Reasoner} & \mbox{\qincorrect{happiness} $[\qcorrect{0.00}, \qincorrect{7.00}]$};\allowbreak\quad \mbox{\qcorrect{neutral} $[\qincorrect{7.00}, \qincorrect{9.00}]$} \\ \addlinespace[2pt]
\end{longtable}
\endgroup

\begingroup
\small
\setlength{\tabcolsep}{6pt}
\renewcommand{\arraystretch}{1.12}
\setlength{\LTpre}{6pt}\setlength{\LTpost}{8pt}
\begin{longtable}{@{}>{\raggedright\arraybackslash}p{0.32\linewidth}>{\raggedright\arraybackslash}p{\dimexpr0.68\linewidth-2\tabcolsep\relax}@{}}
\noalign{\label{tab:iemocap_qualitative_full_000263}}
\toprule
\textbf{Span / emotion / time} & \textbf{Speech and tone} \\
\midrule
\endfirsthead
\multicolumn{2}{@{}l@{}}{\textit{Example 12: annotations (continued)}} \\
\toprule
\textbf{Span / emotion / time} & \textbf{Speech and tone} \\
\midrule
\endhead
\midrule
\multicolumn{2}{r@{}}{\textit{Continued on next page}} \\
\endfoot
\bottomrule
\endlastfoot
\multicolumn{2}{@{}l@{}}{\qualmodel{Reference}} \\*
{\footnotesize\textbf{Reference / 1}}\par{\footnotesize\texttt{<speaker1>}\quad \textbf{neutral}}\par{\footnotesize\texttt{<|0.00|>s}--\texttt{<|4.50|>s}} & \textbf{Speech:} ``Wow. Well, that's really important, obviously. But, um --''\par\smallskip
\textbf{Tone:}  \\ \addlinespace[4pt]
{\footnotesize\textbf{Reference / 2}}\par{\footnotesize\texttt{<speaker1>}\quad \textbf{happiness}}\par{\footnotesize\texttt{<|5.00|>s}--\texttt{<|6.50|>s}} & \textbf{Speech:} ``Yeah. We'll have so much fun.''\par\smallskip
\textbf{Tone:}  \\ \addlinespace[4pt]
{\footnotesize\textbf{Reference / 3}}\par{\footnotesize\texttt{<speaker2>}\quad \textbf{neutral}}\par{\footnotesize\texttt{<|7.25|>s}--\texttt{<|8.75|>s}} & \textbf{Speech:} ``Well We will see them this year.''\par\smallskip
\textbf{Tone:}  \\ \addlinespace[4pt]
\multicolumn{2}{@{}l@{}}{\qualmodel{Emo-TAG}} \\*
{\footnotesize\textbf{Emo-TAG / 1}}\par{\footnotesize\texttt{<speaker1>}\quad \qcorrect{\textbf{neutral}}}\par{\footnotesize\qcorrect{\texttt{<|0.00|>s}}--\qcorrect{\texttt{<|4.50|>s}}} & \textbf{Speech:} ``\qcorrect{Wow}. \qcorrect{Well} \qcorrect{that's} \qcorrect{really} \qcorrect{important} \qcorrect{obviously}, \qcorrect{but} \qcorrect{um}--''\par\smallskip
\textbf{Tone:} Calm and composed with slight hesitation, moderate volume, steady pitch, and steady pace \\ \addlinespace[4pt]
{\footnotesize\textbf{Emo-TAG / 2}}\par{\footnotesize\texttt{<speaker1>}\quad \qcorrect{\textbf{happiness}}}\par{\footnotesize\qincorrect{\texttt{<|3.25|>s}}--\qcorrect{\texttt{<|6.50|>s}}} & \textbf{Speech:} ``\qcorrect{Yeah}, \qcorrect{we'll} \qcorrect{have} \qcorrect{so} \qcorrect{much} \qcorrect{fun}.''\par\smallskip
\textbf{Tone:} Bright, enthusiastic delivery with consistently high pitch, fast pace, and strong volume, conveying eager anticipation through animated delivery \\ \addlinespace[4pt]
{\footnotesize\textbf{Emo-TAG / 3}}\par{\footnotesize\texttt{<speaker2>}\quad \qcorrect{\textbf{neutral}}}\par{\footnotesize\qcorrect{\texttt{<|7.25|>s}}--\qcorrect{\texttt{<|8.75|>s}}} & \textbf{Speech:} ``\qcorrect{Well}, \qincorrect{we'll} \qcorrect{see} \qcorrect{them} \qcorrect{this} \qcorrect{year}.''\par\smallskip
\textbf{Tone:} Steady mid-range pitch with slight rising intonation on 'we'll' and flat dynamics throughout, indicating mild curiosity without emotional emphasis or vocal strain \\ \addlinespace[4pt]
\multicolumn{2}{@{}l@{}}{\qualmodel{FlamingoNext}} \\*
{\footnotesize\textbf{FlamingoNext / 1}}\par{\footnotesize\texttt{<Speaker A>}\quad \qincorrect{\textbf{surprise}}}\par{\footnotesize\qincorrect{\texttt{<|0.50|>s}}--\qincorrect{\texttt{<|3.75|>s}}} & \textbf{Speech:} ``\qcorrect{Wow}''\par\smallskip
\textbf{Tone:}  \\ \addlinespace[4pt]
{\footnotesize\textbf{FlamingoNext / 2}}\par{\footnotesize\texttt{<Speaker A>}\quad \qcorrect{\textbf{neutral}}}\par{\footnotesize\qincorrect{\texttt{<|3.00|>s}}--\qincorrect{\texttt{<|7.00|>s}}} & \textbf{Speech:} ``\qcorrect{well} \qcorrect{that's} \qcorrect{really} \qcorrect{important} \qcorrect{obviously} \qcorrect{but} \qcorrect{um} \qcorrect{yeah} \qcorrect{we'll} \qcorrect{have} \qcorrect{so} \qcorrect{much} \qcorrect{fun} \qincorrect{ever}''\par\smallskip
\textbf{Tone:}  \\ \addlinespace[4pt]
{\footnotesize\qincorrect{\textbf{FlamingoNext / 3}}}\par{\footnotesize\texttt{<Speaker A>}\quad \qincorrect{\textbf{neutral}}}\par{\footnotesize\qincorrect{\texttt{<|7.00|>s}}--\qincorrect{\texttt{<|7.25|>s}}} & \textbf{Speech:} ``\qincorrect{ever}''\par\smallskip
\textbf{Tone:}  \\ \addlinespace[4pt]
{\footnotesize\textbf{FlamingoNext / 4}}\par{\footnotesize\texttt{<Speaker B>}\quad \qincorrect{\textbf{happiness}}}\par{\footnotesize\qincorrect{\texttt{<|8.00|>s}}--\qincorrect{\texttt{<|9.50|>s}}} & \textbf{Speech:} ``\qcorrect{Well} \qincorrect{we'll} \qcorrect{see} \qcorrect{them} \qcorrect{this} \qcorrect{year}''\par\smallskip
\textbf{Tone:}  \\ \addlinespace[4pt]
\multicolumn{2}{@{}l@{}}{\qualmodel{Audio-Reasoner}} \\*
{\footnotesize\textbf{Audio-Reasoner / 1}}\par{\footnotesize\texttt{<speaker1>}\quad \qincorrect{\textbf{happiness}}}\par{\footnotesize\qcorrect{\texttt{<|0.00|>s}}--\qincorrect{\texttt{<|7.00|>s}}} & \textbf{Speech:} ``\qcorrect{Wow}. \qcorrect{Well}, \qcorrect{that's}, \qcorrect{really} \qcorrect{important} \qcorrect{obviously}. \qcorrect{But} \qcorrect{um}, \qcorrect{yeah}, \qcorrect{we'll} \qcorrect{have} \qcorrect{so} \qcorrect{much} \qcorrect{fun}.''\par\smallskip
\textbf{Tone:}  \\ \addlinespace[4pt]
{\footnotesize\textbf{Audio-Reasoner / 2}}\par{\footnotesize\texttt{<speaker2>}\quad \qcorrect{\textbf{neutral}}}\par{\footnotesize\qincorrect{\texttt{<|7.00|>s}}--\qincorrect{\texttt{<|9.00|>s}}} & \textbf{Speech:} ``\qcorrect{Well}, \qincorrect{we'll} \qcorrect{see} \qcorrect{them} \qcorrect{this} \qcorrect{year}.''\par\smallskip
\textbf{Tone:}  \\ \addlinespace[4pt]
\end{longtable}
\endgroup

\clearpage

\section{Limitations}
\label{app:limitations}

Our training data are imbalanced. In MELD, neutral emotion accounts for about 42\%
of training samples, compared with 3\% for fear. The evaluation split is
similarly skewed: 47\% neutral and 1.5\% fear. Filtering preserves much of
this imbalance.
We did not manually verify or evaluate tone descriptions due to
limited resources and the difficulty in defining a consistent quality criterion.
Finally, we focus on eight basic emotions. Extending the label set to
finer distinctions, such as those in emotion wheels
\citep{plutchik1980general}, could make the approach more useful in real-world
settings. We leave these questions for future work.


\clearpage
\section{Prompts}
\label{app:prompts}

The prompts below were used for training, context ablations, evaluation,
and tone extraction and refinement. Full-context training includes speech,
timestamps, tone, and emotion. The partial-context settings in
Table~\ref{tab:context_loss_ablation} vary the inclusion of speech and tone;
all retain timestamps.

\subsection{Model Training and Context Ablations}
\label{app:training_prompts}

\noindent\begin{minipage}{\linewidth}
\Needspace{8\baselineskip}
\subsubsection{Emotion Only}
\label{app:prompt_emotion_only}
\Needspace{5\baselineskip}
\noindent\textit{System prompt.}\par\nopagebreak
\noindent\begin{minipage}{\linewidth}
\begin{lstlisting}[style=appendixprompt]
You are an affective audio expert in emotion recognition.

Guidelines:
- Focus only on the audio.
- Use timestamp tokens exactly as provided.
- Follow the requested output format exactly.
\end{lstlisting}
\end{minipage}\par

\Needspace{5\baselineskip}
\noindent\textit{User prompt.}\par\nopagebreak
\noindent\begin{minipage}{\linewidth}
\begin{lstlisting}[style=appendixprompt]
Analyze the input audio and generate one structured annotation for emotion recognition.

Output the final emotion label for each segement.

Output exactly in this format:
From <|0.00|>s to <|0.25|>s,
[emotion] emotion label.

Requirements:
- Use valid timestamp tokens such as <|0.00|>, <|0.25|>, <|0.50|>, and so on.
- The emotion label must be the final affective interpretation of the audio.
\end{lstlisting}
\end{minipage}\par


\end{minipage}\par\medskip

\noindent\begin{minipage}{\linewidth}
\Needspace{8\baselineskip}
\subsubsection{Speech and Emotion}
\label{app:prompt_speech_emotion}
\Needspace{5\baselineskip}
\noindent\textit{System prompt.}\par\nopagebreak
\noindent\begin{minipage}{\linewidth}
\begin{lstlisting}[style=appendixprompt]
You are an affective audio expert in emotion recognition.

Guidelines:
- Focus only on the audio.
- Infer speech before emotion.
- Use timestamp tokens exactly as provided.
- Follow the requested output format exactly.
\end{lstlisting}
\end{minipage}\par

\Needspace{5\baselineskip}
\noindent\textit{User prompt.}\par\nopagebreak
\noindent\begin{minipage}{\linewidth}
\begin{lstlisting}[style=appendixprompt]
Analyze the input audio and generate one structured annotation for emotion recognition.

First determine the speaker’s speech, then output the final emotion label.

Output exactly in this format:
[speech] <speakerX>: "transcript" from <|0.00|>s to <|0.25|>s,
[emotion] emotion label.

Requirements:
- Use valid timestamp tokens such as <|0.00|>, <|0.25|>, <|0.50|>, and so on.
- The emotion label must be the final affective interpretation of the audio.
\end{lstlisting}
\end{minipage}\par

\end{minipage}\par\medskip

\noindent\begin{minipage}{\linewidth}
\Needspace{8\baselineskip}
\subsubsection{Tone and Emotion}
\label{app:prompt_tone_emotion}
\Needspace{5\baselineskip}
\noindent\textit{System prompt.}\par\nopagebreak
\noindent\begin{minipage}{\linewidth}
\begin{lstlisting}[style=appendixprompt]
You are an affective audio expert in emotion recognition.

Guidelines:
- Focus only on the audio.
- Infer tone before emotion.
- Use timestamp tokens exactly as provided.
- Follow the requested output format exactly.
\end{lstlisting}
\end{minipage}\par

\Needspace{5\baselineskip}
\noindent\textit{User prompt.}\par\nopagebreak
\noindent\begin{minipage}{\linewidth}
\begin{lstlisting}[style=appendixprompt]
Analyze the input audio and generate one structured annotation for emotion recognition.

First determine the speaker’s tone, then output the final emotion label.

Output exactly in this format:
[tone] <speakerX>: "tone description" from <|0.00|>s to <|0.25|>s,
[emotion] emotion label.

Requirements:
- Use valid timestamp tokens such as <|0.00|>, <|0.25|>, <|0.50|>, and so on.
- The emotion label must be the final affective interpretation of the audio.
\end{lstlisting}
\end{minipage}\par

\end{minipage}\par\medskip

\noindent\begin{minipage}{\linewidth}
\Needspace{8\baselineskip}
\subsubsection{Full Context}
\label{app:prompt_full_context}
\Needspace{5\baselineskip}
\noindent\textit{System prompt.}\par\nopagebreak
\noindent\begin{minipage}{\linewidth}
\begin{lstlisting}[style=appendixprompt]
You are an affective audio expert in emotion recognition.

Guidelines:
- Focus only on the audio.
- Infer speech and tone before emotion.
- Use timestamp tokens exactly as provided.
- Follow the requested output format exactly.
\end{lstlisting}
\end{minipage}\par

\Needspace{5\baselineskip}
\noindent\textit{User prompt.}\par\nopagebreak
\noindent\begin{minipage}{\linewidth}
\begin{lstlisting}[style=appendixprompt]
Analyze the input audio and generate one structured annotation for emotion recognition.

First determine the speaker’s speech and tone, then output the final emotion label.

Output exactly in this format:
[speech] <speakerX>: "transcript" from <|0.00|>s to <|0.25|>s, [tone] tone description, [emotion] emotion label.

Requirements:
- Use valid timestamp tokens such as <|0.00|>, <|0.25|>, <|0.50|>, and so on.
- The tone description must describe the speaker’s audible vocal characteristics.
- The emotion label must be the final affective interpretation of the audio.
\end{lstlisting}
\end{minipage}\par
\end{minipage}\par\medskip

\clearpage
\subsection{Model Evaluation}
\label{app:evaluation_prompts}

We keep each model's original system prompt and use the following user
prompt for evaluation.

\noindent\begin{minipage}{\linewidth}
\begin{lstlisting}[style=appendixprompt]
Listen to the audio and identify each spoken utterance.

For each utterance, provide:
- the spoken words
- the start and end time in seconds
- the speaker's emotion

Emotion labels: neutral, happiness, sadness, anger, surprise, fear, disgust, frustration.
\end{lstlisting}
\end{minipage}\par

\Needspace{8\baselineskip}
\subsubsection{Qwen3 Evaluation Extractor}
\label{app:prompt_evaluation_extractor}
\Needspace{5\baselineskip}
\noindent\textit{System prompt.}\par\nopagebreak
\noindent\begin{minipage}{\linewidth}
\begin{lstlisting}[style=appendixprompt]
You are an information extraction system. You convert a model's free-form audio-emotion answer into JSON for evaluation. Extract only information stated or clearly implied by the model answer. Do not use outside knowledge and do not guess missing fields.
\end{lstlisting}
\end{minipage}\par

\Needspace{5\baselineskip}
\noindent\textit{User prompt.}\par\nopagebreak
\noindent\begin{minipage}{\linewidth}
\begin{lstlisting}[style=appendixprompt]
Extract structured fields from the model answer below.

Allowed emotion labels:
neutral, frustration, anger, sadness, happiness, surprise, fear, disgust

Return ONLY valid JSON with this schema:
{
  "segments": [
    {
      "speech": "transcribed speech or null",
      "emotion": "one allowed emotion label or null",
      "start_sec": ,
      "end_sec": 
    }
  ]
}

Rules:
- If there is no explicit or clearly implied emotion, use null for emotion.
- If there is no timestamp or time range, use null for start_sec and end_sec.
- If there are multiple utterances/spans, return multiple segments in order.
- Convert timestamp tokens like <|0.50|> to seconds as numbers.
- Do not invent timestamps, emotions, or speech.
- Do not include markdown fences, comments, or explanations.

Model answer:
{model_output}
\end{lstlisting}
\end{minipage}\par

\clearpage
\subsection{Tone Extraction}
\label{app:tone_extraction_prompts}

\Needspace{8\baselineskip}
\subsubsection{Flamingo Prompt}
\label{app:prompt_flamingo}
\noindent\begin{minipage}{\linewidth}
\begin{lstlisting}[style=appendixprompt]
Reason step by step with timestamps before answering. Listen to the speech and extract the spoken words and vocal tone. Do not infer the emotion label. For the tone be descreptive, mention emotional cues such as pitch, loudness, speaking rate, pauses, tension, laughter, pace, delivery, sighing, uncertainty, crying, shouting, whispering, or breathiness if audible. Focus on how the words are spoken, not just what is spoken. Be descriptive but do not add details not supported by the audio.
\end{lstlisting}
\end{minipage}\par

\Needspace{8\baselineskip}
\subsubsection{Qwen3 Tone Extractor}
\label{app:prompt_tone_extractor}
\noindent\begin{minipage}{\linewidth}
\begin{lstlisting}[style=appendixprompt]
Extract the spoken speech and vocal tone from below.

Return ONLY valid JSON with exactly these keys:
{
  "speech": "...",
  "tone": "..."
}

Rules:
- Use the full text, including any <think> trace.
- The speech is the transcript/spoken words if present.
- The speech is usually scattered in segments between quotation marks; concatenate quoted speech spans without repetition.
- Focus on the main speaker only, ignore background laughter or other speakers if present.
- The tone should summarize the vocal/acoustic delivery, not the semantic content.
- Focus on the emotional and acoustic cues, including pitch, loudness, pace, pauses, tension, laughter, sighing, crying, shouting, whispering, breathiness, hesitation, trembling, or vocal strain, if present.
- Do not add unnecessary explanation or preamble.
- Do not add details that are not supported by the text.
- If the tone is clearly inconsistent with the given emotion, output "wrong tone" as the tone.

Emotion:
{emotion}

Text:
{flamingo_raw}
\end{lstlisting}
\end{minipage}\par

\clearpage
\subsection{Tone Refinement}
\label{app:tone_refinement_prompt}

\Needspace{6\baselineskip}
\noindent\textbf{ChatGPT prompt.}\par
\noindent\begin{minipage}{\linewidth}
\begin{lstlisting}[style=appendixprompt]
You are given a tone description, an emotion label, speaker-specific
reference ranges, and acoustic features from five temporal segments.

Rewrite the tone description as one concrete and informative sentence.

Requirements:

- Ground every stated cue in the supplied acoustic features.

- Interpret pitch, loudness, and other features relative to the speaker’s
   low, mean, and high reference values.

- Focus on emotion-relevant characteristics: pitch level and variation,
   loudness and its variation, speaking pace and pauses, voice quality, and
   changes across the utterance.

- Use the existing tone description only as a semantic reference and correct
   it when it conflicts with the acoustic evidence.

- Do not describe every segment separately. Summarize only meaningful changes
   across the five segments.

- Follow this structure:

   “[overall delivery] with [pitch], [loudness], [pacing], and [voice quality];
   [meaningful acoustic descriptions].”

- Use 15–30 words, do not explicitly name the emotion, and return only the refined tone description.

Emotion label:

{emotion}

Existing tone description:

{original_tone}

Speaker-specific reference values:

{speaker_reference_values}

Segments acoustic features:

{segment_features}
\end{lstlisting}
\end{minipage}\par


\end{document}